\documentclass[traditabstract]{aa}

\usepackage{caption}

\usepackage{graphicx}
\usepackage{txfonts}
\usepackage{natbib}
\usepackage{enumerate}
\usepackage{bm}
\usepackage[breaklinks]{hyperref}
\hypersetup{
  colorlinks=true,
  linkcolor=blue,
  citecolor=blue,
  urlcolor=blue,
}

\usepackage{etoolbox}

\bibpunct{(}{)}{;}{a}{}{,} 

\newcommand{\barolo}{$^\mathrm{3D}$Barolo}

\newcommand{\Nohi}{N0$_\mathrm{HI}$}
\newcommand{\woco}{W0$_\mathrm{CO}$}

\newcommand{\ico}{$I_\mathrm{CO}$}
\newcommand{\xco}{$X_\mathrm{CO}$}
\newcommand{\vsys}{$V_\mathrm{sys}$}

\newcommand{\coone}{$^{12}$CO(1--0)}
\newcommand{\cotwo}{$^{12}$CO(2--1)}

\newcommand{\ci}{C{\sc i}}
\newcommand{\cii}{C{\sc ii}}
\newcommand{\ha}{H$\alpha$}

\newcommand{\kms}{km\,s$^{-1}$}
\newcommand{\jykms}{Jy\,km\,s$^{-1}$}
\newcommand{\jykmsbm}{Jy\,km\,s$^{-1}$\,beam$^{-1}$}

\newcommand{\mjybm}{mJy\,beam$^{-1}$}
\newcommand{\kkms}{K\,km\,s$^{-1}$}
\newcommand{\kkmspc}{K\,km\,s$^{-1}$\,pc$^{2}$}
\newcommand{\htwo}{H$_2$}
\newcommand{\Nhtwo}{$N_\mathrm{H2}$}
\newcommand{\Nhtwodgr}{$N_\mathrm{H2}^\mathrm{DGR}$}

\newcommand{\nee}{$n_{\rm e}$}
\newcommand{\Nco}{$N_\mathrm{CO}$}
\newcommand{\Nhi}{$N_\mathrm{HI}$}
\newcommand{\ngas}{$N_\mathrm{gas}$}

\newcommand{\tb}{$T_{\rm B}$}
\newcommand{\tbg}{$T_\mathrm{BG}$}
\newcommand{\Tex}{$T_\mathrm{ex}$}
\newcommand{\huwco}{$W_{10}$}

\newcommand{\av}{$A_V$}
\newcommand{\avmuse}{$A_V^\mathrm{MUSE}$}
\newcommand{\avhicrit}{$A_V^\mathrm{HIcrit}$}
\newcommand{\avcothresh}{$A_V^\mathrm{CO,thresh}$}
\newcommand{\ebv}{$E(B-V)$}
\newcommand{\spit}{\textit{Spitzer}}
\newcommand{\hers}{\textit{Herschel}}
\newcommand{\hst}{\textit{HST}}

\newcommand{\logoh}{12$+$log(O/H)}

\newcommand{\deltagdr}{$\delta_\mathrm{GDR}$}
\newcommand{\deltadgr}{$\delta_\mathrm{DGR}$}
\newcommand{\mytau}{$\tau_\mathrm{160}$}
\newcommand{\tauv}{$\tau_\mathrm{V}$}
\newcommand{\tc}{$T_\mathrm{cool}$}
\newcommand{\tw}{$T_\mathrm{warm}$}
\newcommand{\td}{$T_\mathrm{dust}$}
\newcommand{\khi}{$k_\mathrm{HI}$}
\newcommand{\kco}{$k_\mathrm{CO}$}
\newcommand{\mhtwo}{M$_{\rm H2}$}

\newcommand{\mstar}{M$_{\rm star}$}

\newcommand{\cmtwo}{cm$^{-2}$}
\newcommand{\cmthree}{cm$^{-3}$}

\newcommand{\hi}{H{\sc i}}
\newcommand{\hii}{H{\sc ii}}

\newcommand{\micron}{$\mu$m}
\newcommand{\zzsun}{${\mathrm Z/Z}_\odot$}
\newcommand{\zsun}{$\mathrm{Z}_\odot$}
\newcommand{\msun}{$M_\odot$}
\newcommand{\msunyr}{$M_\odot$\,yr$^{-1}$}
\newcommand{\uv}{$u\varv$}

\begin{document}

\title{Gas, dust, and the CO-to-molecular gas conversion factor in low-metallicity starbursts
\thanks{This paper makes use of the following ALMA data: ADS/JAO.ALMA\#2018.1.00219.S. ALMA is a partnership of ESO (representing its member states), NSF (USA) and NINS (Japan), together with NRC (Canada), MOST and ASIAA (Taiwan), and KASI (Republic of Korea), in cooperation with the Republic of Chile. The Joint ALMA Observatory is operated by ESO, AUI/NRAO and NAOJ.}}

\author{L.~K. Hunt \inst{\ref{inst:hunt}}
\and
F. Belfiore \inst{\ref{inst:hunt}}
\and
F. Lelli \inst{\ref{inst:hunt}}
\and
B.~T. Draine \inst{\ref{inst:draine}}
\and
A. Marasco \inst{\ref{inst:padova},\ref{inst:hunt}}
\and
S. Garc{\'i}a-Burillo \inst{\ref{inst:santi}}
\and
G. Venturi \inst{\ref{inst:venturi}, \ref{inst:hunt}}
\and
F. Combes \inst{\ref{inst:combes}}
\and
A. Wei\ss \inst{\ref{inst:menten}}
\and
C. Henkel \inst{\ref{inst:menten},\ref{inst:henkel_b}}
\and
K.~M. Menten \inst{\ref{inst:menten}}
\and
F. Annibali \inst{\ref{inst:annibali}}
\and
V. Casasola \inst{\ref{inst:casasola}}
\and
M. Cignoni \inst{\ref{inst:cignoni_a},\ref{inst:cignoni_b},\ref{inst:cignoni_c}}
\and
A. McLeod \inst{\ref{inst:mcleod_a},\ref{inst:mcleod_b}}
\and
M. Tosi \inst{\ref{inst:annibali}}
\and
M. Beltr\'an \inst{\ref{inst:hunt}}
\and
A. Concas \inst{\ref{inst:concas}}
\and
G. Cresci \inst{\ref{inst:hunt}}
\and
M. Ginolfi \inst{\ref{inst:ginolfi}}
\and
N. Kumari \inst{\ref{inst:kumari}}
\and
F. Mannucci \inst{\ref{inst:hunt}}
}

\offprints{L. K. Hunt}
\institute{INAF - Osservatorio Astrofisico di Arcetri, Largo E. Fermi, 5, 50125, Firenze, Italy
\label{inst:hunt}
\email{leslie.hunt@inaf.it}
\and
Princeton University Observatory, Peyton Hall, Princeton, NJ 08544-1001, USA \label{inst:draine}
\and
INAF - Osservatorio Astronomico di Padova, Vicolo dell'Osservatorio, 5, 35122 Padova, Italy
\label{inst:padova}
\and
Observatorio Astron\'omico Nacional (OAN)-Observatorio de Madrid,
Alfonso XII, 3, 28014-Madrid, Spain
\label{inst:santi}
\and
Instituto de Astrof{\'i}sica, Facultad de F{\'i}sica, Pontificia Universidad Cat{\'o}lica de Chile, Casilla 306, Santiago 22, Chile
\label{inst:venturi}
\and
Observatoire de Paris, LERMA, College de France, CNRS, PSL, Sorbonne University, F-75014, Paris, France
\label{inst:combes}
\and
Max-Planck-Institut f\"ur Radioastronomie, Auf dem H\"ugel 69, 53121 Bonn, Germany
\label{inst:menten}
\and
Astronomy Department, King Abdulaziz University, P.O. Box 80203, Jeddah, Saudia Arabia
\label{inst:henkel_b}
\and
INAF - Osservatorio di Astrofisica e Scienza dello Spazio, 
Via Piero Gobetti, 93/3, 40129 Bologna, Italy
\label{inst:annibali}
\and
INAF - Istituto di Radioastronomia, 
Via Piero Gobetti, 93/3, 40129 Bologna, Italy
\label{inst:casasola}
\and
Department of Physics - University of Pisa, Largo B. Pontecorvo 3, 56127, Pisa, Italy 
\label{inst:cignoni_a}
\and
INFN, Largo B. Pontecorvo 3, 56127, Pisa, Italy
\label{inst:cignoni_b}
\and
INAF - Osservatorio Astronomico di Capodimonte, Via Moiariello 16, 80131, Napoli, Italy
\label{inst:cignoni_c}
\and
Centre for Extragalactic Astronomy, Department of Physics, Durham University, South Road, Durham DH1 3LE, UK
\label{inst:mcleod_a}
\and
Institute for Computational Cosmology, Department of Physics, University of Durham, South Road, Durham DH1 3LE, UK
\label{inst:mcleod_b}
\and
European Southern Observatory, Karl-Schwarzschild-Strasse 2, D-85748 Garching bei M\"unchen, Germany
\label{inst:concas}
\and
Dipartimento di Astronomia e Scienza dello Spazio, Universit\`a degli Studi di Firenze, Largo E. Fermi 2, 50125 Firenze, Italy
\label{inst:ginolfi}
\and
AURA for the European Space Agency, Space Telescope Science Institute, 3700 San Martin Drive, Baltimore, MD 21218, USA
\label{inst:kumari}
}

   \date{draft version \today}

   \titlerunning{\xco\ in low-metallicity starbursts}
   \authorrunning{Hunt et al.}

\abstract{The factor relating CO emission to molecular hydrogen column density, \xco, is still subject to uncertainty, 
in particular at low metallicity.
Here, to quantify \xco\ at two different spatial resolutions,
we exploit a dust-based method together with ALMA 12-m and ACA data and \hi\ maps of 
three nearby metal-poor starbursts, NGC\,625, NGC\,1705, and NGC\,5253.
Dust opacity at 250\,pc resolution is derived based on
dust temperatures estimated by fitting two-temperature modified blackbodies to \hers\ PACS data.
By using the \hi\ maps, we are then able to estimate dust-to-gas ratios in the atomic-gas dominated regions,
and infer total gas column densities and \htwo\ column densities as the difference with \hi.
Finally, from the ACA CO(1--0) maps, we derive \xco. 
We use a similar technique with 40\,pc ALMA 12-m data for the three galaxies, but instead derive
dust attenuation at 40\,pc resolution from reddening maps based on VLT/MUSE data. 
At 250\,pc resolution, we find \xco\ 
$\sim\,10^{22} - 10^{23}$\,\cmtwo/\kkms, 5-1000 times the Milky Way value,
with much larger values than would be expected from a simple metallicity dependence. 
Instead at 40\,pc resolution, \xco\ again shows large variation, but is roughly consistent 
with a power-law metallicity dependence, given the $Z\,\sim\,$1/3\,\zsun\ metal abundances of our targets.
The large scatter in both estimations could imply additional parameter dependence, that we have investigated
by comparing \xco\ with the observed velocity-integrated brightness temperatures, \ico, as
predicted by recent simulations.
Indeed, larger \xco\ is significantly correlated with smaller \ico,
but with slightly different slopes and normalizations than predicted by theory.
Such behavior can be attributed to the increasing fraction of CO-faint (or dark) \htwo\ gas with 
lower spatial resolution (larger beams).
This confirms the idea the \xco\ is multi-variate, depending not only on metallicity
but also on CO brightness temperature and beam size.
Future work is needed to consolidate these empirical results by sampling galaxies with different metal
abundances observed at varying spatial resolutions.
}


\maketitle


\section{Introduction}
\label{sec:intro}

From early cosmic times, molecular clouds are the cradles of star formation.
However, molecular gas, \htwo, is not easily traced observationally. 
The rotational levels of \htwo\ are widely separated, with negligible
thermal excitation in cold clouds.  
In addition, \htwo\ is a
homonuclear diatomic molecule with no permanent dipole moment so
that there are no rovibrational electric dipole transitions.

Luckily, molecular gas is not pure \htwo, but also contains heavy, trace elements.
The 
abundances of oxygen and carbon in the interstellar
medium (ISM) enable the formation of carbon monoxide (CO) within dense, cold, molecular clouds.
Due to its low excitation energy in the ground-state rotational transition
($h\nu/k_{B}\,\approx\,5.5$\,K), and its low critical density
($\sim 1400$\,\cmthree, for a kinetic temperature $T\,=\,25$\,K),
CO emission can be easily excited even in cold molecular clouds\footnote{This assumes
an optically-thin line; if optically thick, as common for CO, the effective critical density would be even lower.}.
Thus, the lowest-order CO transition $J\,=\,1\rightarrow0$ at 2.6\,mm, fortunately
located within a fairly transparent atmospheric window,
has become a common tracer of \htwo\ in the Milky Way and external galaxies. 

Nevertheless, CO has its own problems.
Under typical ISM conditions, CO emission is optically thick, so that even when its abundance
relative to \htwo\ is known, CO cannot be used to directly trace molecular column density or mass
along the line of sight.
Thus, it is standard to connect the column (or mass) density of molecular gas \Nhtwo\ 
to observed \coone\ velocity-integrated brightness temperature, \ico, via a CO-to-\htwo\ conversion factor, \xco:
\begin{equation}
N_\mathrm{H2}\,=\,X_\mathrm{CO}\ I_\mathrm{CO}\ ,
\label{eqn:xcodef}
\end{equation}
where the \htwo\ column density \Nhtwo\ is in units of \cmtwo,
and \ico\ in units of \kkms.
The existence of a nearly constant \xco\ at solar-like 
metallicities where \htwo\ and CO coexist to a large extent 
relies on the assumption that the optically-thick CO emission comes from virialized clouds that do not overlap.
Thus \xco\ is an empirical shortcut essentially to count clouds along the line of sight. 

Although the conversion factor \xco\ relating CO emission to \htwo\ mass \mhtwo\
is relatively constant across the Galaxy \citep[e.g.,][]{sanders84,bolatto13},
at low metallicity and in starburst galaxies, its value is still uncertain.
Inferring \xco\ in metal-poor environments has been the subject of an enormous
effort, both observational and theoretical \citep[see][for an extensive review]{bolatto13}.
The problem lies in the complexity of the relation between CO emission and \htwo.
Simulations and observational work have suggested that 
the amount of CO and its emission 
depends primarily on shielding, i.e., the protection of CO molecules from dissociating 
far-ultraviolet (FUV) radiation.
Self-shielding of \htwo\ is highly efficient, meaning that the photodissociation
rate of \htwo\ is typically very small wherever there is a significant amount
of \htwo.
On the other hand, 
self-shielding of CO and cross-shielding of CO by \htwo\ are inefficient processes
\citep[e.g.,][]{gong18}.
Consequently, CO needs 
additional protection by dust. 

Photodissociation of CO tends to erode the CO-emitting regions of molecular clouds,
even at solar metallicity.
At sub-solar abundances, 
stars are hotter and have higher hard photon fluxes, enhancing the effect;  
thus the regions where \htwo\ resides 
are much larger than those where 
CO and \htwo\ coexist \citep[e.g.,][]{vandishoeck88}.
Consequently, there are extended regions of
so-called ``CO-faint'' (or dark) gas in which the dominant
forms of carbon are 
ionized (C$^+$) or neutral carbon \citep[e.g.,][]{wolfire10}.
The CO-emitting regions in low-metallicity environments are expected to be much smaller (a few pc) than
they would be in more typical, metal-rich, conditions
\citep[e.g.,][]{bolatto99,bolatto13}.
Indeed, the small size of CO-emitting clouds at low metallicity has 
been verified observationally in the Magellanic clouds
\citep[e.g.,][]{pak98,indebetouw13,tokuda21},
and in nearby dwarf galaxies \citep[e.g.,][]{rubio15,schruba17,shi20}.

Early attempts at estimating \xco\ theoretically relied on models of photodissociation regions 
which are able to resolve  the thermal and chemical structure within an individual cloud
\citep[e.g.,][]{wolfire93,bell06}.
Later cloud models \citep[e.g.,][]{glover11,shetty11,glover16}, 
based on hydrodynamical simulations with more complex geometries and density structure, 
confirmed that \xco\ varies with visual extinction \av, or depth within the cloud.
In particular, there was evidence for an extinction threshold above which clouds
become CO bright \citep[see also][]{bell06}.
Metallicity plays a fundamental role
because 
lower dust abundance provides less dust shielding against CO photodissociation,
resulting in more extended regions of CO-faint gas.
Cosmic-ray ionization is also emerging as an important parameter \citep[e.g.,][]{bisbas21}.

Global models of \xco\ in galaxies using hydrodynamical simulations 
\citep[e.g.,][]{feldmann12,nara12}
suggest that \xco\ depends on metallicity, dust extinction, and \htwo\ column density \Nhtwo;
in addition, there may be a dependence on observational spatial scale \citep[see also][]{rubio93} 
and velocity-integrated CO brightness temperature (\tb), \ico.
These two latter trends reflect the fact that 
\xco\ tends to be lower when estimated from regions with higher CO surface brightness; thus
from an observational perspective, smaller beams generally give lower \xco. 
In fact, CO observations with beam sizes of a few pc that resolve 
the central regions containing CO in low-metallicity molecular clouds
tend to result in \xco\ values that are roughly consistent with \xco\ for the Milky Way 
\citep[e.g.,][]{bolatto08}.

The main problem with these global models for \xco\ is the determination of the physical conditions
and line-emission properties on sub-grid scales. 
More recent models overcome this limitation through
a better match of large-scale simulations and the physics
and chemistry of small-scale resolved cloud structure.
In the solar metallicity models of \citet{gong17,gong18,gong20},
the average \xco\ increases by a factor of $\sim$2 as the observational beam size
grows from 1 to 100\,pc.
Moreover, the CO-dark \htwo\ fraction ranges from $\sim 30-80$\%, and is anticorrelated
with visual extinction.
With models over a range of metallicities, \citet{hu22}
find that the parsec-scale \xco\ is roughly the Galactic value,
independently of metallicity once dust shielding becomes effective.
\citet{hu22} also find that CO emission becomes more optically thin at lower 
metallicity \citep[see][for a tentative observational verification]{hunt17}.
The \citet{hu22} simulations also show that although
most CO emission originates from dense gas with low \xco, 
most of the cloud area is filled by diffuse gas with high \xco.
This leads to an anti-correlation of \xco\ and \ico, also implying
that \xco\ is leveraged by beam averaging, and tends to be biased
toward the 
highest density regions of the molecular gas. 

Here we attempt to test theoretical predictions of \xco\ at low metallicity by observationally quantifying
 trends of \xco\ with dust content and spatial resolution.
Using CO maps of three iconic low-metallicity dwarf galaxies observed
with 
the ALMA 12-m and compact (ACA, 7-m) arrays, at $\sim$40\,pc and $\sim$250\,pc resolutions, respectively,
we first estimate dust content at $\sim$250\,pc resolution.
To do this, we calculate the dust opacity at 160\,\micron\ \mytau\ from \hers\ observations, 
and compare it with \hi\ and CO maps at the same resolution.
Then, we infer dust content at $\sim$40\,pc resolution from VLT/MUSE maps of \ebv\
and compare it with the higher resolution CO maps.
Finally, we compute \xco\ at both resolutions, and compare the measured values
with theoretical predictions.
Section \ref{sec:sample} describes the targets,
and Sect. \ref{sec:observations} the ALMA observations and our ancillary data.
The determination of dust optical depth, gas content, and inference of
\xco\ on $\sim$250\,pc scales is given in Sect. \ref{sec:largescale},
while Sect. \ref{sec:smallscale} reports an analogous assessment for \xco\ at $\sim$40\,pc resolution.
The two sets of results are compared in Sect. \ref{sec:comparison},
and Sect. \ref{sec:conclusions} provides a brief discussion and our conclusions.

\begin{table*}
\begin{center}
      \caption[]{Parameters for observed galaxies} 
\label{tab:sample}
\resizebox{\linewidth}{!}{
\addtolength{\tabcolsep}{1pt}
\begin{tabular}{lrrcccccc}
\hline
\\
\multicolumn{1}{c}{Name} &
\multicolumn{2}{c}{Optical center (J2000)$^\mathrm{a}$} &
\multicolumn{1}{c}{Redshift$^\mathrm{a}$} &
\multicolumn{1}{c}{Distance} &
\multicolumn{1}{c}{Distance} &
\multicolumn{1}{c}{log($M_\star$/\msun)$^\mathrm{b}$} &
\multicolumn{1}{c}{log(SFR/\msunyr)$^\mathrm{b}$} &
\multicolumn{1}{c}{\logoh} \\
& \multicolumn{1}{c}{RA} 
& \multicolumn{1}{c}{Dec.} 
&
& \multicolumn{1}{c}{(Mpc)} 
& \multicolumn{1}{c}{method} 
&&& \multicolumn{1}{c}{(direct $T_e$ method)}\\
\\
\hline
\\
NGC\,625  & 01:35:04.63  & $-$41:26:10.3  & 0.00132 & 3.90$^\mathrm{c}$ & TRGB$^\mathrm{e}$ 
	& $8.58\,\pm\,0.19$ & $-1.22\,\pm\,0.14$ & 8.08$^\mathrm{f}$ \\ 
NGC\,1705  & 04:54:13.50  & $-$53:21:39.8  & 0.00211 & 5.22$^\mathrm{d}$ & TRGB$^\mathrm{e}$ 
	& $8.08\,\pm\,0.24$ & $-1.35\,\pm\,0.09$ & 8.21$^\mathrm{f}$ \\ 
NGC\,5253  & 13:39:55.96 & $-$31:38:24.4 & 0.00136 & 3.31$^\mathrm{d}$ & TRGB$^\mathrm{e}$ 
	& $8.59\,\pm\,0.24$ & $-0.31\,\pm\,0.15$ & 8.18$^\mathrm{g}$ \\ 
\\
\hline
\end{tabular}
} 
$^{\mathrm{a}}$~Taken from NED.  
$^{\mathrm{b}}$~Stellar masses and star-formation rates (SFRs) taken from \citet{marasco23}
reported to the distances used here;
$^{\mathrm{c}}$~\citet{lelli14}; 
$^{\mathrm{d}}$~\citet{sabbi18}; 
$^{\mathrm{e}}$~Tip of the Red Giant Branch;
$^{\mathrm{f}}$~\citet{berg12}; 
$^{\mathrm{g}}$~\citet{lopezsanchez12}.
We have adopted O/H for \hii-region A since it is closer to the brightest SSC,
and more consistent with the previous determination by \citet{kobulnicky97} of \logoh\,=\,8.16.
\end{center}
\end{table*}

\begin{table*}
\begin{center}
      \caption[]{Observational CO parameters} 
\label{tab:co}
\resizebox{\linewidth}{!}{
\addtolength{\tabcolsep}{1pt}
\begin{tabular}{lcccccccc}
\hline
\\
\multicolumn{1}{c}{Name} &
\multicolumn{1}{c}{\vsys$^\mathrm{a}$} &
\multicolumn{1}{c}{FWZI$^\mathrm{b}$} &
\multicolumn{1}{c}{Imaged} &
\multicolumn{1}{c}{12\,m beam} &
\multicolumn{1}{c}{12\,m rms} &
\multicolumn{1}{c}{12\,m total$^\mathrm{c}$} &
\multicolumn{1}{c}{12\,m total$^\mathrm{d}$} &
\multicolumn{1}{c}{12\,m total$^\mathrm{e}$} \\
& \multicolumn{1}{c}{(\kms)} &
\multicolumn{1}{c}{(\kms)} &
\multicolumn{1}{c}{area} & &
\multicolumn{1}{c}{(\mjybm)} &
\multicolumn{1}{c}{(\jykms)} &
\multicolumn{1}{c}{(\jykms)} &
\multicolumn{1}{c}{($10^5$\,\kkmspc)} \\
\multicolumn{1}{c}{(1)} &
\multicolumn{1}{c}{(2)} &
\multicolumn{1}{c}{(3)} &
\multicolumn{1}{c}{(4)} &
\multicolumn{1}{c}{(5)} &
\multicolumn{1}{c}{(6)} &
\multicolumn{1}{c}{(7)} &
\multicolumn{1}{c}{(8)} &
\multicolumn{1}{c}{(9)} \\
\hline
\\
NGC\,625  &  395.50 & [$-45,\,-\,5$]  & 46\arcsec$\,\times\,$30\arcsec
& 2\farcs08$\,\times\,$1\farcs49 ($85.7^\circ$) & 1.4 & $15.2\,\pm\,1.5$ & $14.19\,\pm\,1.4$ & 5.64\\
& & & & 39\,pc$\,\times\,$28\,pc \\
NGC\,1705 &  631.93 & [$-50,\,-5;\ 80,\,102] $  & 40\arcsec$\,\times\,$30\arcsec 
& 2\farcs49$\,\times\,$1\farcs82 ($-83.9^\circ$) & 1.2 & $0.95\,\pm\,0.05$ & $0.86\,\pm\,0.05$ & 0.63\\ 
& & & & 63\,pc$\,\times\,$46\,pc  \\
NGC\,5253 &  406.56 & [$-50,\,40$]  & 50\arcsec$\,\times\,$30\arcsec
& 2\farcs68$\,\times\,$2\farcs41 ($85.3^\circ$) & 1.3 & $12.5\,\pm\,0.04$ & $12.1\,\pm\,0.1$ & 3.35\\ 
& & & &  43\,pc$\,\times\,$39\,pc \\
\hline
\\
& & & &
\multicolumn{1}{c}{7\,m beam} &
\multicolumn{1}{c}{7\,m rms} &
\multicolumn{1}{c}{7\,m total$^\mathrm{c}$} &
\multicolumn{1}{c}{7\,m total$^\mathrm{d}$} &
\multicolumn{1}{c}{7\,m total$^\mathrm{e}$} \\
& & & & &
\multicolumn{1}{c}{(\mjybm)} &
\multicolumn{1}{c}{(\jykms)} &
\multicolumn{1}{c}{(\jykms)} &
\multicolumn{1}{c}{($10^5$\,\kkmspc)} \\
\hline
\\
NGC\,625  & & & & 12\farcs1$\,\times\,$7\farcs8 ($89.0^\circ$) & 6.4 & $14.85\,\pm\,1.5$ & $14.7\,\pm\,1.5$ & 5.52 \\ 
&&&& 229\,pc$\,\times\,$147\,pc & \\
NGC\,1705 & & & & 12\farcs4$\,\times\,$9\farcs1 ($89.9^\circ$) & 7.0 & $0.55\,\pm\,0.05$ & $0.44\,\pm\,0.05$ & 0.37 \\ 
&&&& 314\,pc$\,\times\,$230\,pc & \\
NGC\,5253 & & & & 13\farcs0$\,\times\,$7\farcs9 ($89.7^\circ$) & 6.4 & $14.1\,\pm\,1.3$ & $13.8\,\pm\,1.2$ & 3.77 \\ 
&&&& 209\,pc$\,\times\,$127\,pc & \\
\hline
\end{tabular}
} 
\begin{flushleft}
$^{\mathrm{a}}$\,Radio convention systemic velocity from observed sky frequency;
$^{\mathrm{b}}$\,Full-width zero intensity (FWZI) as shown in Fig. \ref{fig:cospectra};
$^{\mathrm{c}}$\,Flux in all available pixels in the velocity-weighted line-intensity (zero moment) maps;
$^{\mathrm{d}}$\,Flux with S/N$\geq$3 in zero moment maps. 
$^{\mathrm{e}}$\,CO line luminosity.
\end{flushleft}
\end{center}
\end{table*}

\section{The targets\label{sec:sample}}

For our ALMA study, we selected three nearby ($\sim$3-5\,Mpc) metal-poor (\zzsun$\sim$\,0.3) dwarf starbursts 
having \textit{(a)} \hi\ observations necessary to quantify their total gas content \citep{lelli14}; 
\textit{(b)} \hst\ color-magnitude diagrams (CMDs)
to constrain their star-formation histories (SFHs) and `burstiness' 
\citep[][]{cannon03,tosi01,annibali09,mcquinn10,cignoni18,cignoni19}; 
\textit{(c)} archival MUSE data, to infer high-resolution extinction maps; and
\textit{(d)} archival \hers\ data, to infer color temperatures for dust opacity maps.
These are the only southern galaxies in the parent sample of \citet{lelli14} 
and enable the start of a statistical approach; their properties are given in Table \ref{tab:sample}. 

\noindent
$-$ \textbf{NGC\,625} hosts a massive starburst with 
several luminous \hii\ regions and Wolf-Rayet (W-R) spectral features \citep[][]{skillman03a,skillman03b}.
The current starburst is relatively long-lived \citep[$\ga$50\,Myr][]{cannon03,mcquinn10}, and
the extended episode of star formation, possibly caused by an interaction or merger
\citep{cote00}, has apparently disrupted the \hi\ disk \citep{cannon04a,lelli14},
and caused an outflow detected in UV absorption lines \citep{cannon05}.

\noindent
$-$ \textbf{NGC\,1705} had more recent starburst episodes than NGC\,625,
with the older of two bursts of star formation having occurred $\sim10-15$\,Myr ago,
and the younger one much more recently, $\sim$3\,Myr ago \citep{annibali09,cignoni18}. 
NGC\,1705 is rich in star clusters \citep{billett02}, with the most massive one being
a Super Star Cluster \citep[SSC,][]{oconnell94,vazquez04,martins12}, that has roughly the $\sim 15$\,Myr age of the oldest starburst
event.
There is evidence for a low-velocity outflow seen through UV absorption \citep{heckman01}, 
but, unlike in NGC\,625, the \hi\ is configured in a regularly rotating disk \citep{elson13,lelli14}.
With \spit\ observations of NGC\,1705, \citet{cannon06} find that
the far-infrared dust morphology differs dramatically from
the optical, with two dust clouds $\sim$250\,pc approximately east and west of the central SSC, 
and apparently unrelated to it.
These two dust clouds also show 8\,\micron\ emission, and the central SSC is coincident 
with an 8\,\micron\ peak. 

\noindent
$-$ \textbf{NGC\,5253} harbors an extremely young starburst with many massive
star clusters \citep[e.g.,][]{calzetti97,cresci05}, 
with the majority of them in the central regions having ages of $\sim1-10$\,Myr
\citep[e.g.,][]{calzetti15}.
NGC\,5253 is overall more active than NGC\,1705.
Nevertheless, its SFH shows little evidence
of a burst over the last 10-20\,Myr, but this could result from the 
extreme crowding and incompleteness of the central region where most of the current SF is concentrated \citep{cignoni19}.
NGC\,5253's CMD suggests that star-formation
activity has been occurring for $\ga$450\,Myr, similar to the dynamical time of the galaxy \citep{mcquinn10}.
There is also a substantial population of W-R stars \citep{schaerer97,westmoquette13},
consistent with the young age of the central clusters.
The most massive cluster lies within a radio nebula \citep[e.g.,][]{turner00},
and is enshrouded by a dust cloud with \av$\sim$50\,mag \citep[e.g.,][]{calzetti15}.
The \hi\ in NGC\,5253 is highly perturbed \citep[e.g.,][]{kobulnicky08,lopezsanchez12,lelli14},
with infalling neutral gas that is apparently triggering the powerful current starburst
\citep[e.g.,][]{lopezsanchez12,turner15}.
In fact, the \hi\ kinematics is consistent with a disk-like structure dominated by a radial inflow motion of 
25 \kms\ \citep{lelli14}.

Although our study does not depend on specific metallicities, relative trends could be important,
and affected by abundance gradients.
Metallicity gradients in late-type dwarf irregulars or blue compact dwarfs tend to be generally negligible 
\citep[$\leq$\,0.1\,dex, e.g.,][]{croxall09}, less severe than those in more massive spirals \citep[e.g.,][]{pilyugin15}.
Resolved abundance investigations of our targets \citep[e.g.,][]{westmoquette13,annibali15,monreal17} show this to be
the case; there is no evidence for strong metallicity gradients in any of the dwarf starburst
targets, although some of the O/H estimates in the literature may differ from the ones
adopted here \citep[e.g.,][]{annibali15}.
 
\section{The data \label{sec:observations}}

To perform our analysis,
we have combined 
ALMA observations with three sets of ancillary 
publicly-available data as described below.

\subsection{ALMA observations \label{sec:alma}}

We observed the \coone\ line and tried
to detect the 3\,mm continuum emission 
in the three targets, NGC\,625, NGC\,1705, and NGC\,5253
with the ALMA 12-m and the ACA 7-m arrays during Cycle 6 using the Band 3 receivers (project-ID: 
\#2018.1.00219.S; PI: Hunt).
Other transitions including CN and SO$_2$ were placed in different sidebands,
and their analysis will be presented in future work.
To cover the entire emitting regions of the galaxies, we adopted
3-point mosaics with the ALMA 12-m array and a single pointing with ACA, obtaining a field-of-view of $\sim$40\,\arcsec.

Data calibration and imaging for the 12-m and ACA data were done using the Common Astronomy Software 
Applications\footnote{\url{http://casa.nrao.edu/}} 
\citep[CASA,][]{casa} 
version 6.2.1.7.
The visibility data were calibrated in the standard way.
We retained the native velocity resolution of 976.6\,kHz channels, corresponding to $\sim$2.5\,\kms.
Imaging was performed with the {\tt tclean} task using a Hogbom deconvolver and Briggs weighting 
with a robust parameter of 0.5.

For ACA, we set the pixel size of the cubes to 2\farcs0, corresponding to $\sim$1/6 of the
synthesized beams. 
The ACA cubes have typical rms noise levels $\sigma$ of (6.4, 7.0, 6.4)\,\mjybm\ for
(NGC\,625, NGC\,1705, NGC\,5253), respectively in velocity channels of 2.5\,\kms.
The pixel size of the ALMA 12-m cubes was set to 0\farcs18, $\sim$1/11 of the synthesized beams.
The final 12-m CO cubes have typical rms $\sigma$ of (1.4, 1.2, 1.3) \mjybm\ in 
velocity channels of 2.5\,\kms. 
Table \ref{tab:co} gives details on the CO observations for each galaxy.

Examining the \uv\ data showed that continuum emission is present only in NGC\,5253;
the continuum subtraction was performed by fitting and subtracting a first-order polynomial 
to line-free channels, in the \uv\ as well as in the image planes, giving 
similar results.
Ultimately, our final cube for NGC\,5253 subtracts the continuum estimated from the image plane.


The 12-m and ACA CO data were analyzed separately, so that there are two data cubes for each galaxy.
We derive total-intensity (moment-zero) maps using the task \texttt{Makemask} in 
\barolo\ \citep{diteodoro15}. This task sums the signal inside a Boolean mask, which is created by smoothing the cube 
in space and velocity and applying specific noise thresholds 
(see \barolo\ documentation for details). 
For this purpose, we use emission-line cubes without correction for primary beam attenuation, 
so the noise level is uniform and well defined across each channel map. 
The primary-beam correction is then applied directly to the moment-zero map. 
When using a Boolean mask to create a moment-zero map, the noise in the map varies from pixel to pixel because 
a different number of channels is summed up at each spatial position. 
We build a signal-to-noise (S/N) map using the relevant equation in the Appendix of \citet{lelli14b}.

\begin{figure*}[h!]
\hbox{
\centerline{
\includegraphics[angle=0,height=0.28\linewidth]{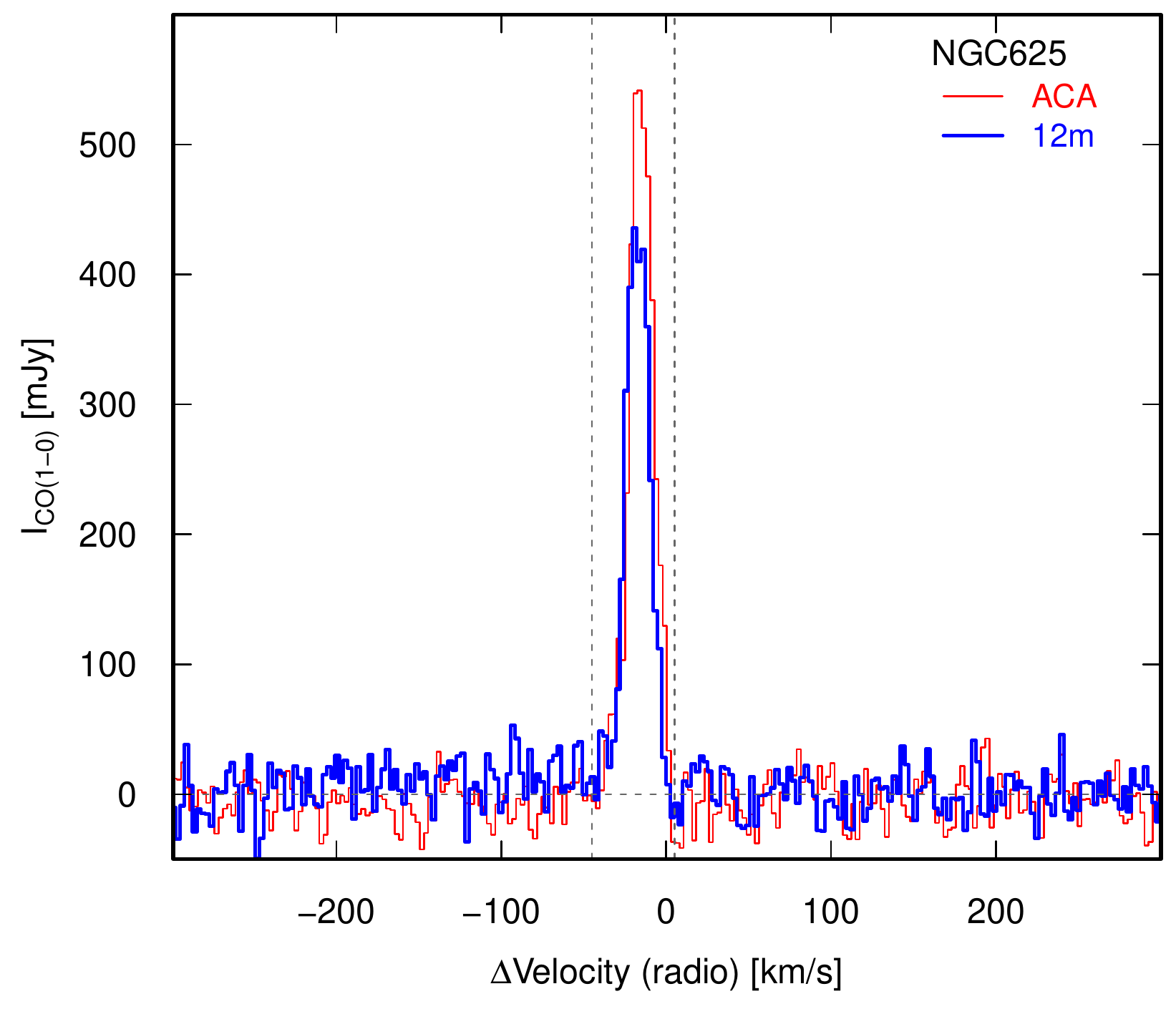}
\includegraphics[angle=0,height=0.28\linewidth]{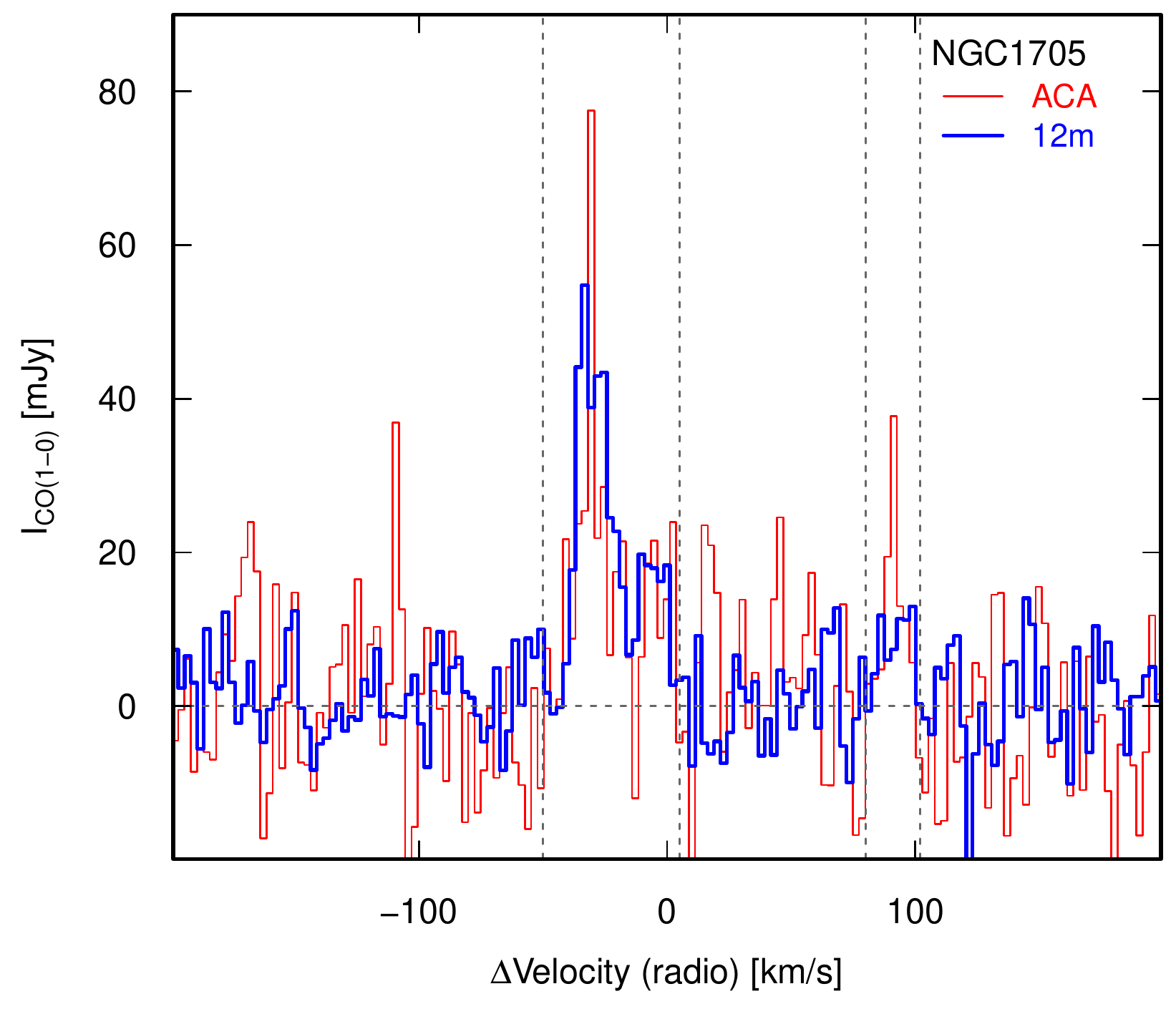}
\includegraphics[angle=0,height=0.28\linewidth]{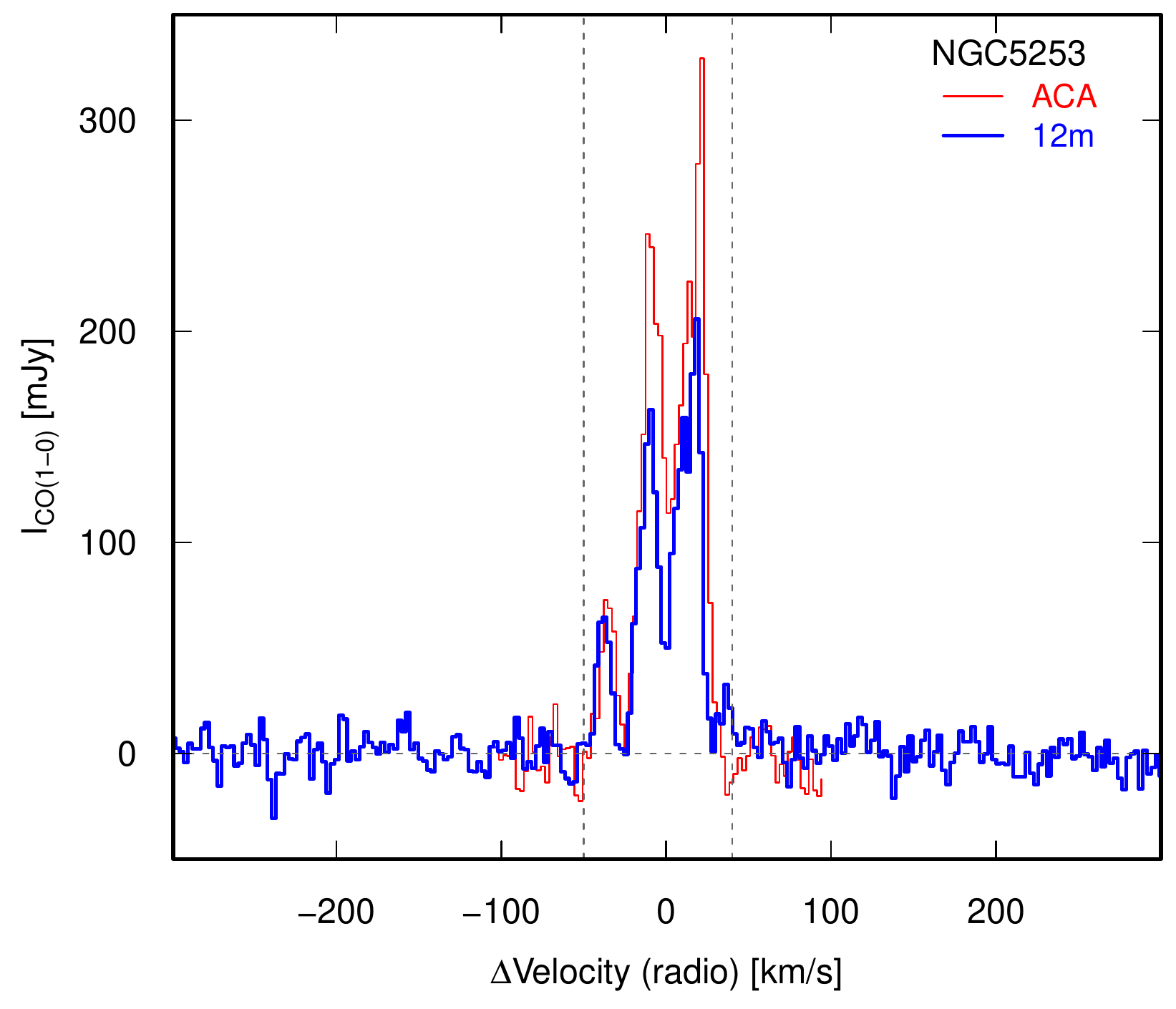}
}
}
\caption{CO spectra of the target galaxies within an aperture containing the 
totality of CO emission regions plotted against velocity, where \vsys\ 
is given in Table \ref{tab:co}. 
The full-width at zero intensity (FWZI) corresponds to
the vertical dotted lines (see Table \ref{tab:co}, Col. (2)). 
The 12-m spectra are shown in blue, and the ACA in red.  
In NGC\,5253, these are both continuum-subtracted spectra.
More details are given in the text.
\label{fig:cospectra}
}
\vspace{-\baselineskip}
\end{figure*}

Total fluxes 
are estimated
by summing the signal in the 
line-intensity (moment-zero) maps,
or by summing the emission within a certain aperture in the data cubes (performed by {\tt CASA/imview}).
The two techniques give similar results.
Figure \ref{fig:cospectra} shows the ACA and 12-m spectra taken from this sum using polygonal apertures.
Total fluxes for both the ACA and the 12-m array are given in Table \ref{tab:co}.
The ACA and 12-m fluxes are comparable, showing 
that we are not losing significant flux through interferometric filtering.
For NGC\,1705, the 12-m integrated flux is higher than for the ACA, by almost a factor of two,
suggesting that the CO emission is intrinsically compact and the ACA data suffers from 
low signal-to-noise. 


\enlargethispage{\baselineskip}

\begin{figure*}[h!]
\hbox{
\centerline{
\includegraphics[angle=0,height=0.38\linewidth]{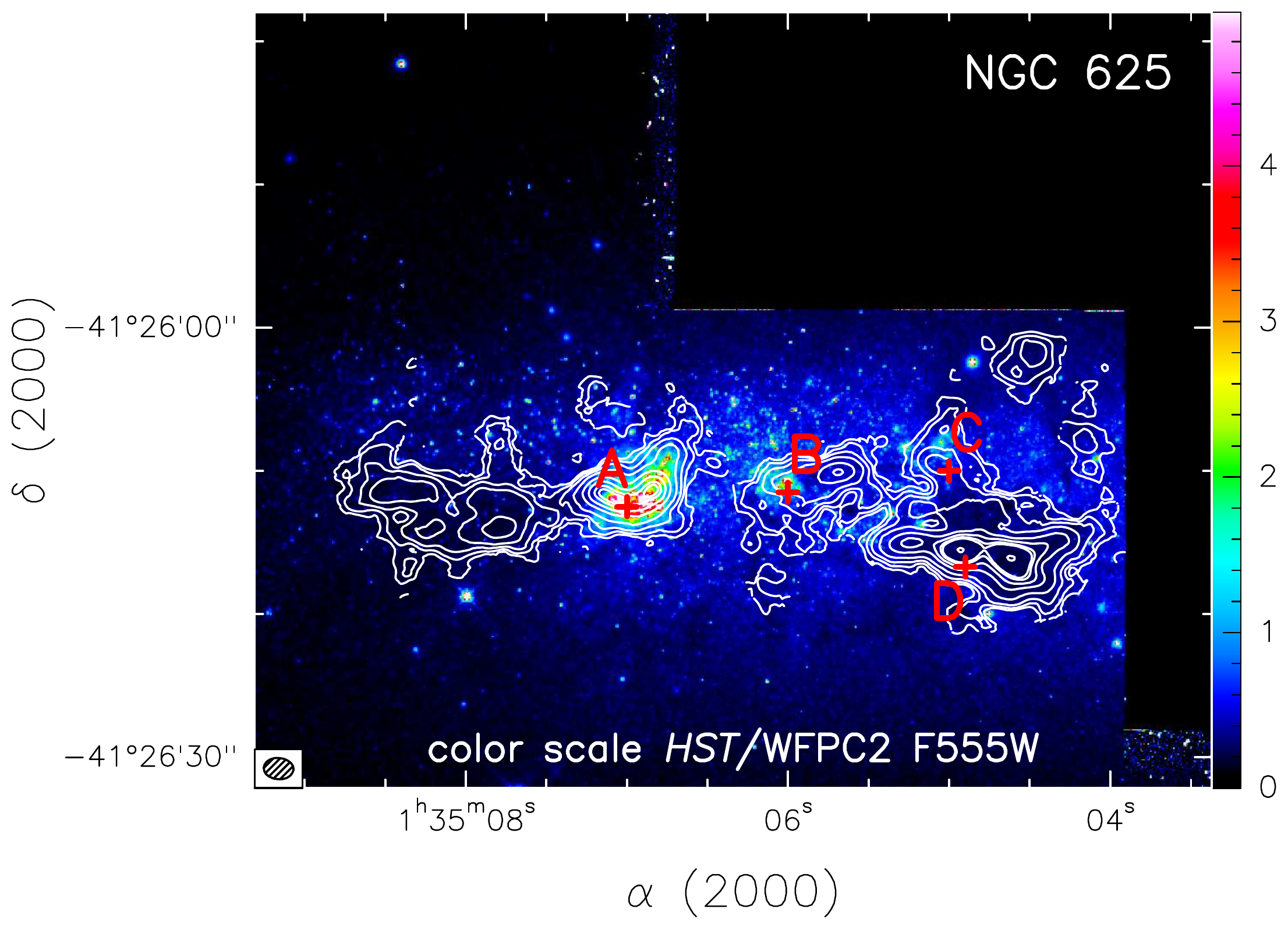}
\includegraphics[angle=0,height=0.38\linewidth]{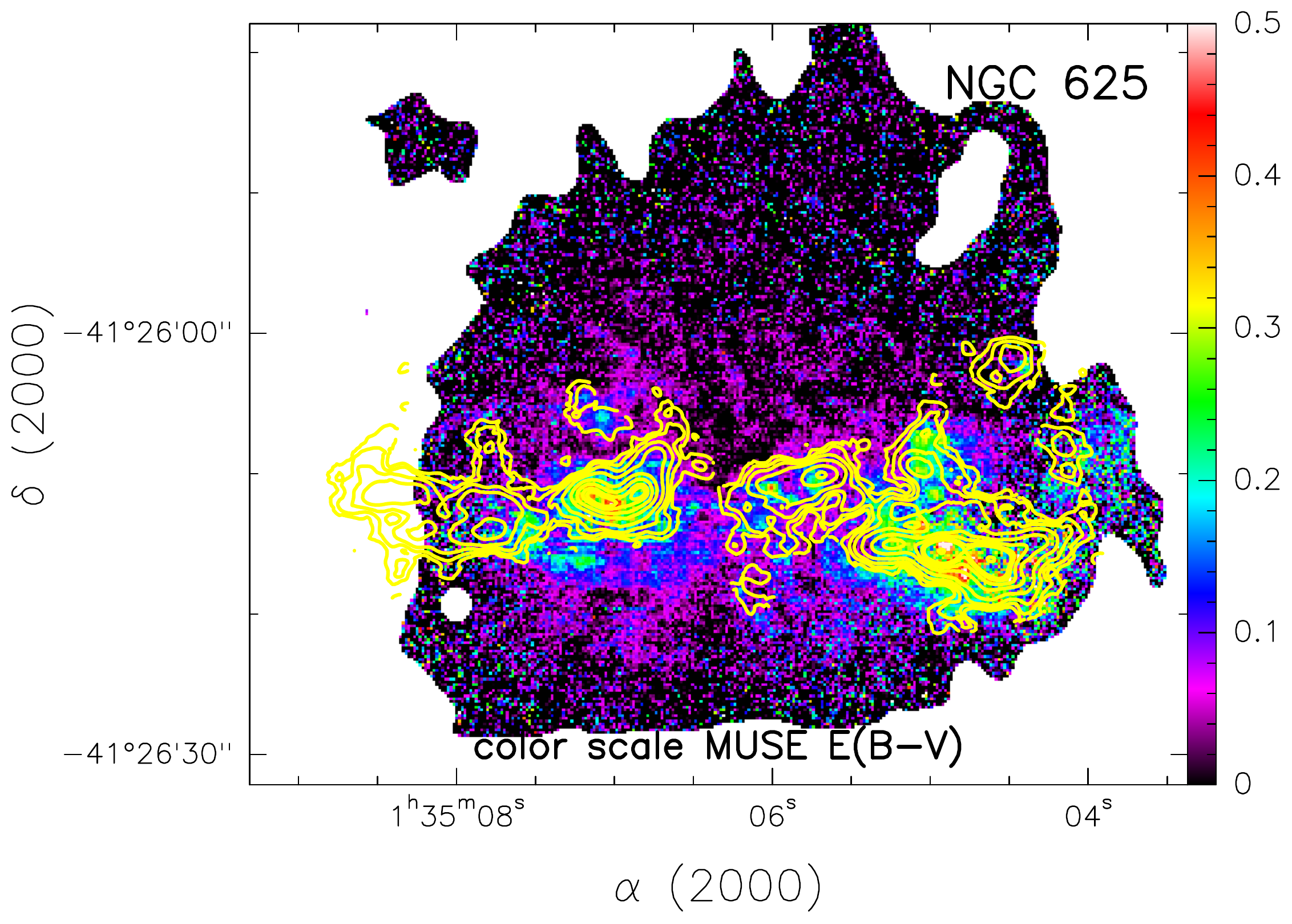}
}
}
\hbox{
\centerline{
\includegraphics[angle=0,height=0.40\linewidth]{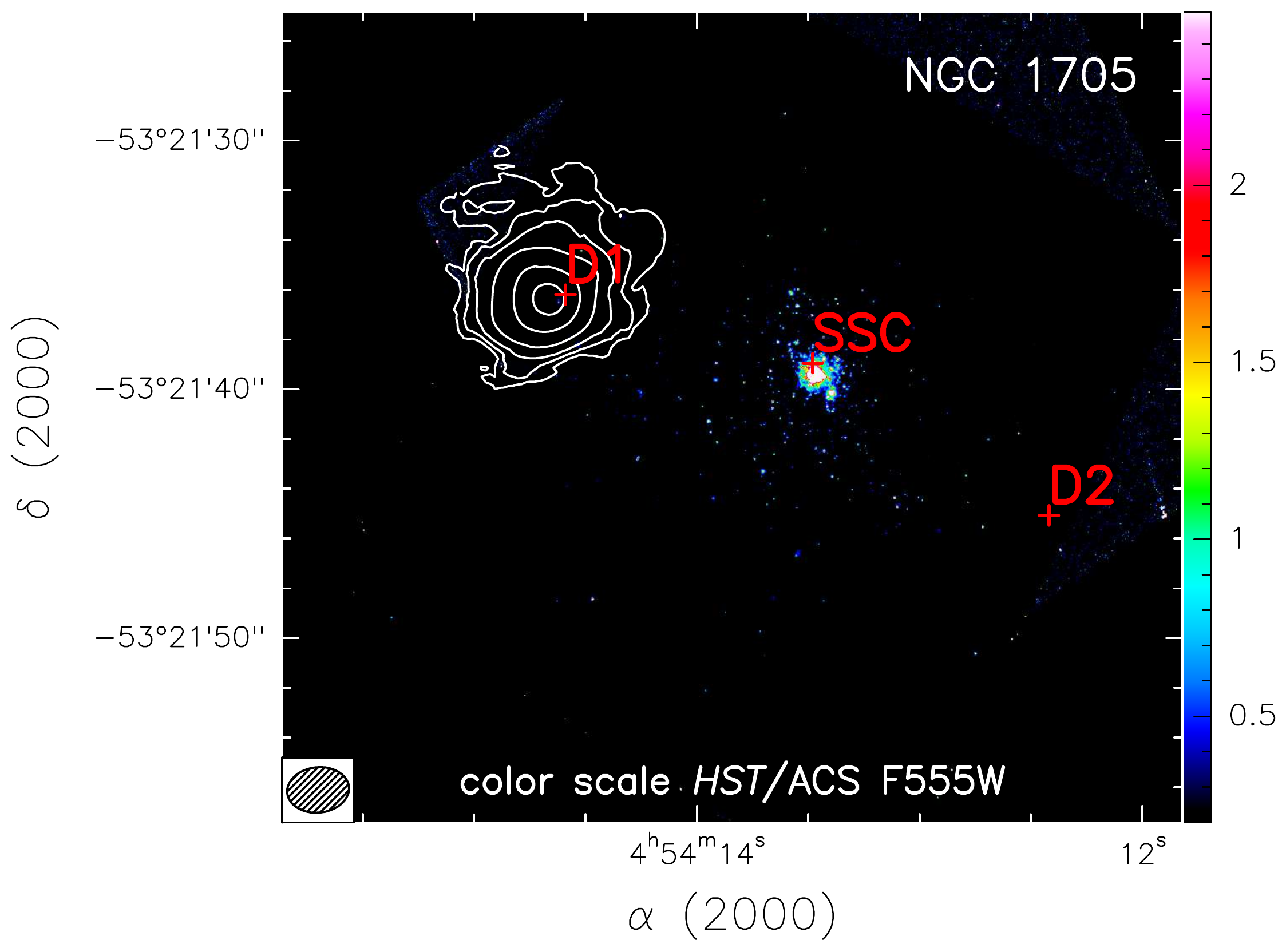}
\includegraphics[angle=0,height=0.40\linewidth]{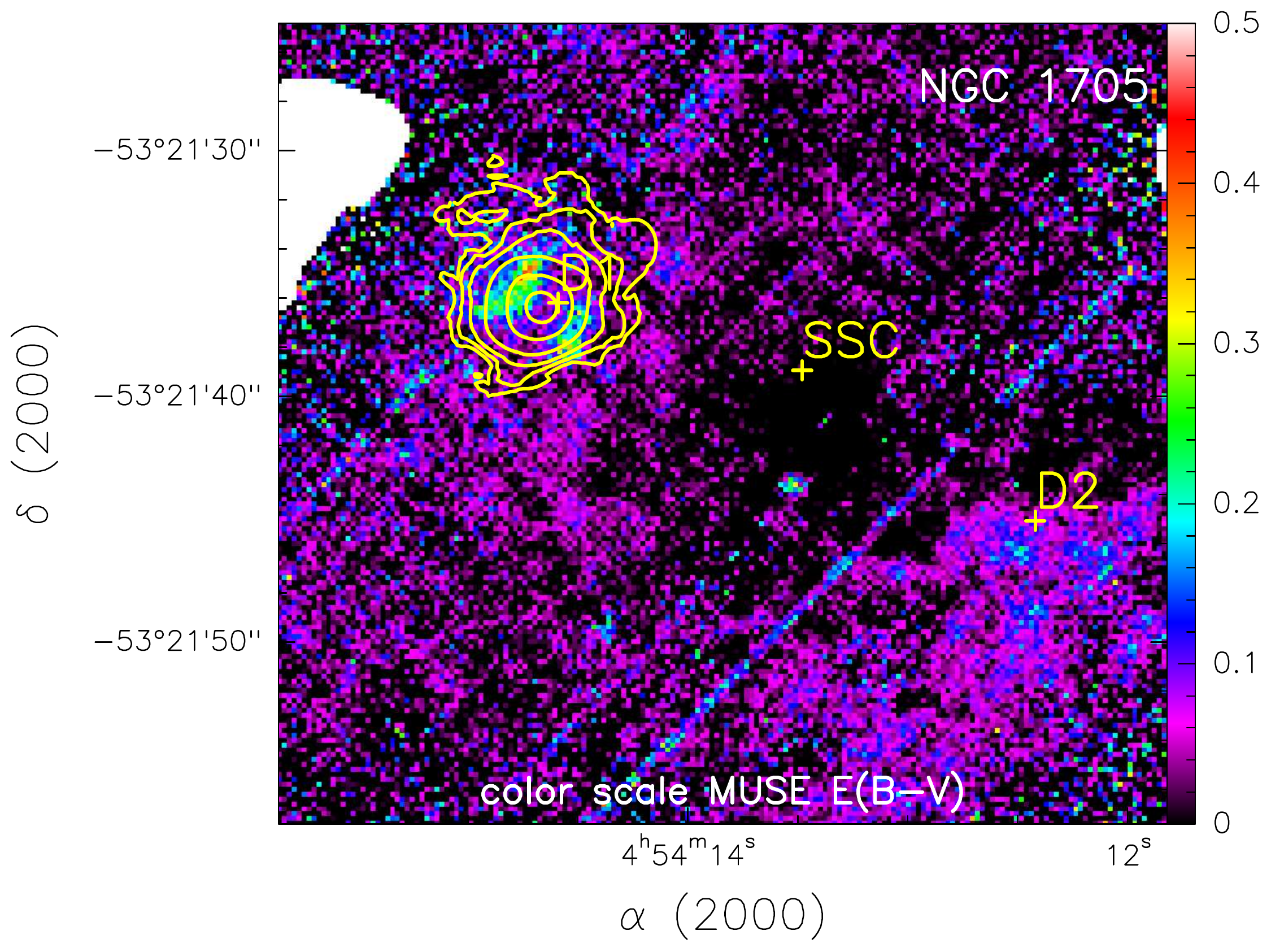}
}
}
\hbox{
\centerline{
\includegraphics[angle=0,height=0.4\linewidth]{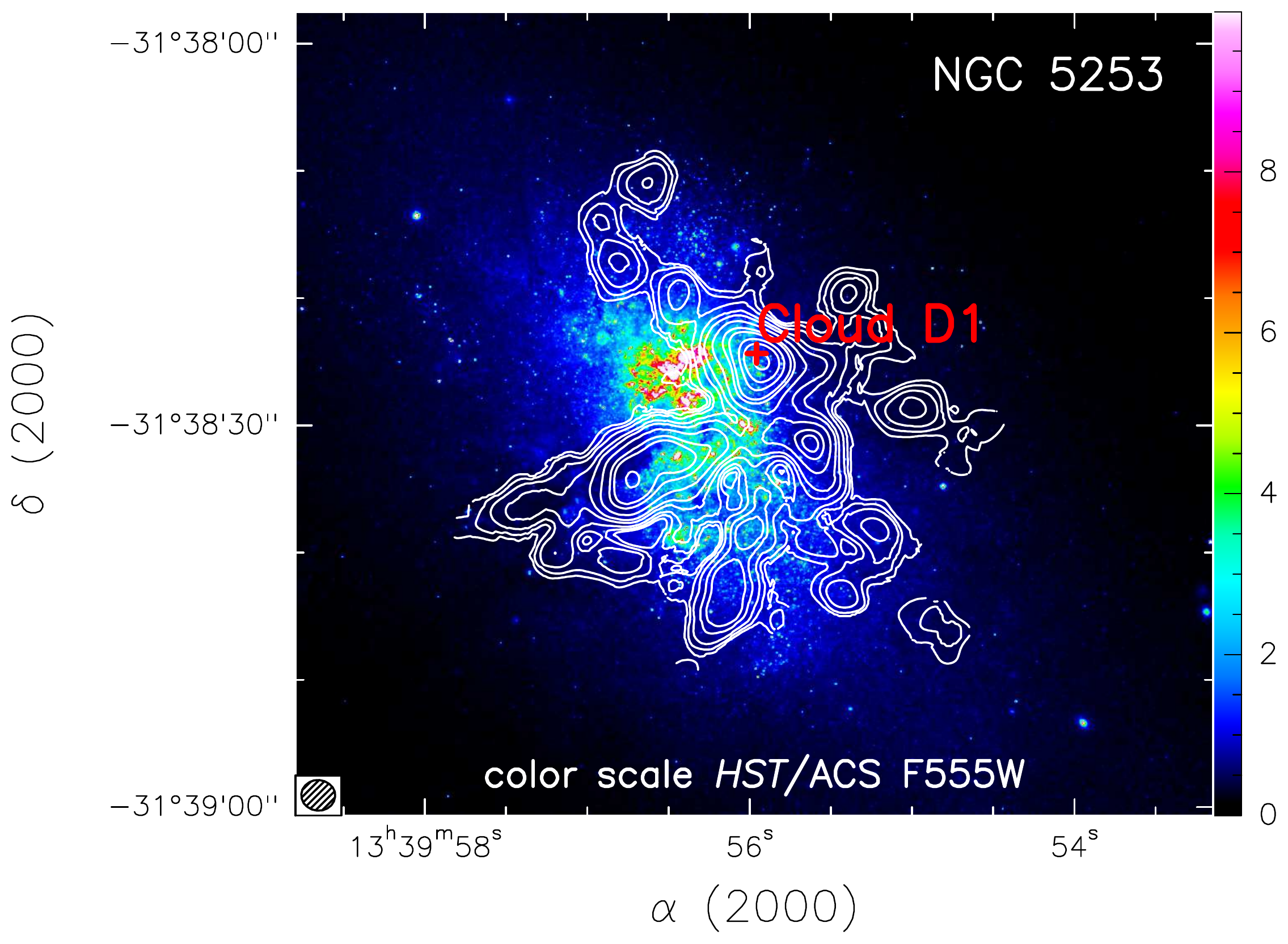}
\includegraphics[angle=0,height=0.4\linewidth]{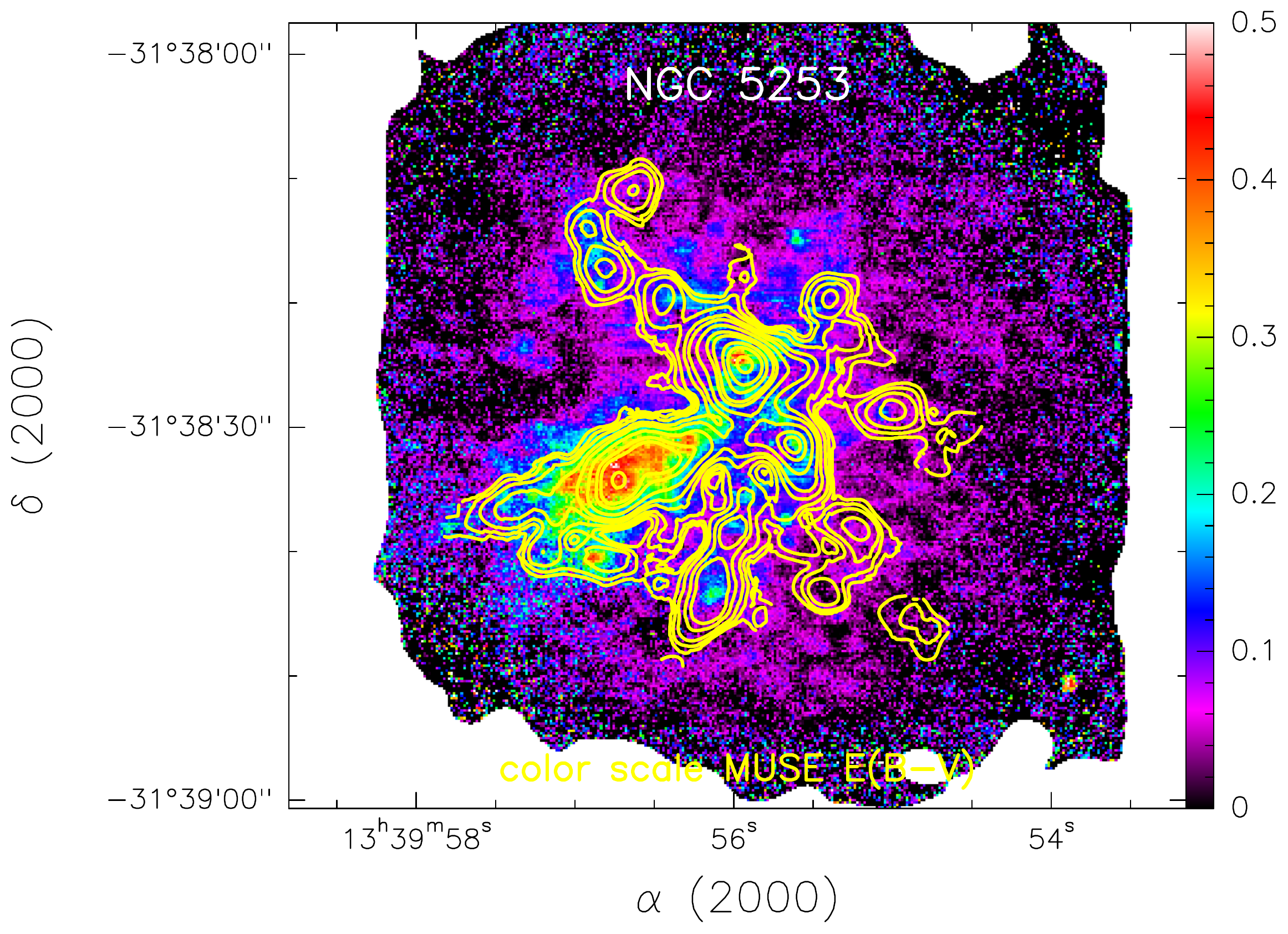}
}
\vspace{-\baselineskip}
}
\caption{Contours of \coone\ (12-m) overlaid on \hst\ F555W images (left panels),
and MUSE \ebv\ maps (right).
The \coone\ contour levels correspond to 
(0.01, 0.025, 0.05, 0.1, 0.2, 0.3, 0.4, 0.450, 0.750)\,\jykmsbm\ for NGC\,625; 
(0.01, 0.025, 0.05, 0.1, 0.2, 0.3)\,\jykmsbm\ for NGC\,1705; 
(0.01, 0.025, 0.05, 0.1, 0.15, 0.25, 0.4, 0.5, 0.75, 1.0, 1.5) \,\jykmsbm\ for NGC\,5253. 
Also marked A--D in the left panels are 
the \hii\ regions identified with radio continuum maps of NGC\,625 by \citet{cannon04a};
the dust clouds and SSC in NGC\,1705 identified from \spit\ observations \citep{cannon06};
and the unusual CO$+$dust cloud ``D1" in NGC\,5253 that coincides with the brightest,
highly dust embedded SSC \citep{turner17}.
The ALMA 12-m beam is shown in the lower-left corner of the left panels.
\label{fig:12moverlays}
\vspace{-3\baselineskip}
}
\end{figure*}

\subsection{Ancillary data \label{sec:ancillary}}

We have used ancillary data at different resolutions to estimate dust column density
and compare it to our CO maps.
The VLT/MUSE \ebv\ images are compared with the 12-m ALMA CO ($\sim$40\,pc),
and the \hers/PACS and \hi\ with the ACA CO ($\sim$250\,pc).

\subsubsection{VLT/MUSE \label{sec:muse}}

Data from the MUSE optical integral field spectrograph at the ESO Very Large Telescope (VLT) was 
obtained from the \textsc{Dwalin} (DWarf galaxies Archival Local survey for Interstellar medium investigatioN) 
survey database \citep{marasco23}. 
Briefly, the \textsc{Dwalin} sample consists of 40 nearby galaxies 
with archival MUSE observations
selected either from \textit{(a)} the \hers\ Dwarf Galaxy Survey \citep{madden13}, or 
\textit{(b)} the \citet{karachentsev13} catalog, 
with distance D $<$ 11 Mpc and 
log(\mstar/\msun)$\,<\,$9.0.

For NGC\,1705, the original data were obtained from program 094.B-0745 (PI: Garc{\'i}a-Benito). The two available MUSE pointings were stitched together to form a mosaic. For NGC\,5253, original data were taken from programs 094.B-0745 (PI: Garc{\'i}a-Benito) and 095.B-0321 (PI: Vanzi). In producing the final cube some exposures were discarded due to poor seeing. For NGC\,625, we used data from 094.B-0745 (PI: Garc{\'i}a-Benito) and considered only 3 of the 4 exposures due to a guiding failure during the first exposure.

The MUSE data reduction was performed using the MUSE pipeline \citep{weilbacher20} v2.8.1, with the ESO Recipe flexible execution workbench \citep[Reflex,][]{freudling13}, which gives a graphical and automated way to execute with EsoRex the Common Pipeline Library (CPL) reduction recipes, within the Kepler workflow engine \citep{altintas06}. The absolute astrometry of the final mosaics is fixed to that of archival \hst\ imaging. In the case of NGC\,625 and NGC\,5253, \ha\ emission is saturated in the regions of brightest line emission. These regions and the surrounding area were masked by hand.

Emission line maps were derived using the PHANGS data analysis pipeline\footnote{https://gitlab.com/francbelf/ifu-pipeline} described in \cite{emsellem22}. In short, the pipeline fits the stellar continuum using a set of E-MILES \citep{vazdekis16} simple stellar population templates. Spectral fitting is carried out twice, first to derive the stellar kinematics, and a second time simultaneously with Gaussian templates for modelling the emission lines. Line maps are corrected for Milky Way foreground extinction.

Dust attenuation intrinsic to the galaxies is calculated using the Balmer decrement (H$\alpha$/H$\beta$) and assuming Case B recombination, temperature $T=10^4$ K and density 
\nee\,=\,$10^2$\,\cmthree,
leading to  $L_{\rm H\alpha}/L_{\rm H\beta}\,=\,2.86$. We also assume the extinction law of \cite{odonnell94} with total-to-selective extinction $R_V = 3.1$, which is identical to the canonical \cite{cardelli89} parameterization. 
In the regions where H$\alpha$ is saturated we have sufficient signal-to-noise to detect the hydrogen Paschen 10 
(P10, at $\lambda = 9017.5\,\AA$) line, which is then used in combination with H$\beta$ to calculate the dust 
attenuation using the predicted ratio from Case B 
$L_{\rm P10}/L_{\rm H\beta}\,=\,0.0184$.
The CCD artifacts for the \ebv\ map of NGC\,1705 do not impact our analysis.
The global values of \ebv\ for our targets \citep{cignoni19} are consistent with 
the spread of spatially resolved \ebv\ resulting from our calculations.

\begin{figure*}[t!]
\hbox{
\centerline{
\includegraphics[angle=0,height=0.28\linewidth]{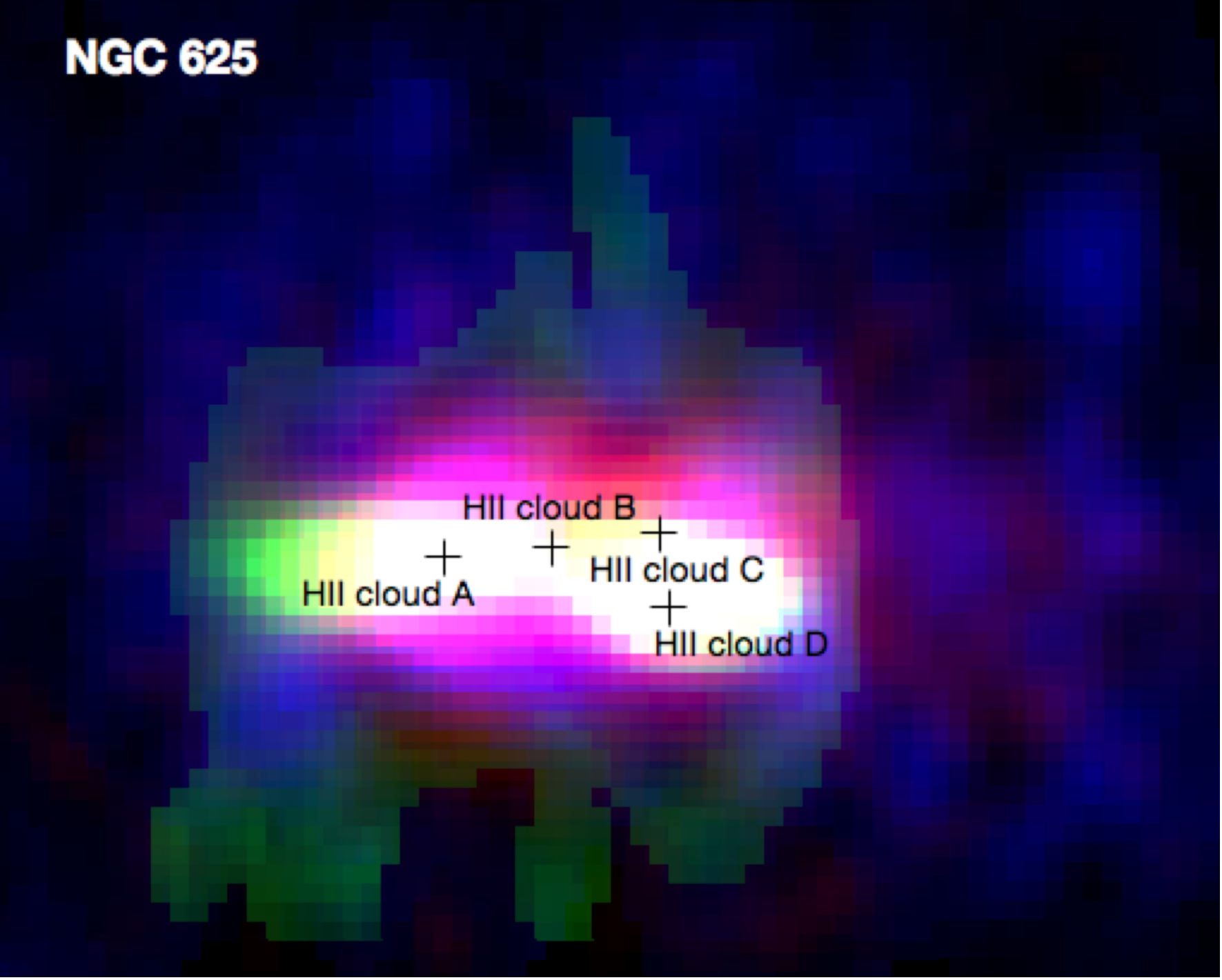}
\includegraphics[angle=0,height=0.28\linewidth]{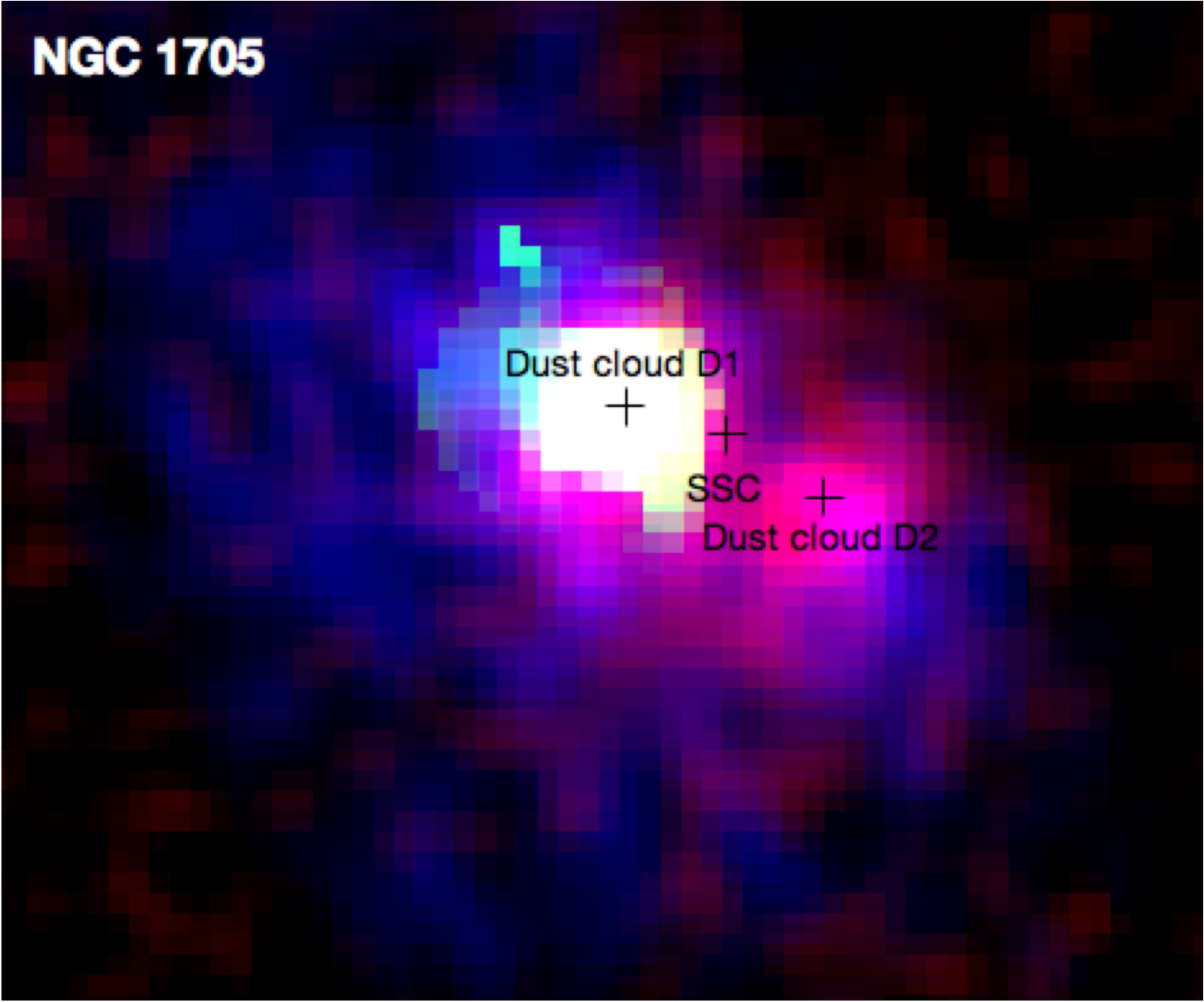}
\includegraphics[angle=0,height=0.28\linewidth]{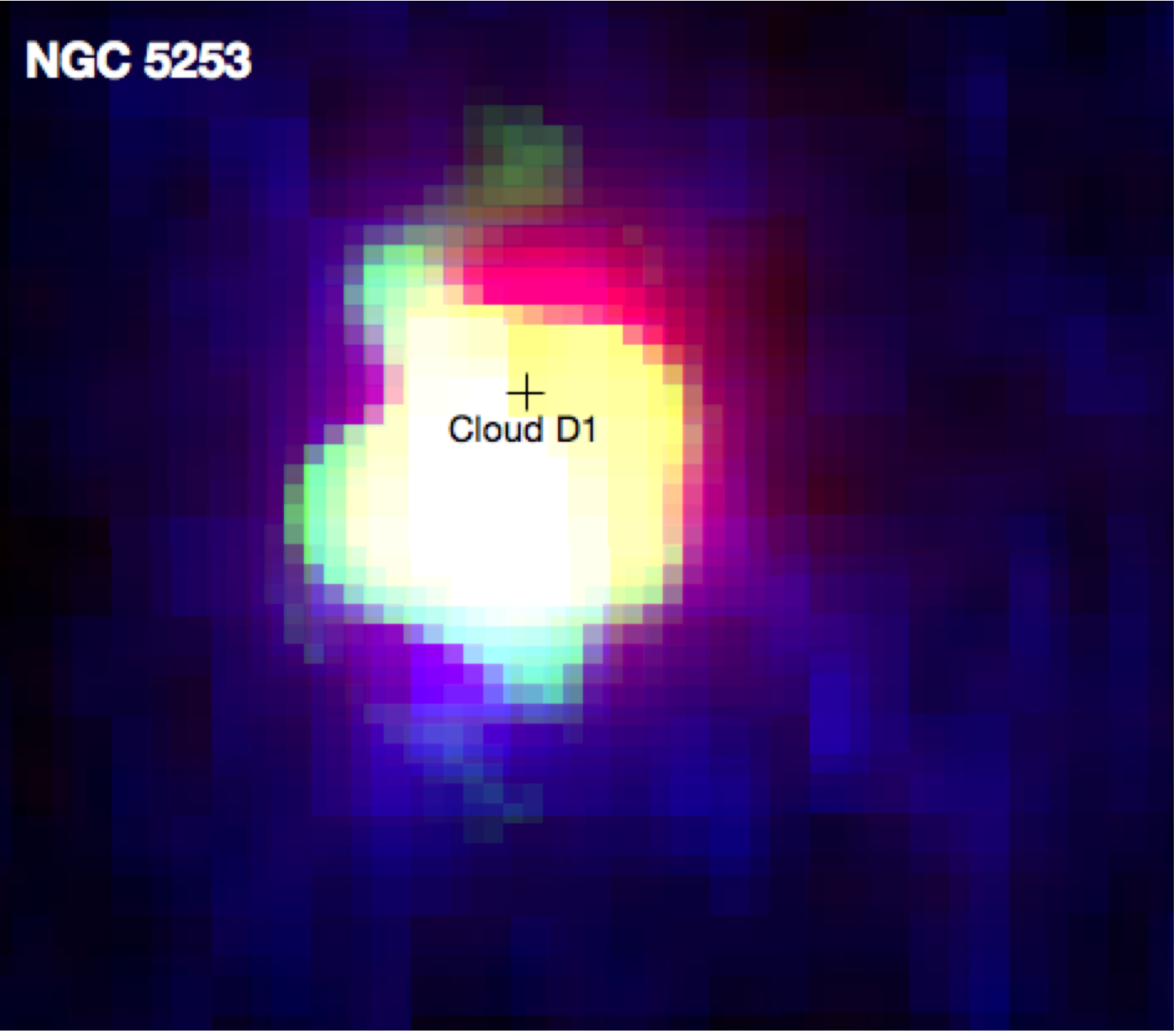}
}
}
\caption{Color RGB images for NGC\,625 (left panel), NGC\,1705 (middle), and NGC\,5253 (right)
with \hers/PACS-R as red, ACA \coone\ as green, and \hi\ as blue; North up, East left.
The image dimensions are $130\times105$\,arcsec$^2$ for NGC\,625, NGC\,1705, and 
$120\times105$\,arcsec$^2$ for NGC\,5253.
All observations have roughly spatial resolutions of $\sim$250\,pc (see text for more details).
The white color indicates where \hi, CO, and dust are co-spatial. 
As in Fig. \ref{fig:12moverlays}, 
also indicated are the \hii\ regions A--D of NGC\,625 \citep{cannon04a};
the dust clouds and SSC in NGC\,1705 \citep{cannon06};
and the unusual CO$+$dust cloud ``D1" in NGC\,5253 that coincides with the brightest,
highly dust embedded SSC \citep{turner17}.
\label{fig:aca}
}
\end{figure*}

\subsubsection{\hers\ PACS \label{sec:pacs}}

We acquired from the \hers\ archive
the raw scans taken with the
Photodetector Array Camera and Spectrometer \citep[PACS,][]{poglitsch10} in the context
of the \hers\ Dwarf Galaxy Survey \citep{madden13}.
The data available for our three targets at 70, 100, and 160\,\micron\
have full widths at half maximum (FWHM) of 5\farcs6, 6\farcs8, and 11\farcs4,
respectively.
We reduced the raw scans using (the final) version 15.0.1 of the 
\hers\ Interactive Processing Environment \citep[HIPE,][]{ott10}
with the PACS\_CAL\_78\_0 photometric calibration,
and adopting the {\tt scanam} procedure to optimize extended
emission.
Sky subtraction was performed within the procedure but also subsequently checked
to ensure roughly zero background.
The reduced, calibrated images were generated with 2\farcs0 pixels, and to
be on the same resolution scale, were
then convolved to the PACS 160\,\micron\ FWHM using kernels from \citet{aniano11}.

\subsubsection{Atomic gas, HI \label{sec:hi}}

We re-analyze \hi\ data from \citet{lelli14} 
using \hi\ cubes with the highest available angular resolution;
beam sizes are (16\farcs9$\times$10\farcs6, 8\farcs5$\times$6\farcs6, 13\farcs6$\times$7\farcs5) for
(NGC\,625, NGC\,1705, NGC\,5253), respectively. 
The \hi\ data were obtained with the Very Large Array (VLA) for NGC\,625 \citep{lelli14} 
and the Australia Telecope Compact Array (ATCA) for NGC\,1705 \citep{elson13} 
and NGC\,5253 \citep{lopezsanchez12}. 
We refer to  those papers for details on observations and data reduction. 
Total-intensity (moment-zero) \hi\ maps are derived using the same strategy as for CO maps (Sect. \ref{sec:alma}).

\subsection{Image alignment and rebinning \label{sec:alignment}}

The ACA CO, \hi, and (convolved) PACS images have been aligned to a common center
and pixel size (2\arcsec) using the routines within the GNU Astronomy Utilities
\citep[{\tt gnuastro}\footnote{https://www.gnu.org/software/gnuastro/},][]{gnuastro}.
The 12-m and MUSE \ebv\ maps have been aligned with the same techniques, but instead
with an intrinsic pixel size of 0\farcs4.
The alignment relies on the internal astrometric keywords, and thus implicitly assumes
that the nominal astrometry is correct.

As mentioned above, the nominal MUSE astrometry was checked against \hst\ and corrected as necessary.
For NGC\,625, there is a known discrepancy of the \hst\ astrometry of the image used
here \citep{cannon03}, and this was taken into account.
For NGC\,1705, a southward shift of of 2\farcs5 was imposed on the nominal 
MUSE pipeline astrometry solution in order to align with \hst\ and ALMA.

The rebinning to large pixels was performed within the \textit{R} statistical package\footnote{For all statistical 
analysis, we rely on the
\textit{R} statistical package: 
R Core Team (2018), R: A language and environment for statistical
  computing, R Foundation for Statistical Computing, Vienna, Austria
  (\url{https://www.R-project.org/}).}.
For the 12-m comparison of CO with \ebv\ we used resolution elements (pixels) of 2\arcsec\ on a side,
and for the ACA comparison with \hi\ and PACS, 
pixels of (13\arcsec, 10\arcsec, 15\arcsec) for (NGC\,625, NGC\,1705, and NGC\,5253), respectively.
The 12-m rebinning corresponds to $\sim$40\,pc (see Table \ref{tab:co}) and
the ACA rebinning to $\sim$250\,pc; the aim of the pixel sizes for ACA, \hi, and PACS
is to account approximately for the differences in distance of the targets.

Figure \ref{fig:12moverlays} shows the results of the small spatial scale ($\sim$40\,pc) alignments, 
where the 12-m CO zero-moment maps are overlaid on \hst\ F555W images (left panel) 
and on the MUSE \ebv\ images (right).
Also shown in Fig. \ref{fig:12moverlays} are specific features
identified by previous work \citep[e.g.,][]{cannon04a,cannon06,turner17}:
\hii\ regions (NGC\,625) and dust clouds (NGC\,1705, NGC\,5253). 
Interestingly, in NGC\,1705, 
PAH emission with \spit/IRS is only present in the eastern dust cloud D1 \citep{cannon06}
which is also the only location of CO emission.
In all three galaxies, 
CO emission is not always co-located with the stars traced by the \hst\ images;
in contrast, it almost perfectly coincides with \ebv\ from MUSE.

Figure \ref{fig:aca} reports ``false-color'' RGB images at $\sim$250\,pc resolution
for the three target galaxies, using \hi\ as blue, CO integrated intensity (zero-moment) maps 
as green, and PACS 160\,\micron\ as red.
As in Fig. \ref{fig:12moverlays}, the salient features identified in previous work
\citep{cannon04a,cannon06,turner17} are also shown.
The green ``spur'' to the east in NGC\,625 shows that CO emission is present even without 
significant dust emission (traced by PACS);
the diffuse \hi\ emission is not well traced by this higher-resolution \hi\ map \citep[see][]{lelli14}.
In NGC\,1705, the white colors of dust cloud D1 \citep{cannon06} show that CO is co-spatial with \hi\
and dust, but, as also illustrated in Fig. \ref{fig:12moverlays}, there is no CO in the second dust
cloud D2, nor at the central SSC.
On the other hand, the SSC embedded within cloud D1 in NGC\,5253 \citep{turner17}, is co-spatial with
CO emission, as also shown at higher resolution in Fig. \ref{fig:12moverlays}.
In all three galaxies, \hi\ is ubiquitous, more extended than the dust and the CO emission.

\subsection{Comparison with previous CO flux measurements \label{sec:cocomparison}}

Although NGC\,1705 was not previously observed in CO, the two other targets have prior CO observations. 
NGC\,625 was observed in \coone\ with the MOPRA 22-m single dish by \citet{cormier14}
and with ALMA by \citet{imara20}.
The total CO velocity-integrated \tb\ \ico\ given by \citet{cormier14} is $4.3\,\pm\,0.6$\,\kkms, 
or a CO(1--0) luminosity of $1.1\times10^6$\,\kkmspc.
With ALMA, in a $\sim$1\farcs31 $\times$1\farcs08 beam, \citet{imara20} find \ico\,=\,$3.0\,\pm\,0.2$\,\kkms, that
over a region of 57\farcs5\,$\times$\,20\arcsec, gives a luminosity of
$1.2\times10^6$\,\kkmspc.
Since \citet{cormier14} and \citet{imara20} give total CO in units of velocity-integrated \tb, 
as averages over regions much larger than the beams, 
we compare our measurements as given in Table \ref{tab:co} with their luminosities;
15.20\,\jykms\ corresponds to a luminosity of $0.6\times10^6$\,\kkmspc, about half that measured by previous work.
Since we find roughly the same velocity-integrated fluxes with the 12-m and ACA 7-m arrays, we would
argue that the problem is not so much missing flux as possible differences in reduction and analysis procedures.
The similarity of our integrated values with both arrays implies that this possible discrepancy will not significantly impact
our results.

\citet{wiklind89} and \citet{turner97} have observed NGC\,5253 with the 15m Swedish-ESO Submillimeter
Telescope (SEST, 45\arcsec\ beam) and the Owens Valley Radio Observatory (OVRO), respectively.
\citet{wiklind89} report a total velocity-integrated \tb\ 
\ico\,=1.3\,\kkms, with a velocity width of 43\,\kms.
With OVRO, using a 14\farcs9$\times$11\farcs4 beam, \citet{turner97} 
find a velocity-integrated flux of 14.0\,\jykms, 
comparable to the flux recovered by our observations. 
NGC\,5253 was also observed with ALMA in \cotwo\ by \citet{miura18}, 
with the 12-m and 7-m arrays, and also with Total Power (TP).
Within a box of 90\arcsec$\times$84\arcsec, \citet{miura18} find  $96.6\,\pm\,0.2$\,\jykms\ (TP),
and within a region of 22\arcsec\ radius $92.6\,\pm\,0.1$\,\jykms\ (combined TP$+$12m$+$7m).
To compare their CO(2--1) observations with CO(1--0) from \citet{wiklind89}, \citet{miura18} derive
a TP velocity-integrated flux in CO(2--1) of 42.3\,\jykms, centered on the Wiklind et al. SEST beam
with a radius of 22\arcsec.
If CO is thermalized, as \citet{miura18} assume, this latter measurement would translate roughly into \ico\,$\approx$10.6\,\jykms\ for 
\coone, between 75\% and 85\% of our values in Table \ref{tab:co} over a similarly-sized area.
Given the variation of the velocity-integrated fluxes found by \citet{miura18} in the different regions,
our measurement is roughly consistent with previous work.
In any case, as for NGC\,625, the difference in our velocity-integrated CO flux with the 12-m and the 7-m arrays for NGC\,5253
is not so large as to adversely affect our results for \xco.


\begin{figure}[!t]
\hbox{
\centerline{
 \includegraphics[angle=0,width=0.99\linewidth]{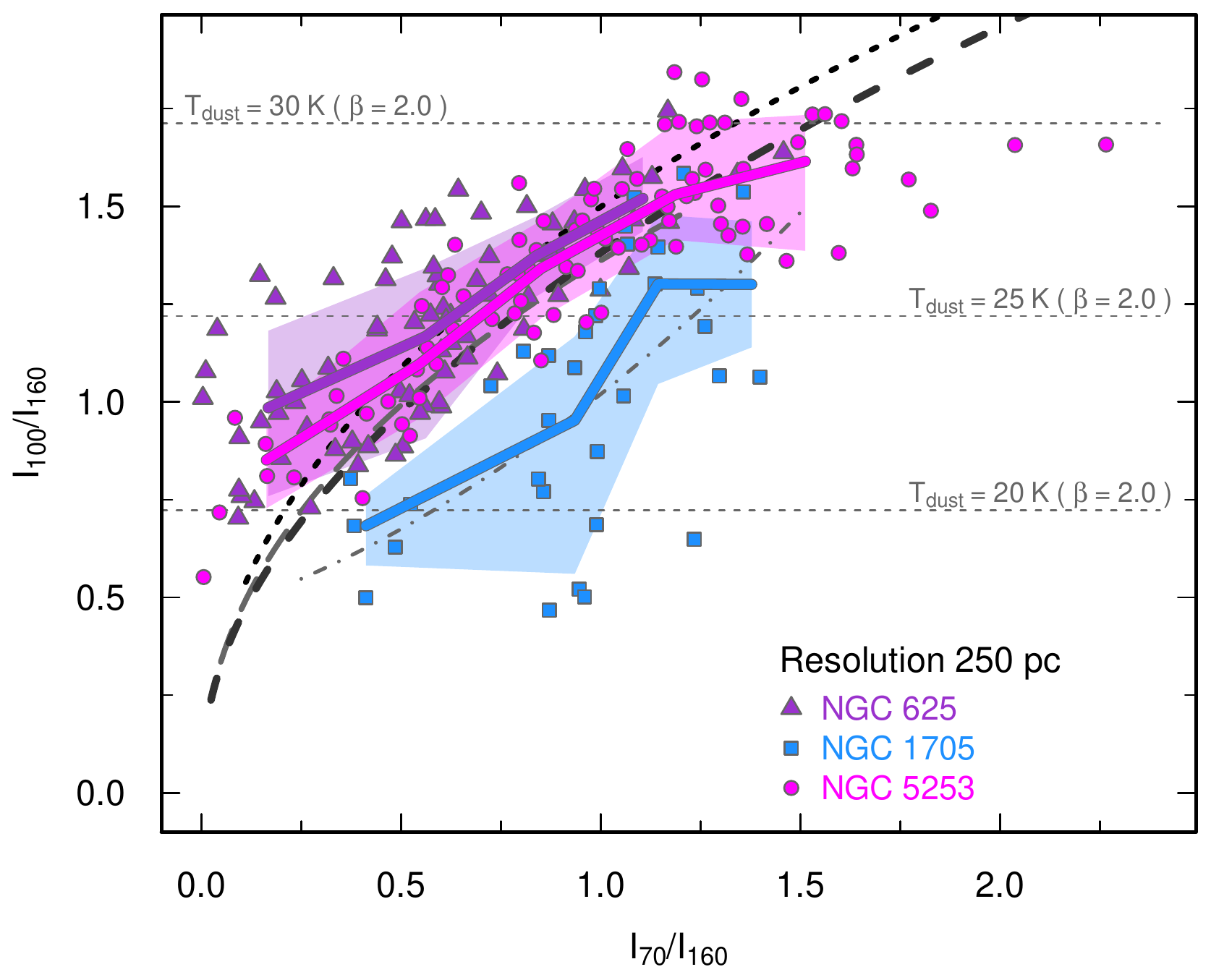}
 }
 }
\caption{PACS flux ratios $I_{100}/I_{160}$ versus $I_{70}/I_{160}$.
The PACS data have been rebinned to $\sim$250\,pc resolution in order to match the \hi\
and ACA CO maps.
Heavy curves show the medians in $I_{100}/I_{160}$ bins
for each galaxy, and the shaded regions indicate 1$\sigma$ standard deviations.
Also shown are the model MBB-T2 flux ratios with $\gamma$\,=\,0.14 (the median value
for NGC\,1705), indicated by the heavy black long-dashed curve.
The IR ratios from two MBB models ($\gamma\,=\,0$) are illustrated as a long-dashed (grey) curve with $\beta$\,=\,1
(virtually indistinguishable from the MBB-T2 curve),
and a short-dashed (black) curve for $\beta$\,=\,2.
The dot-dashed curve corresponds to the best fit polynomial for the SMC by \citet{leroy09}. 
The horizontal lines illustrate the $I_{100}/I_{160}$ ratios that would be expected
for \td\,=\,20\,K, 25\,K, and 30\,K.
The line ratios for NGC\,625 and NGC\,5253 are fairly well approximated by MBBs or MBB-T2 fits,
while the polynomial fit by \citet{leroy09} is a better approximation for NGC\,1705.
This fit is roughly equivalent to assuming that only half of the observed 70\,\micron\ emission
comes from large, ``classical'' grains, with the other half originating from stochastic heating of small grains. 
\label{fig:pacs}
}
\end{figure}

\section{Large-scale dust opacity, gas content, and the \xco\ conversion factor \label{sec:largescale}}

We trace the dust content along a line of sight at $\sim$250\,pc resolution 
using the optical depth at 160\,\micron, \mytau\
\citep[e.g.,][]{leroy09,bolatto11}. 
We then compare the distribution of \mytau\ with the 
distributions of \hi\ and ACA \coone\ at an angular resolution of $\sim$250\,pc. 
Finally, we infer the \htwo\ column density from the estimated dust-to-gas column density ratios,
and derive the \xco\ conversion factor.

\subsection{Infrared dust opacity from \hers/PACS}
\label{sec:tau}

With the exception of extreme ultra-luminous infrared galaxies
\citep[e.g., Arp\,220,][]{wilson14},
the assumption that dust emission in galaxies is optically thin at 160\,\micron\ is 
well established \citep[e.g.,][]{misselt01,whelan11}.
For an optically thin dust grain population with an equilibrium temperature, \td, 
\mytau\ is related to the measured 160\,\micron\ 
flux, $I_{160}$, by:
\begin{equation}
\tau_{160}\,=\,\frac{I_{160}}{B_\nu\,(T_\mathrm{dust}, 160\,\mu\mathrm{m})}\quad ,
\end{equation}
where $B_\nu$ corresponds to the intensity of a blackbody emitting at a temperature
\td\ at wavelength $\lambda$.
Thus, to compute \mytau, we need to estimate the equilibrium dust temperature, \td.

To determine \td, the PACS 70, 100, and 160\,\micron\ data have been fit with two functions:
\begin{enumerate}[(1)]
\item
``MBB fits'': a single-temperature modified blackbody (MBB) with emissivity index $\beta\,=\,2.0$ having two fitted parameters:
\mytau\ and \td. 
The choice of $\beta=2$ is consistent with current estimates for
    the $70-160$\,\micron\ opacity of dust in the diffuse ISM in the solar
    neighborhood \citep{hensley21}.
\item
``MBB-2T fits'': 
the sum of two MBBs with emissivity index $\beta\,=\,2.0$ with dust temperatures \tc\ and 
\tw$\,=\,1.5\times$\,\tc .  
The two-temperature model is intended as a first
    approximation to the distribution of dust temperatures in a
    star-forming galaxy, where dust near newly-formed star clusters
    will be heated by more intense radiation.  
The choice of \tw/\tc\,=\,1.5 is somewhat arbitrary, but is motivated by
the factor of $\sim$2.3 range of wavelengths (70\,\micron$-$160\,\micron) used in the fit.
These fits have three fitted parameters: \mytau, \tc, and $\gamma$ defined 
as the fraction of the dust emission due to the warmer component.
For these fits, there are zero degrees of freedom since the fits are limited to
the three PACS cameras because we want to preserve the 250\,pc resolution
(and the longer \hers/SPIRE wavelengths give, at best, a resolution 50\% worse).
\end{enumerate}

\begin{figure*}[!t]
\hbox{
 \includegraphics[angle=0,width=0.33\linewidth]{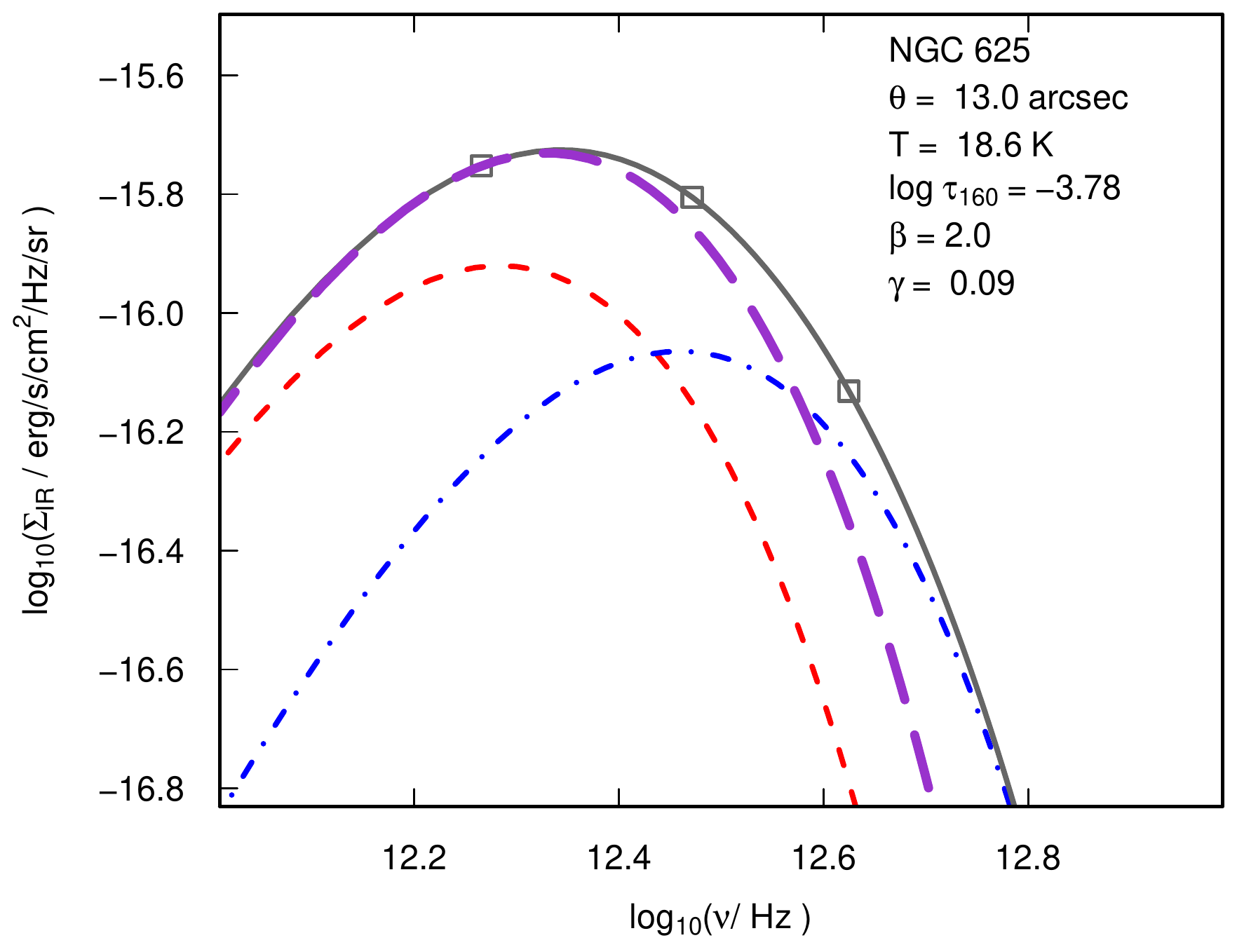}
 \includegraphics[angle=0,width=0.33\linewidth]{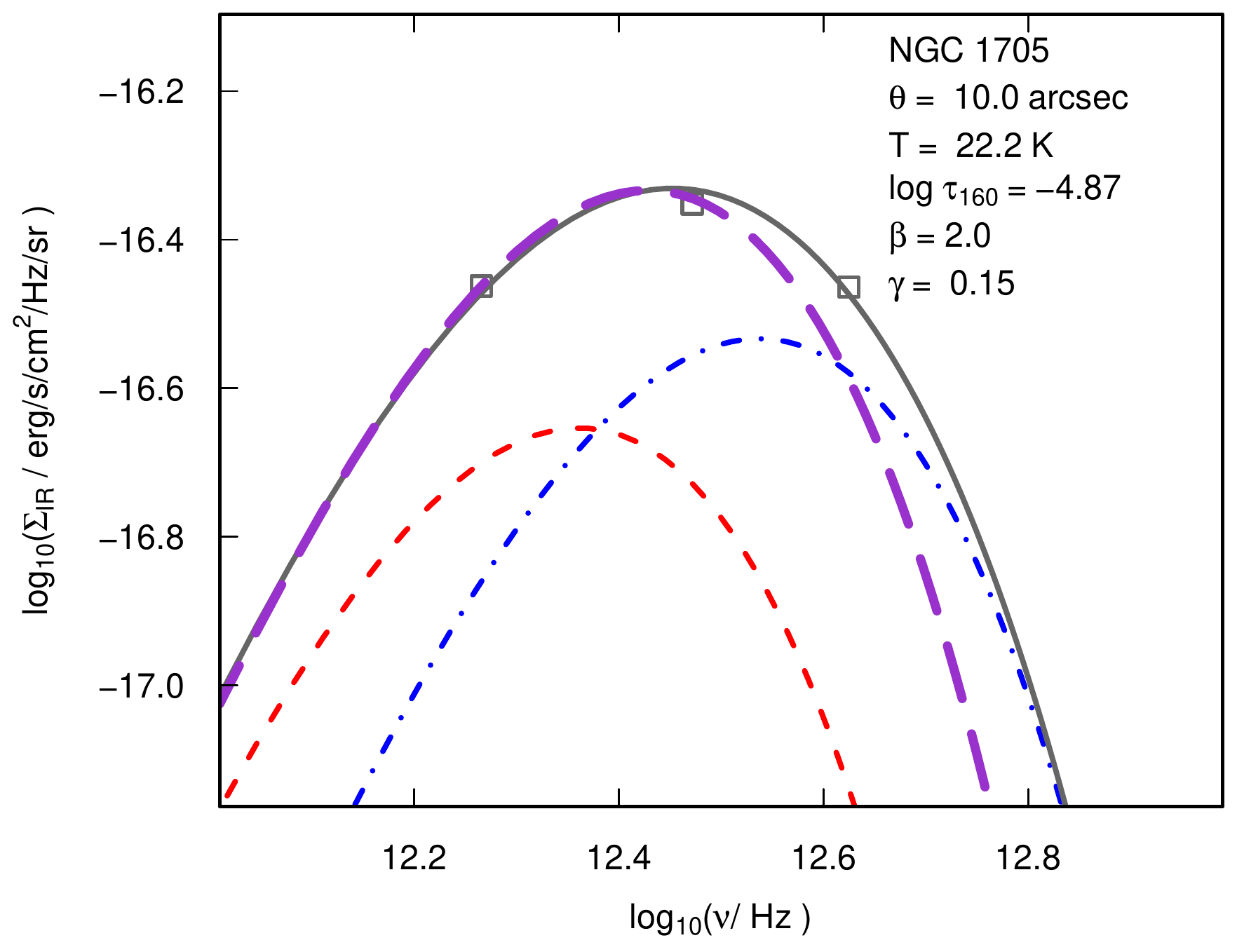}
 \includegraphics[angle=0,width=0.33\linewidth]{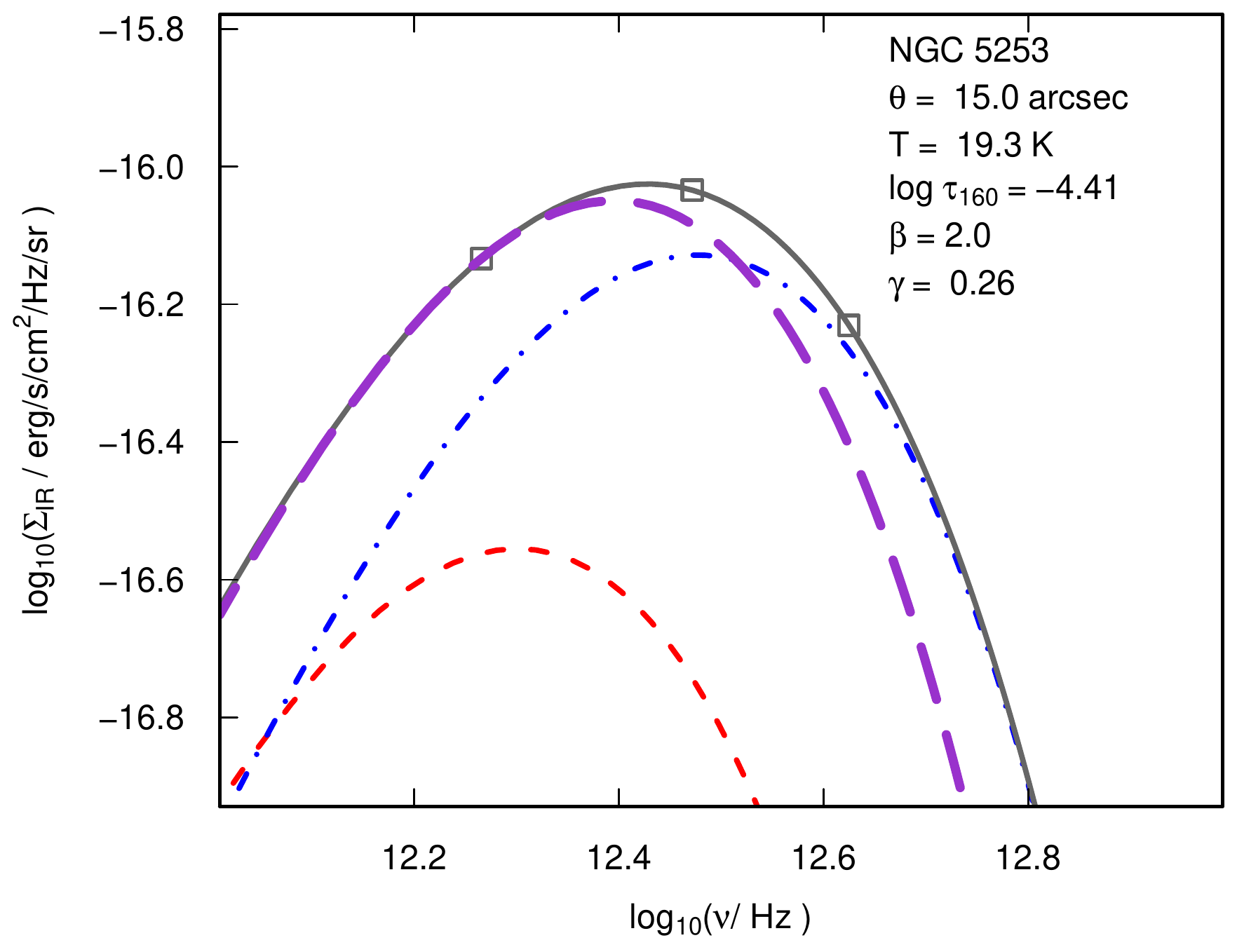}
 }
\caption{Representative fits of Eq. \eqref{eqn:btd} with significant $\gamma$
values to illustrate the importance of two temperatures in the fitting function.
The long-dashed red curves show the MBB with \tc, the dot-dashed blue curves \tw,
and the grey (solid) ones the overall best fit.
The (heavy) purple long-dashed curve corresponds to a single-temperature MBB with \td,
normalized to the 160\,\micron\ data point.
For $\gamma > 0$,
the simple MBB is clearly missing the contribution toward higher frequencies caused by warm dust at \tw.
\label{fig:seds}
}
\end{figure*}

\begin{figure*}[!h]
\hbox{
\includegraphics[angle=0,width=0.495\linewidth]{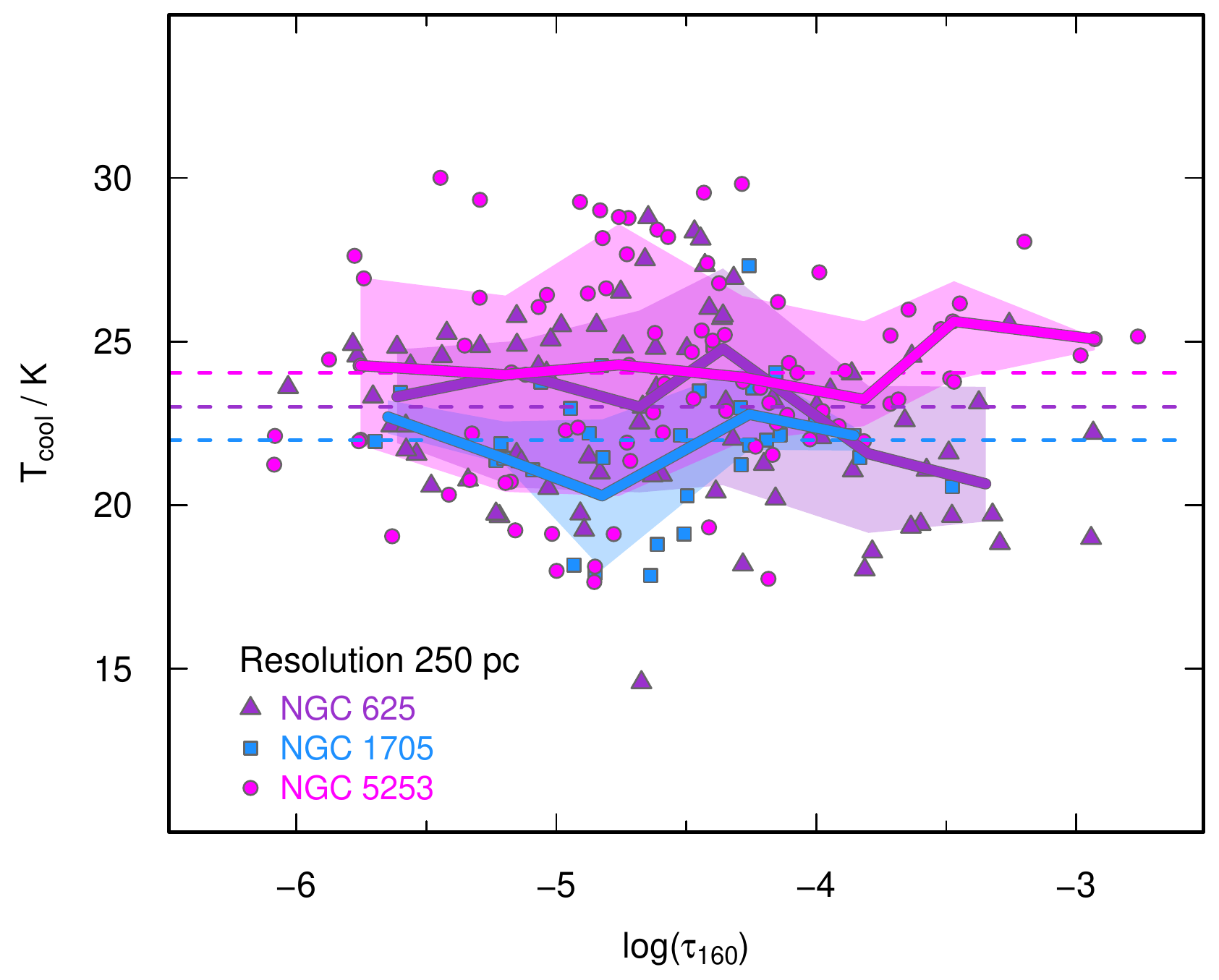}
\hspace{0.01\linewidth}
\includegraphics[angle=0,width=0.495\linewidth]{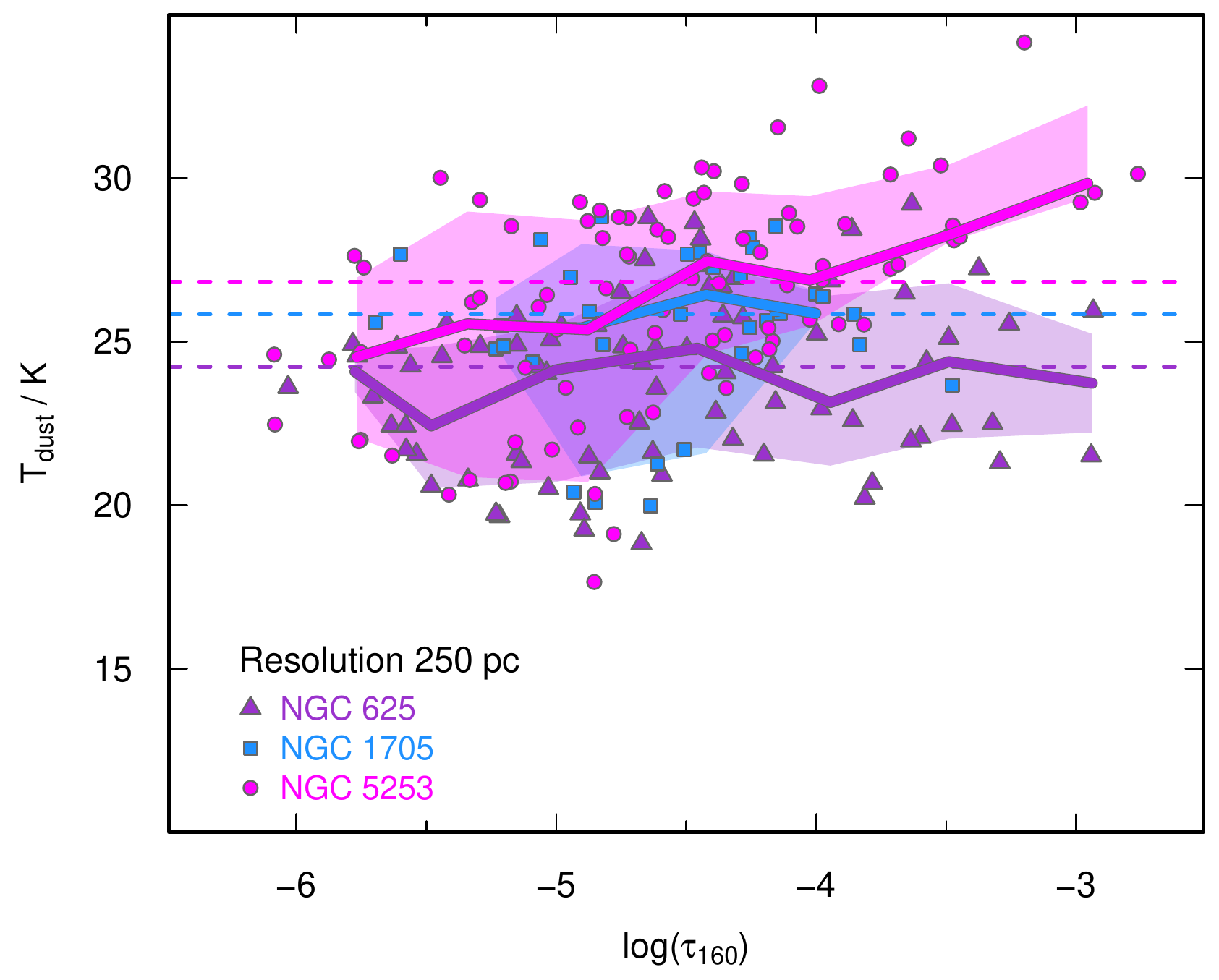}
}
\caption{Dust temperatures 
plotted against \mytau\ for the 250\,pc regions within the target galaxies: in the
\textit{left panel} best-fit \tc\ and in the \textit{right panel} effective temperature \td. 
The medians binned in log$_{10}$(\mytau) are shown as heavy lines characterized by different colors, 
with the 1$\sigma$ variation by shaded regions.
The horizontal dashed lines correspond to the overall \tc\ galaxy medians (left panel)
and \td\ (right). 
Only in the case of NGC\,5253 does the effective temperature \td\ present a trend with \mytau,
in the sense that more optically-thick regions show higher temperatures.
\label{fig:tdust}
}
\end{figure*}

The MBB-2T fitting function is given by: 
\begin{equation}
I_\nu\,=\,\tau_\mathrm{160}\ \left( \frac{2\,h\,\nu^3}{c^2} \right) \left( \frac{\nu}{\nu_0} \right)\ 
\left[ \frac{1\ -\ \gamma}{e^{h \nu/k T_\mathrm{cool}} - 1} + \frac{\gamma}{e^{h \nu/k T_\mathrm{warm}} - 1} \right] 
\label{eqn:btd}
\end{equation}
where \mytau\ is the PACS 160\,\micron\ opacity, $\nu_0$ corresponds to PACS 160\,\micron,
$\gamma$ is the fraction of the dust emission due to the warmer component, and
\tw\ and \tc\ are the warm and cool dust temperatures defined above. 
The MBB model is merely Eq. \eqref{eqn:btd} with $\gamma\,=\,0$  and \td\,=\,\tc.
Since it is the cool dust with T\,=\,\tc\ that defines \mytau, the MBB-T2 model is better for our purposes.
We know that there is temperature mixing along the line-of-sight, and thus
almost certainly a component of warmer dust that would skew the determination of \mytau.
Hence, we also define an effective temperature: 
\begin{equation}
T_\mathrm{dust} = [\,(1-\gamma) + \gamma\,(1.5)^6\,]^{1/6}\ T_\mathrm{cool}
\end{equation}


The observed PACS flux ratios are shown in Fig. \ref{fig:pacs}, together with 
individual MBB and MBB-T2 models.
The comparison of IR ratios from two MBB models, with $\beta$\,=\,1, 2
and the MBB-T2 model with $\gamma\,=\,0.14$, the median value for NGC\,1705,
shows that the two-temperature MBB-T2 model with $\beta\,=\,2$ mimics the MBB curve with lower 
emissivity, $\beta\,=\,1$.
This is a crude, although effective, example of temperature mixing
being able to reproduce flat spectral emissivity with $\beta\la\,1$
\citep[e.g.,][]{hunt15a}, which however does not represent grain physical properties.

Figure \ref{fig:seds} shows representative MBB-2T fits of the three targets, with non-zero $\gamma$ values.
The grey (solid) curve shows the best MBB-2T fits, while the long-dashed (purple) curves the single-temperature MBB fit
with T$\,=\,$\td.
The advantage of the two-temperature fits is evident by comparing the dotted MBB with \td\ to the
combined MBB-2T solid curve;
the single-temperature fit misses the high-frequency contribution from the warmer dust.

\begin{figure*}[ht!]
\hbox{
\includegraphics[angle=0,width=0.495\linewidth]{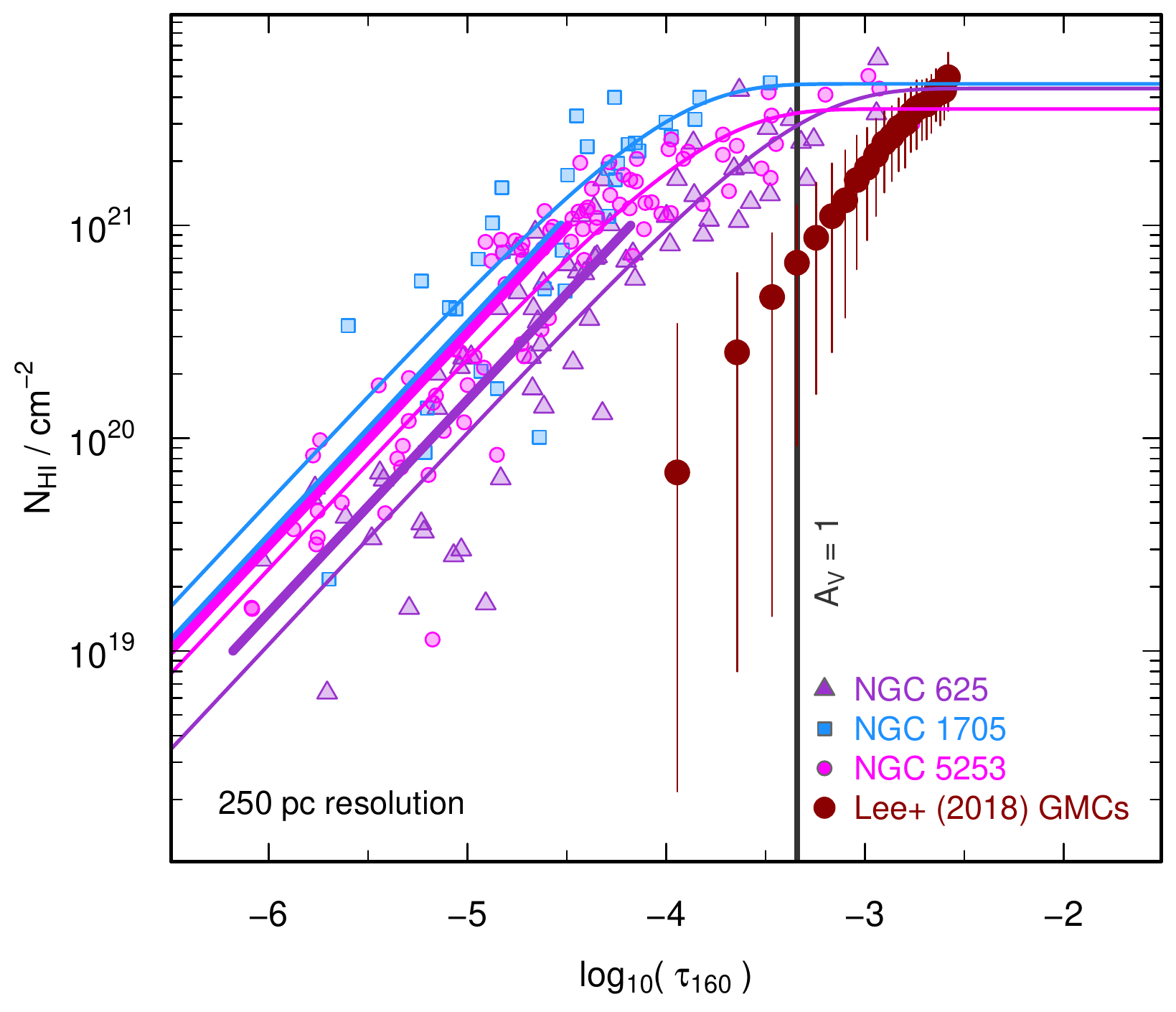}
\hspace{0.01\textwidth}
\includegraphics[angle=0,width=0.495\linewidth]{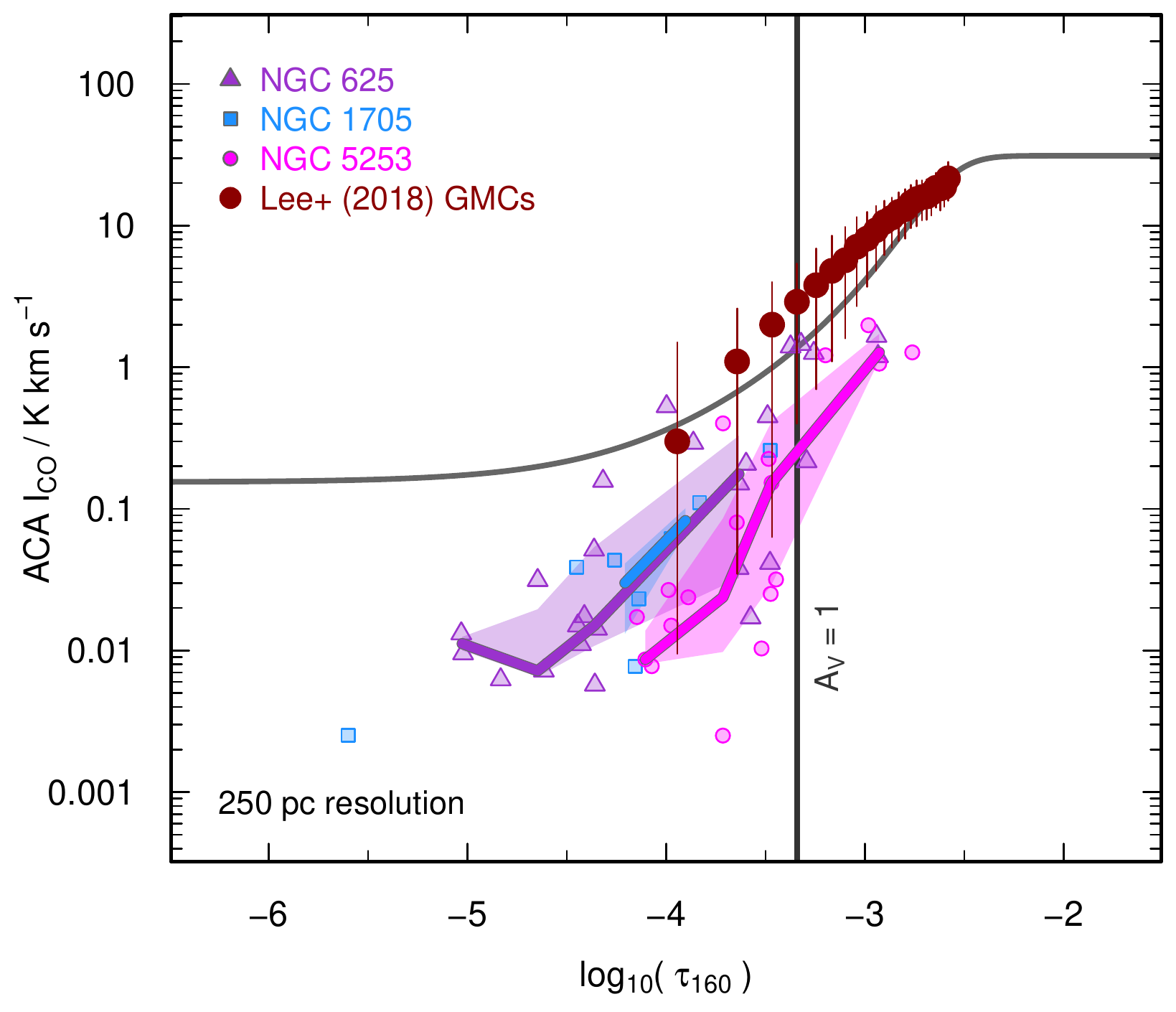}
}
\caption{Atomic gas column density \Nhi\ versus 160\,\micron\ optical depth \mytau\ (\textit{left panel})
and CO ACA (velocity-integrated) \tb\ versus \mytau\ (\textit{right}).
Only regions with S/N\,$\geq$\,3 are shown (and included in the fits).
The best-fit linear relations [Eq. \eqref{eqn:taudgr}] for \mytau\ as a function of \Nhi\ are shown as heavy curves (in log space),
and only consider the points with $10^{19}\,\leq\,$\Nhi$\,\leq\,10^{21}$\,\cmtwo\ because of the
need to consider the regions dominated by \hi.
The lighter curves show the fit to Eq.\,\eqref{eqn:radtranhi} as described in the text,
and are used to determine \avhicrit. 
Also shown are the GMCs from \citet{lee18} assuming \av/\mytau\,=\,2200 for consistency with their paper.
The vertical line shows the equivalent of \av\,=\,1 assuming this conversion.
The heavy (colored) curves in the right panel correspond to \ico\ binned in log(\mytau), and
the grey curve is the fit from \citet{lombardi06} 
for CO in the MW Pipe Nebula GMC versus \av\ converted to \mytau\ as above 
\citep[see also the study][on the Perseus molecular cloud complex]{pineda08}.
\label{fig:gastau}
}
\end{figure*}

The MBB-T2 fits are numerically non-trivial. 
We left $\gamma$ unconstrained, and 
in some cases, the fitted $\gamma\,<\,0$ or $\gamma\,>\,1$.
In those cases, an MBB fit (with $\gamma\,=\,0$, by definition)
was applied with $\beta\,=\,2.0$ as for the other MBB-T2 fits. 
This condition occurred for (45, 0, 33) individual resolution elements 
in  NGC\,(625, 1705, 5253).
Since rebinning to (13\arcsec, 10\arcsec, 15\arcsec) for NGC\,(625, 1705, 5253), respectively,
gives (77, 33, 90) resolution elements 
with PACS fluxes greater than the background, there are (58\%, 0, 37\%) of the fitted resolution
elements with single-T MBB fits.

Given that the expected uncertainty in the individual PACS photometry is $\sim$10\%,
the rms values of both MBB and MBB-2T fits are quite good, and, except for NGC\,1705, well within the expected
uncertainties.
Moreover, as shown in Appendix \ref{sec:taudetails},
the MBB-2T fits are overall superior to MBB, possibly due to the 
extra fitted parameter and the consequent zero degrees of freedom. 
More details of the fits to the PACS data to derive \mytau\ are given in Appendix \ref{sec:taudetails}.

Figure \ref{fig:tdust} compares the best-fit \tc\ with dust optical depth, \mytau\ (left panel),
and with \td\ versus \mytau\ (right).
We initially expected that \tc\ might correlate with \mytau;
however as shown in the left panel of Fig. \ref{fig:tdust}, it does not.
The range of \tc\ is relatively narrow, a standard deviation of 2-3\,K, 
over a 3-order of magnitude range of \mytau.
Moreover, the median cool-dust temperatures of the three targets do not differ significantly, given the 2-3\,K spread:
we find \tc\,=\,(23.0, 22.0, 24.1)\,K for NGC\,(625, 1705, 5253), respectively.
These temperatures are toward the high end of the temperature range of cool dust in normal, star-forming spirals
\citep[e.g.,][]{galametz12}.

\citet{kirkpatrick14} found that \tc\ is partly driven by star-formation
activity, so that higher temperatures would be found for more intense SFRs,
consistent with the starburst nature of our targets.
Moreover, as shown in the right panel of Fig. \ref{fig:tdust},
in NGC\,5253,
the effective temperature \td\ is correlated with \mytau\ although 
the other two galaxies show little trend.
Thus, in extreme dusty starbursts such as NGC\,5253, warmer dust is associated with higher \mytau.

\subsection{Infrared opacity and atomic gas surface density}
\label{sec:dgrtau}

Atomic gas is an important part of the overall gas budget,
especially in dwarf galaxies \citep[e.g.,][]{hunt20}.
Comparing atomic gas surface density with a measurement of
dust opacity enables an estimate of the dust-to-gas ratio (DGR),
as well as an assessment of the overall association of dust content with \hi.
Figure \ref{fig:gastau} (left panel) shows the trend of \hi\
column density \Nhi\ with \mytau\ determined in the previous
section; only points with signal-to-noise S/N$\geq$ 3 are shown.
The data show a fairly tight correlation, but with an inflection or flattening
at 
\mytau$\,\sim\,10^{-4}$.
This is due to the ``saturation'' of \Nhi\ at high column densities
where most of the hydrogen has transitioned to molecular form, traceable by CO.
Such a correlation between spatially-resolved dust content and \hi\ is common in \hi-dominated
dwarf galaxies \citep[e.g.,][]{leroy09}, but is not 
generally observed in spirals \citep[e.g.,][]{hughes14,casasola22}.

\begin{figure*}[t!]
\hbox{
\includegraphics[angle=0,width=0.495\linewidth]{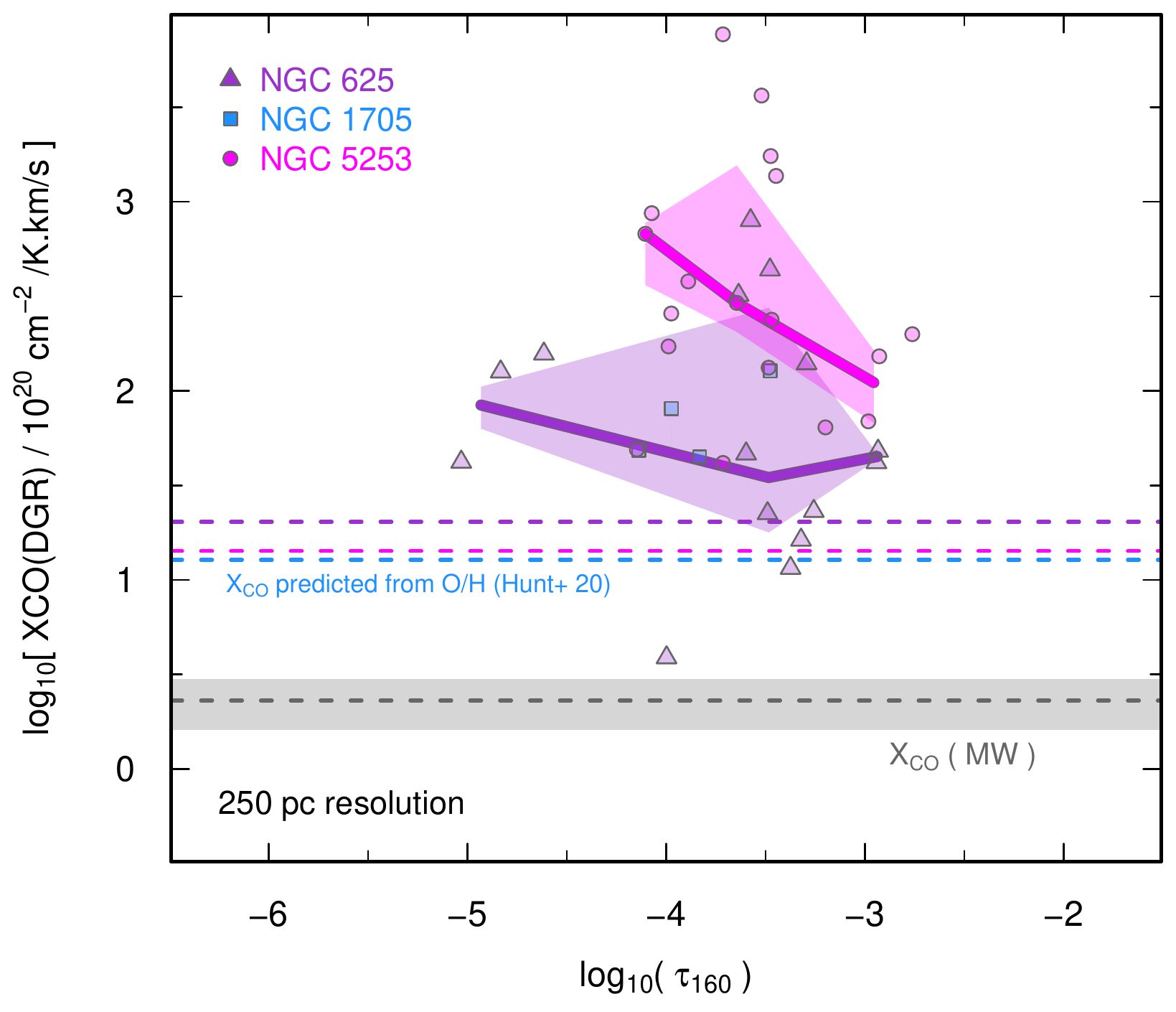}
\hspace{0.01\linewidth}
\includegraphics[angle=0,width=0.495\linewidth]{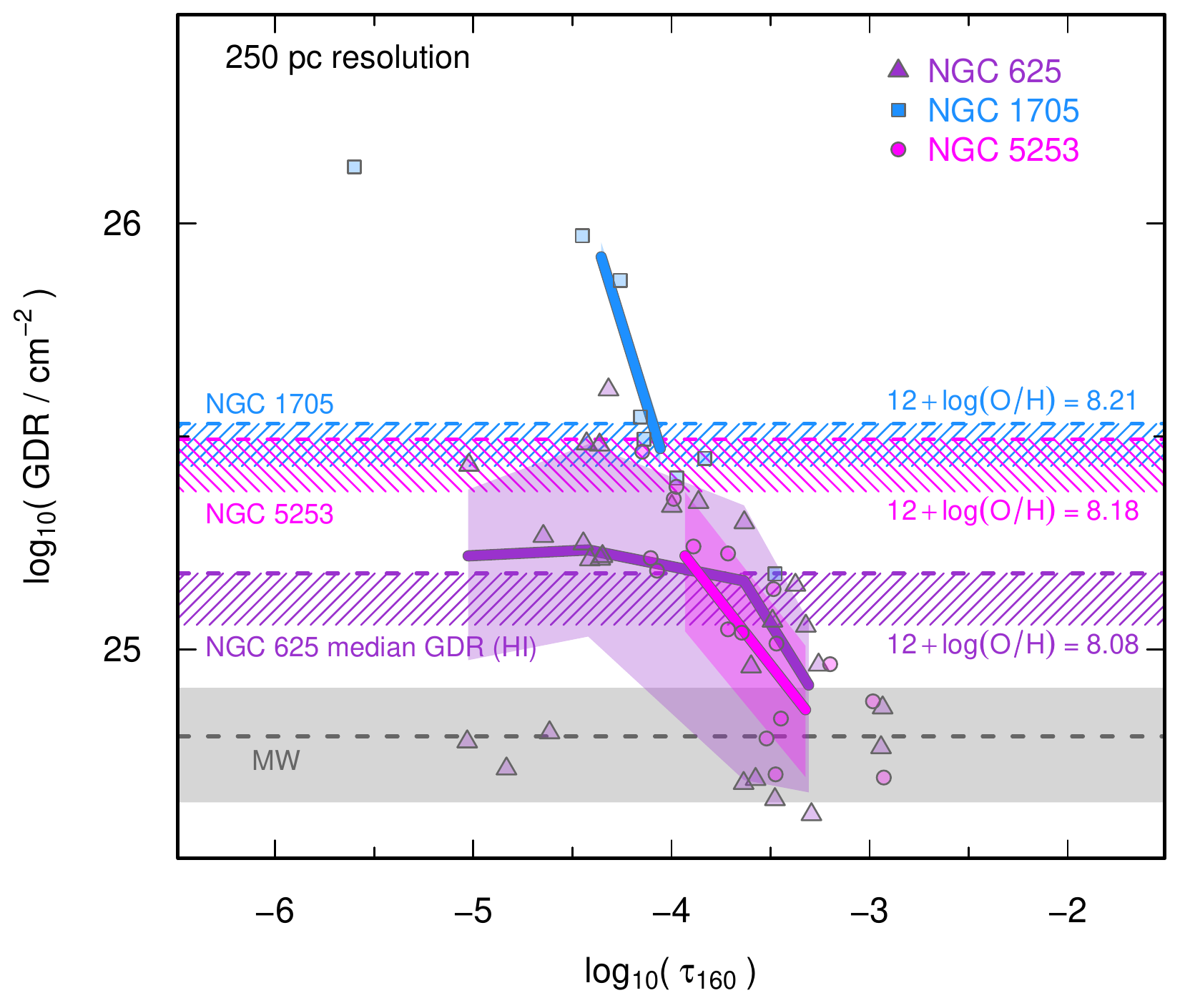}
}
\caption{\textit{Left panel:} Inferred \xco\ conversion factor from Eq.\,\eqref{eqn:h2dgr} as a function of \mytau.
Also shown are values that would be estimated from the metallicity dependence found by \citet{hunt20}
as \xco\,$\propto$\,$Z^{-1.55}$ \citep[see also][]{accurso17}.
Heavy curves show the medians in log$_{10}$(\mytau) for each galaxy 
(except for NGC\,1705 for which there are too few points), 
and the shaded regions indicate 1$\sigma$ standard deviations. 
\textit{Right panel:} GDR (here we are using the inverse of \deltadgr)
that would be inferred assuming the metallicity dependence of the \xco\ conversion factor 
found by \citet{hunt20}.
The range of GDRs estimated by different methods is shown as the hatched horizontal strips. 
As in the left panel, 
heavy curves show the medians, and shaded regions the 1$\sigma$ standard deviation.
\label{fig:xcotau}
}
\end{figure*}

We assume that for dark clouds with large \mytau\
there is a finite amount of \hi\ in the cloud ``skin'',
and that the cloud chemistry limits the amount of \hi\ 
\citep[see also][]{pineda08}. 
Thus, we can write:

\begin{equation}
N_\mathrm{HI}\,=\,\mathrm{N0}_\mathrm{HI}\ \left( 1 - e^{-k_\mathrm{HI}\,\tau_\mathrm{160}} \right), 
\label{eqn:radtranhi}
\end{equation}
where \Nohi\ is the \hi\ column density at saturation, and \khi\ defines the 
transition threshold for the dust optical depth \mytau\ where saturation onsets.
The best fits to this function are shown as curves in the left panel of Fig. \ref{fig:gastau};
only points with \hi\ signal-to-noise (S/N)\,$\geq$\,3 are included in the fits.
We will incorporate \khi\ to estimate \xco\ from \av\ in Sect. \ref{sec:smallscale}.

Following \citet{leroy09} and \citet{bolatto11}, we estimate
the DGR from \mytau\ and the total gas column density, \ngas\ (\ngas\,$\equiv$\,\Nhi\,$+$\,2\,\Nhtwo),
by fitting a linear relation: 
\begin{equation}
\tau_{160}\,=\,\delta_\mathrm{DGR}\,N_\mathrm{gas} ,
\label{eqn:taudgr}
\end{equation}
where \deltadgr\ is the slope of the relation, namely the DGR, and
\ngas\ is in units of column density \cmtwo. 
Here the
underlying assumption is that this relation holds at all column densities. 
Thus, we can first calibrate it on \hi, 
by restricting the fit to low column densities where \ngas\ is dominated by \hi,
and then extrapolate it to \hi$+$\htwo.
The fit of Eq. \eqref{eqn:taudgr} was applied to \hi\ columns with 19$\,\leq\,$log(\Nhi/\cmtwo)$\,\leq\,21$;
only points with signal-to-noise (S/N)\,$\geq$\,3 are included in the fit.
We also performed fits with a non-zero intercept $\tau_0$, and found that it is always 0 to within the uncertainties;
thus we fixed $\tau_0$ to 0 to ensure a more reliable estimation of \deltadgr. 

As shown in Fig. \ref{fig:gastau},
these linear fits 
well approximate the trend of \mytau\ and \hi\ at low column densities where \hi\ dominates the gas budget.
The best-fit \deltadgr\ in Eq.\,\eqref{eqn:taudgr} corresponds to $\langle$\,\mytau/\Nhi\,$\rangle$
and differs slightly for the three galaxies, 
in qualitative agreement with what we would expect considering
the albeit small metallicity differences
\citep[e.g.,][see Table\,\ref{tab:sample}]{remyruyer14}:
\begin{equation}
\delta_\mathrm{DGR}\,=\,\begin{cases}
(6.6\,\pm\,0.4)\,\times\,10^{-26}\ \mathrm{cm}^{2} & \text{NGC\,625} \\
(3.0\,\pm\,0.5)\,\times\,10^{-26}\ \mathrm{cm}^{2} & \text{NGC\,1705} \\
(3.2\,\pm\,0.2)\,\times\,10^{-26}\ \mathrm{cm}^{2} & \text{NGC\,5253} 
\end{cases}
\label{eqn:dgr}
\end{equation}
These values are 3$-$5 times lower 
than \mytau/$N_H\,=\,1.61\times10^{-25}$\,\cmtwo/H,
found in the local diffuse ISM by \citet{hensley21}, 
roughly consistent with the linear trend with the 
$\sim\,\frac{1}{3}-\frac{1}{4}$\,\zsun\ metallicities of our targets.
The values for our dwarf starbursts are also comparable with \deltadgr\ for N\,83,
with O/H of $\sim\frac{1}{5}$\,\zsun\ \citep[e.g.,][]{berg12}, 
in the Small Magellanic Cloud (SMC): 
$\tau_\mathrm{160}\,=\,2.3-4.3\,\times\,10^{-26}\,N_\mathrm{gas}$/\cmtwo\ \citep{bolatto11},
where we have reported their 36\% correction for helium to our convention without the helium correction
\citep[see also][]{leroy09}.

In the limit of the \hi-dominated regime at low \mytau, Eq. \eqref{eqn:radtranhi} should converge to Eq. \eqref{eqn:taudgr} with
\deltadgr\,=\,1/($\mathrm{N0}_\mathrm{HI}\ k_\mathrm{HI}$).
Thus, in principle, we could have used the fits of Eq. \eqref{eqn:radtranhi} to infer \deltadgr, rather than Eq. \eqref{eqn:taudgr}.
However, Fig. \ref{fig:gastau} shows offsets of 30-40\% between the two estimates, so it is unclear which to prefer.
After error propagation, the uncertainties on \deltadgr\ from the fits of Eq. \eqref{eqn:radtranhi}
are larger than those given in Eq. \eqref{eqn:dgr}, so 
we prefer to adopt the results 
from fitting Eq. \eqref{eqn:taudgr} with only one free parameter and over the range in column
density dominated by \hi.

The DGR \deltadgr\ is given in units of optical depth per H, 
because we want to compare dust columns with gas columns.
Although it would be preferable to determine the DGR for individual regions
\citep[e.g.,][]{bolatto11,sandstrom13},
this is difficult for our dwarf targets. 
The small number of individual resolution elements within each galaxy 
effectively precludes a statistically relevant spatially-resolved DGR determination.
Thus, we use an average DGR determined from the data 
as shown in the left panel of Fig. \ref{fig:gastau}.
This point is further discussed in Sect. \ref{sec:conclusions}.

\subsection{Molecular gas surface density, \mytau, and \xco\ at 250\,pc resolution}
\label{sec:xco250}

The trend of CO velocity-integrated brightness temperature observed with ACA at $\sim$250\,pc resolution
is shown in the right panel of Fig. \ref{fig:gastau}.
Also shown are the giant molecular clouds (GMCs) in the Milky Way (MW) studied by \citet{lee18},
where they estimate \av\ from $\tau_{850}$.
As for \hi, \ico\ in units of \kkms\ 
correlates with dust opacity, although with somewhat larger scatter than for \hi.
At a given \mytau, our low-metallicity starbursts show lower surface brightness
CO emission than the MW GMCs.
Part of this may be due to the much higher resolution in the MW observations \citep[1\,pc,][]{lee18},
relative to the 250\,pc regions studied here.
However, another part of this stems from the \xco\ factor needed to bring observed CO brightness temperatures
to gas column densities.

We now have the necessary ingredients to infer \Nhtwo\ from \Nhi\ and \deltagdr,
using the following expression:

\begin{equation}
N_\mathrm{H2}^\mathrm{DGR}\,=\,\frac{1}{2}\,\left( \frac{\tau_{160}}{\delta_\mathrm{DGR}} - N_\mathrm{HI} \right)\quad ,
\label{eqn:h2dgr}
\end{equation}
where \deltadgr\ is determined from the fit of Eq.\,\eqref{eqn:taudgr}.
For each 250\,pc resolution element, we can derive \Nhtwodgr, and then use \ico\ to 
find the local \xco\,=\,\ico/\Nhtwodgr. 
To include a resolution element in this estimate, we require that both \hi\ and CO binned data points have S/N$\geq$3. 
Thus, instead of the (70, 33, 90) 250\,pc PACS resolution elements for the 
determination of \mytau\ for NGC\,(625, 1705, 5253), respectively,
the numbers are now reduced to (26, 8, 18) usable elements for the three galaxies.
However, a small fraction of these gives negative values of \Nhtwo\ in Eq.\eqref{eqn:h2dgr}, because
the \hi\ column density slightly exceeds what would be inferred from the ratio,
\mytau/\deltadgr. 
This could be a consequence of our assumption of a global DGR, as also discussed in Sect. \ref{sec:conclusions}.
In any case, the number of regions
useful for the calculation of \xco\ dwindles further to (20, 4, 17) for  NGC\,(625, 1705, 5253), respectively.

The results are shown in the left panel of Fig. \ref{fig:xcotau}, where
also illustrated are the metallicity-dependent predictions for global \xco\ [\xco$\,\propto\,(Z/Z_\odot)^{-1.55}$]
according to \citet{hunt20}.
The DGR-inferred \xco\ values 
sometimes exceed the expectation from the metallicity dependence, in some cases by 
an order of magnitude or more.
The large scatter precludes definitive statements, but the main point that emerges from
Fig. \ref{fig:xcotau} is that a simple metallicity dependence does not explain the \xco\ factors
derived from the DGR on 250\,pc scales.

To investigate this further, and check for consistency, the right panel of Fig. \ref{fig:xcotau}
shows the gas-to-dust ratios (GDR) that would be inferred if we had assumed a metallicity-dependent \xco.
The plotted GDRs are simply the inverse of \deltadgr\ that were derived in Eq. \eqref{eqn:taudgr}, and
the hatched areas 
show the range of GDRs estimated by different methods. 
At \mytau$\la\,10^{-4}$, where the gas budget is dominated by \hi,
the inferred GDRs more or less conform to expectations.
However, for higher \mytau, where \htwo\ inferred from CO starts to impact the gas content,
the GDRs are lower, reaching MW values at \mytau$\sim 10^{-3}$.
Such behavior is consistent with the left panel of Fig. \ref{fig:xcotau} where \xco\
computed from the DGRs are higher than what would be expected from the 
global metallicity dependence found by \citet{hunt20}.
This point will be explored in Sect. \ref{sec:comparison}.

\begin{figure*}[t!]
\hbox{
\includegraphics[angle=0,width=0.495\linewidth]{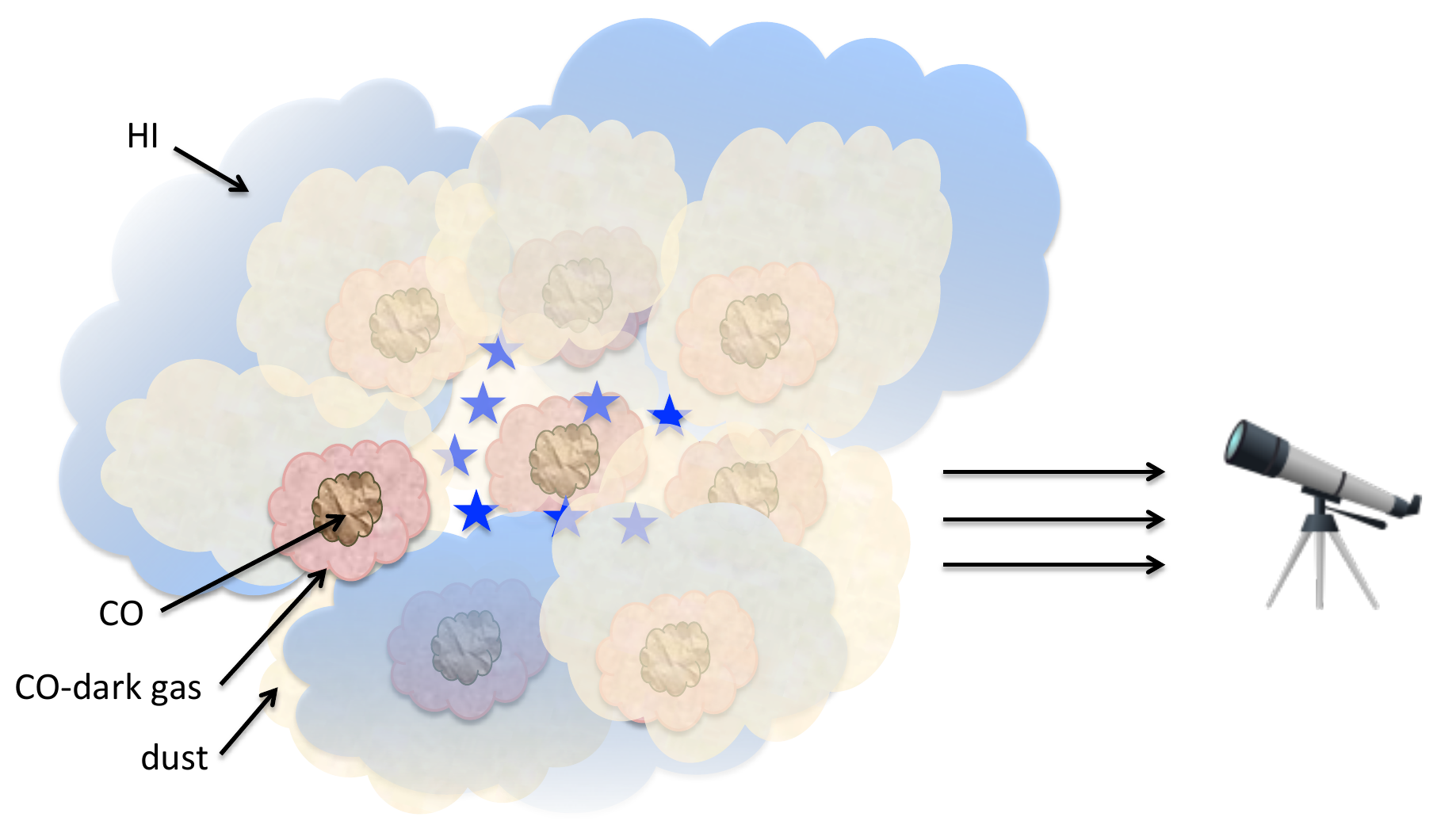}
\hspace{0.01\linewidth}
\includegraphics[angle=0,width=0.495\linewidth]{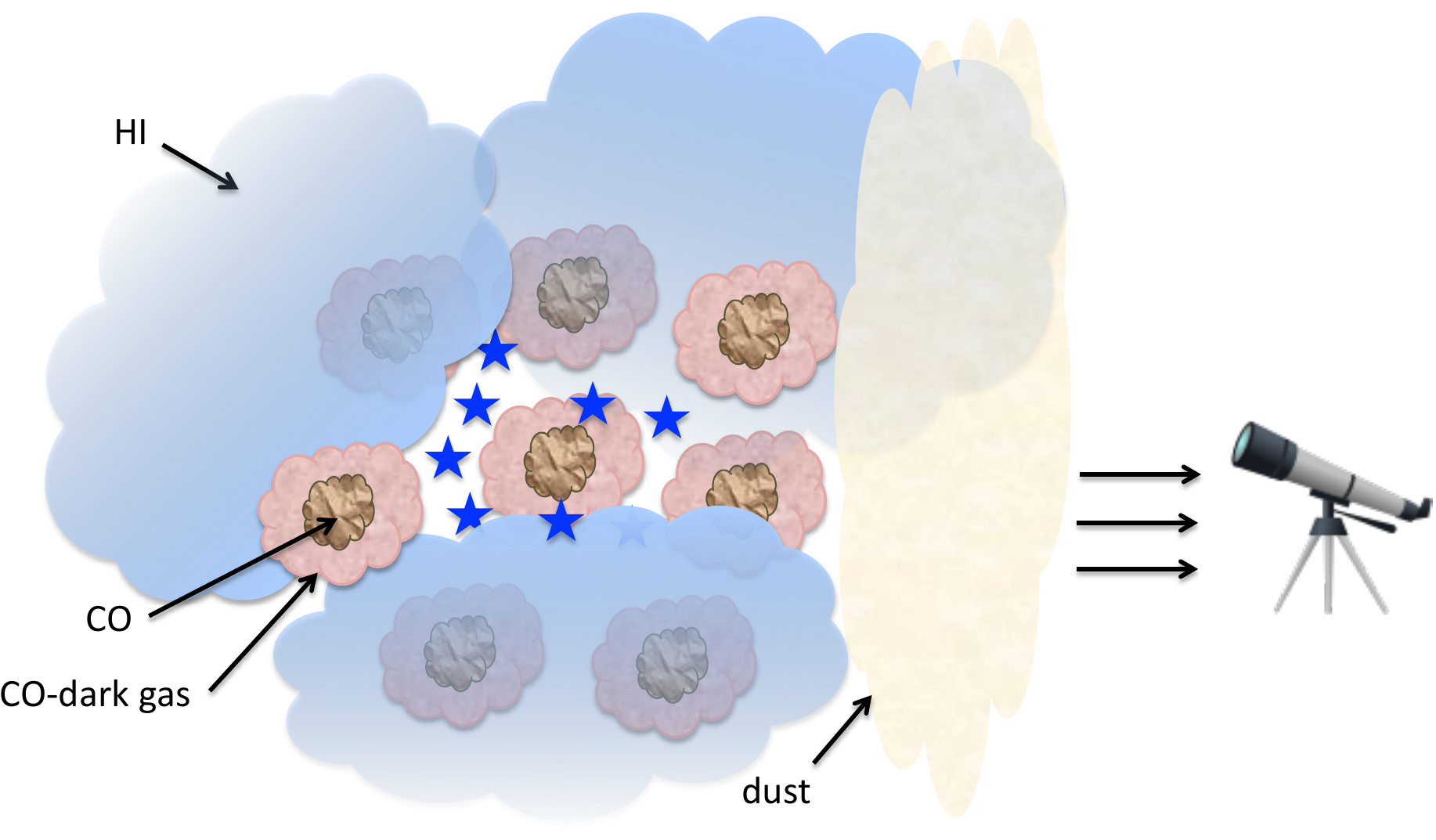}
}
\caption{\textit{Left panel:} Idealized cartoon illustrating the mixed configuration of dust, stars, and gas.
\textit{Right panel:} The same but with the foreground screen configuration.
The amount of dust in the screen in the right panel is assumed to be roughly half of the 
total dust shown in the left panel.
While the infrared \mytau\ penetrates the full dust column along the line of sight,
the optical depth inferred from the Balmer decrement only captures the dust between
the stars that ionize the gas and the observer.
\label{fig:mixedscreen}
}
\end{figure*}
\begin{figure*}[h!]
\hbox{
\includegraphics[angle=0,width=0.495\linewidth]{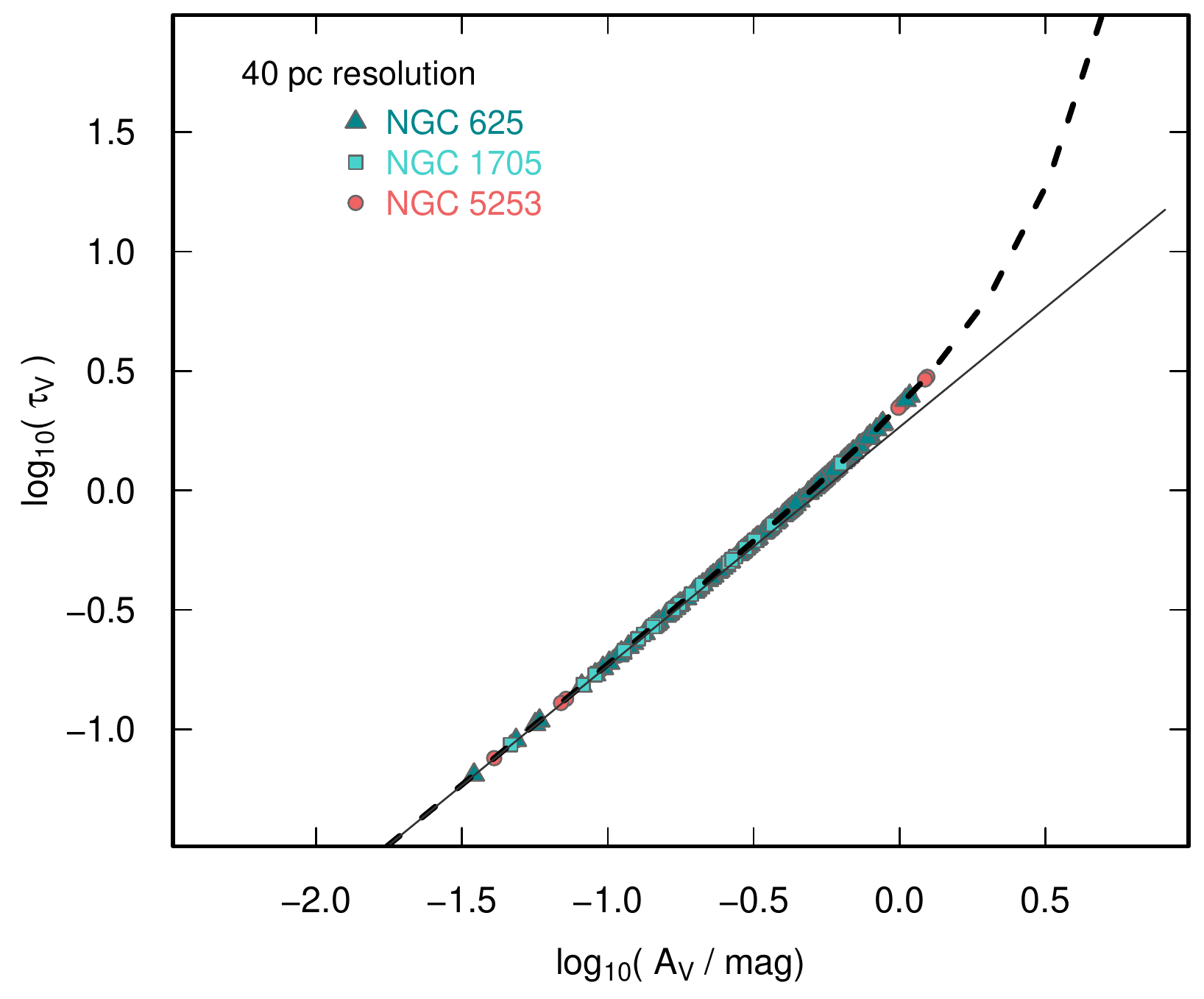}
\hspace{0.01\linewidth}
\includegraphics[angle=0,width=0.495\linewidth]{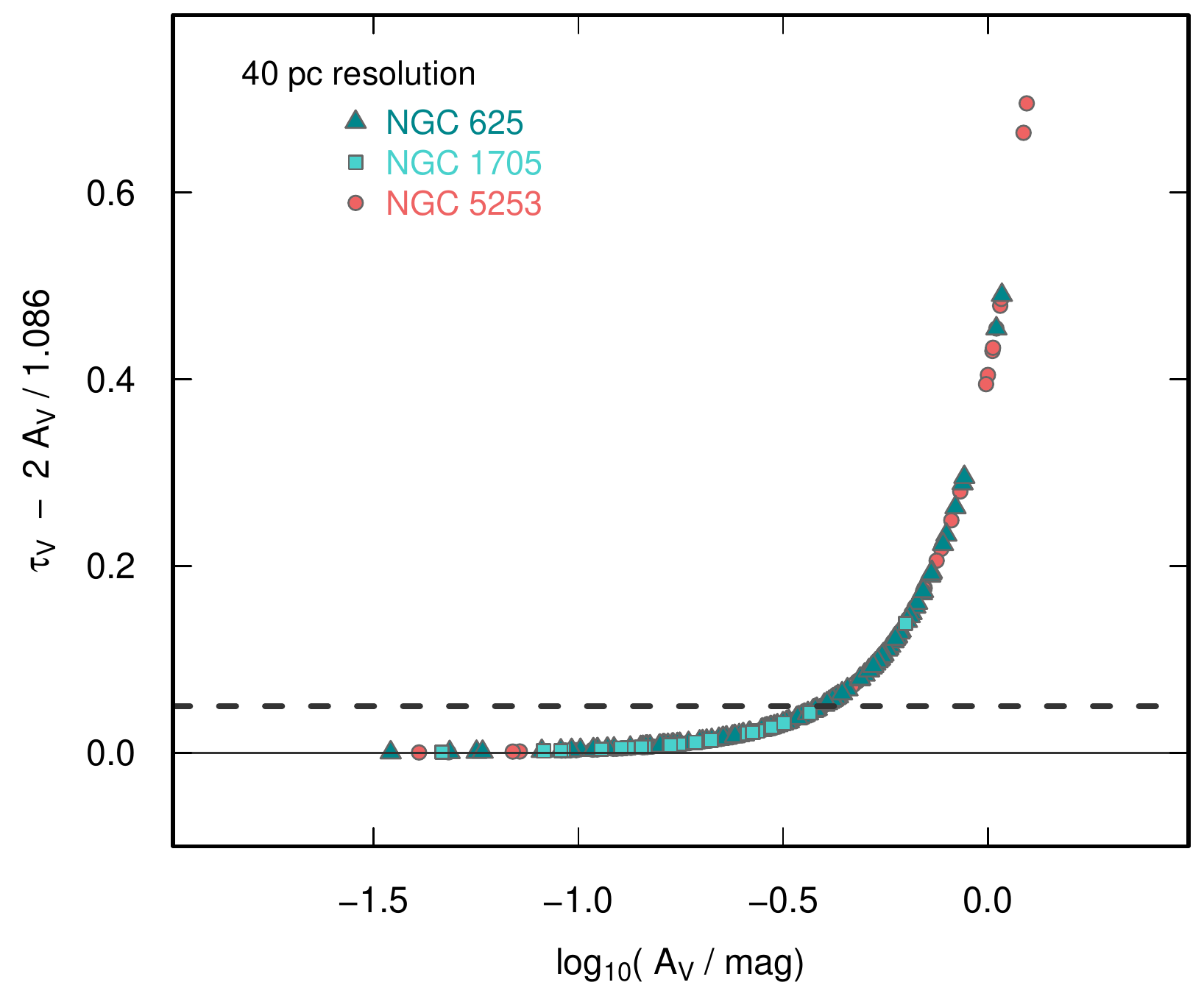}
}
\caption{\textit{Left panel}: \tauv\ inferred from Eq.\eqref{eqn:lambert} for the targets plotted against log$_{10}$(\avmuse).
The heavy dashed curve shows the true solution, and the lighter weight solid curve
shows not identity, but rather $y\,=\log_{10}(\tau_V)\,=\,x - \log_{10}(1.086) + \log_{10}(2)$,
where $x\,=\,$log$_{10}$(\avmuse).
The factor of 2 takes into account that \avmuse\ only probes the extinction along the 
line of sight, i.e., the front of the foreground screen, and the other factor is
from Eq.\,\eqref{eqn:avscreen}. 
The difference of the two quantities \tauv\ and \avmuse\ is negligible at low \tauv,
but becomes significant for \avmuse$\gtrsim 0.3-0.5$\,mag.
At \avmuse$>$1\,mag, \tauv\ ``runs'' away, relative to \av. 
\textit{Right panel}: Residuals of \tauv\ - 2\,\avmuse/1.086 versus log(\av).
Although difficult to appreciate in the left panel, 34\% of the data points have residuals $>$0.05,
marked by the horizontal dashed line. 
\label{fig:tauv_vs_av}
}
\end{figure*}

\section{Small-scale dust extinction, molecular gas content, and the \xco\ conversion factor
\label{sec:smallscale}}

We now turn to the ALMA 12-m data with a higher resolution, $\sim$40\,pc.
Dust columns at this resolution are inferred from foreground visual reddening, \ebv, estimated
from the Balmer decrement from the MUSE maps.
We convert \ebv\ to \av\ by assuming \av/\ebv\,=\,3.1 \citep[e.g.,][]{draine03b}.
The problem with tracing dust extinction through \avmuse\ from the Balmer decrement
is that \av\ gives attenuation rather than the true visual optical depth \tauv.
Moreover, the attenuation \avmuse\ is due only to dust in front
of the ionized gas, whereas \mytau\ measures all of the dust along the line of sight. 
More generally, we need to distinguish between dust \textit{attenuation} from \avmuse\
and true visual \textit{optical depth} \citep[e.g.,][]{boquien13}.
In the following, we quantify this distinction and show that 
\avmuse\ from the Balmer decrement does not always directly trace \tauv\ in these low-metallicity starbursts.

\begin{figure*}[t!]
\hbox{
\includegraphics[angle=0,width=0.495\linewidth]{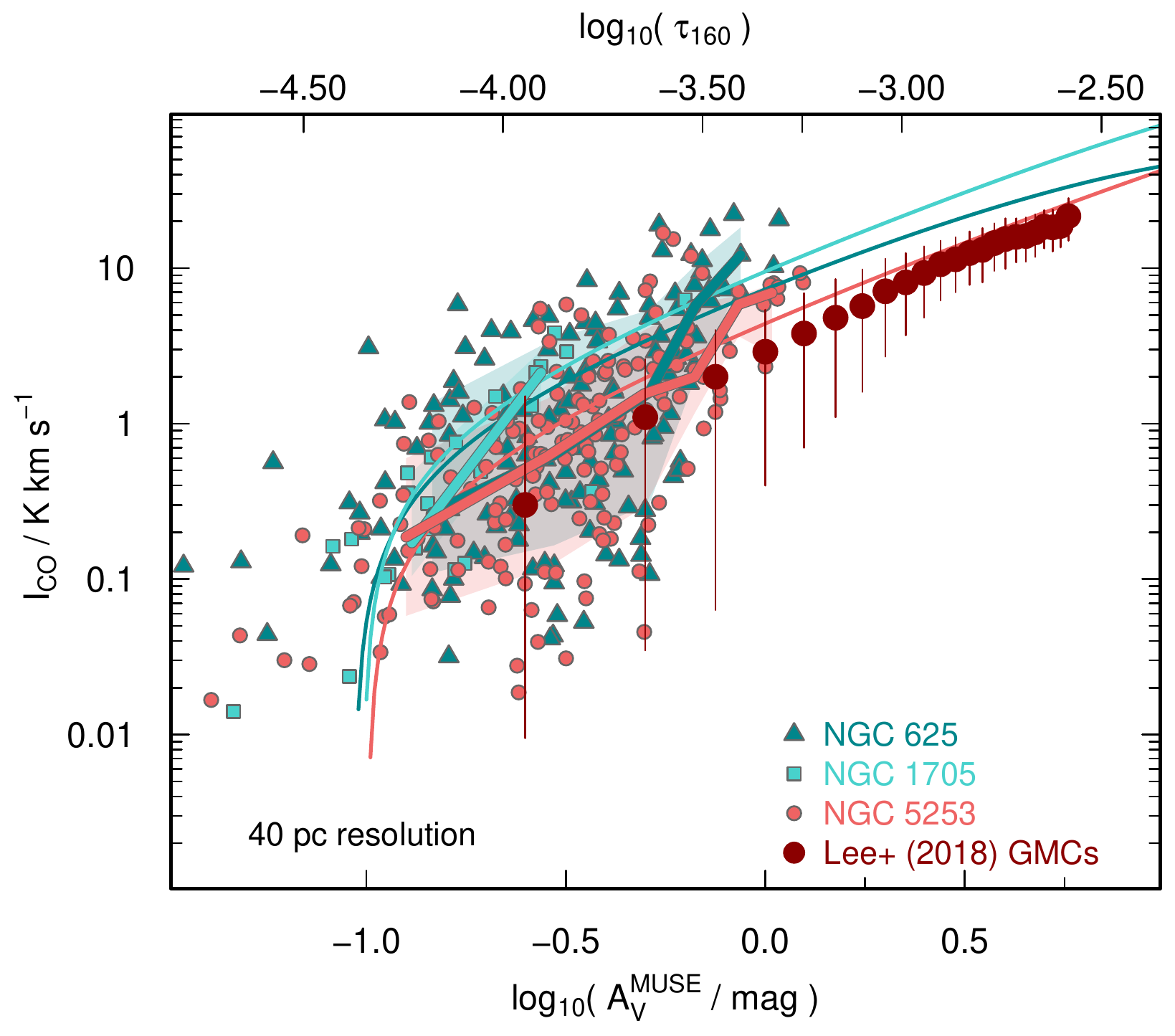}
\hspace{0.01\linewidth}
\includegraphics[angle=0,width=0.495\linewidth]{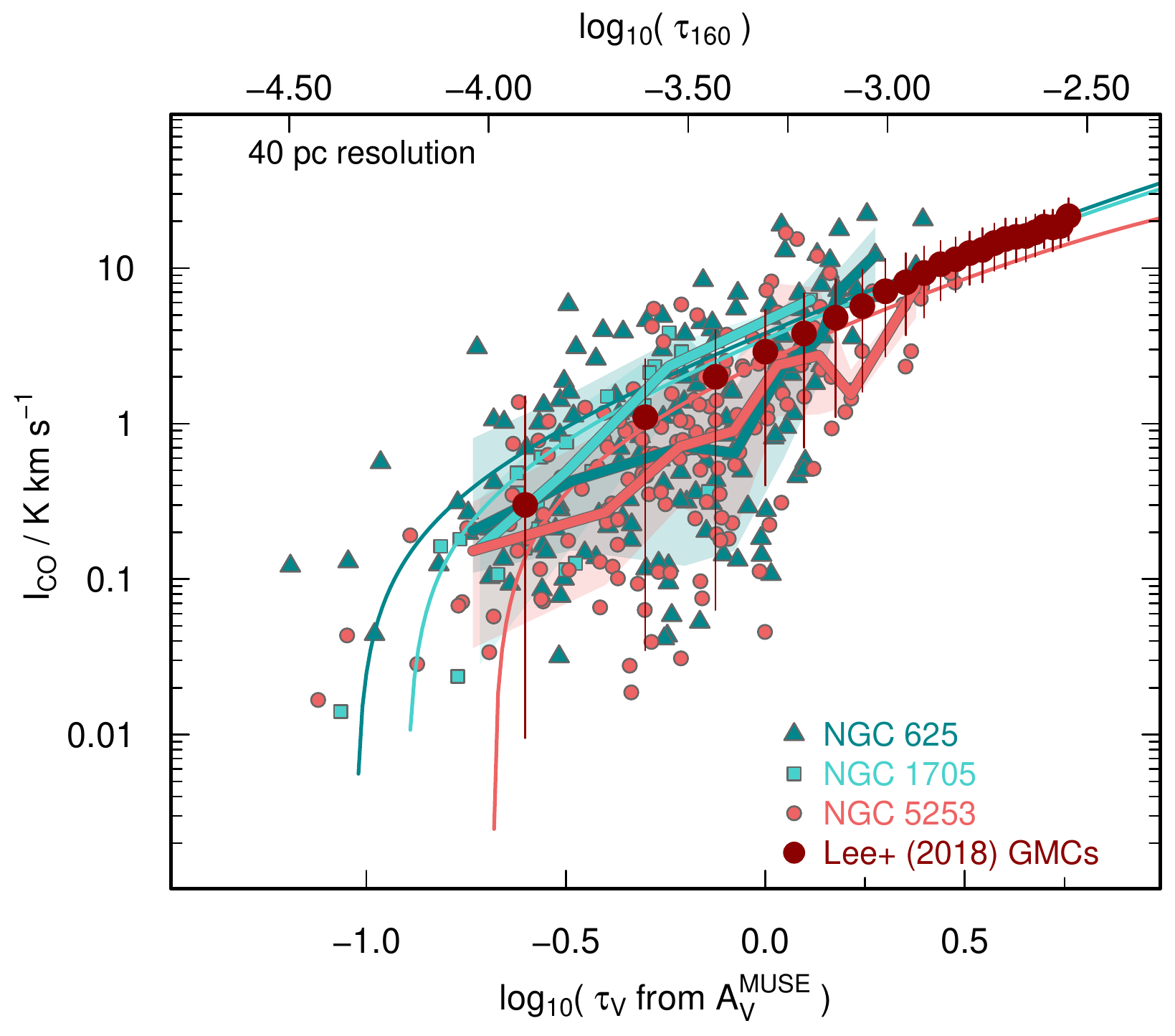}
}
\caption{CO velocity-integrated \tb\ \ico\ plotted versus (logarithm) visual extinction \av\ (\textit{left panel}) and (logarithm) 
visual dust optical depth \tauv\ (\textit{right}).
The top axes show the scale for \mytau\ assuming that \av/\mytau\,=\,2180, as described in Sect. \ref{sec:xco_tauv}.
The light curves show fits to Eq. \eqref{eqn:radtranco}. 
Heavy curves show the medians for each galaxy binned in log$_{10}$(\av) (left), log$_{10}$(\tauv) (right), 
with the shaded regions corresponding to $\pm\,1\sigma$ deviations.
Also shown in both panels are the MW GMCs studied by \citet{lee18}, with \av\ inferred from \textit{Planck};
see 
Sect. \ref{sec:smallscale}
for more details.
\label{fig:avco}
}
\end{figure*}

\subsection{Visual attenuation versus extinction and dust optical depth}
\label{sec:av_tauv} 

Assuming that the extinguishing dust is placed in a foreground screen,
the relation between attenuation and dust optical depth is simple:

\begin{equation}
A_V\,=\,-2.5\,\log_{10}(e^{-\tau_V})\,=\,1.086\,\tau_V
\label{eqn:avscreen}
\end{equation}
Thus, the attenuation \av\ is exactly proportional to the dust optical depth \tauv.
However, assuming that the dust is uniformly mixed with the emitters gives a slightly
more complicated relation:
\begin{equation}
A_V\,=\,-2.5\,\log_{10}\left( \frac{1-e^{-\tau_V}}{\tau_V} \right).
\label{eqn:avmixed}
\end{equation}

Figure \ref{fig:mixedscreen} illustrates these two configurations of the dust.
The infrared optical depth traced by \mytau\ penetrates the full dust column along the line of
sight, while extinction given by the Balmer decrement only probes the dust between the
observer and the ionized gas.
The assumption that the dust is uniformly mixed with the ionized gas is an approximation,
but, as shown in the figure, has the advantage of being slightly more realistic from a physical point of view.

As described by \citet{natta84}, \citet{disney89}, and others \citep[e.g.,][]{witt96,witt00}, 
solutions can be obtained for even more complex assumptions on the relative geometry of the dust and
the emitting regions.
In any case, the problem that emerges is the derivation of \tauv\ from \av\ in
more complex geometries of the dust relative to the emission source. 
Equation \ref{eqn:avmixed} provides a fortunate example because it is
invertible using the Lambert W function as shown by \citet{boquien13}\footnote{The Lambert W function
provides the solution $y$ of $x\,=\,y\ e^y$, where $x\,=\,-10^{0.4\,A_V}\ e^{-10^{0.4\,A_V}}$ 
and $y\,=\,\tau_V - 10^{0.4\,A_V}$.}:

\begin{equation}
\tau_V\,=\,W\ (-10^{0.4\,A_V}\ e^{-10^{0.4 A_V}})\ +\ 10^{0.4 A_V} \ .
\label{eqn:lambert}
\end{equation}

Here we make the assumption that the dust is uniformly mixed with the ionized gas,
namely that Eq.\,\eqref{eqn:avmixed} is a better description
of the dust column \tauv\ than Eq.\,\eqref{eqn:avscreen}, at least for our
use of the Balmer-decrement derived \avmuse\ to infer dust columns. 
We calculate \tauv\ under this assumption and compare it with what would
be inferred by simply equating \tauv\ with \av\ (with a small proportional factor)
as in Eq.\,\eqref{eqn:avscreen}.

The comparison of \av\ and \tauv\ [Eqs.\,\eqref{eqn:avmixed}, \eqref{eqn:lambert}] 
under the assumption of dust uniformly mixed with the 
ionized gas emission is shown in Fig. \ref{fig:tauv_vs_av}.
The dashed curve shows the analytical solution, while the solid line indicates
$y\,=\log_{10}(\tau_V)\,=\,x - \log_{10}(1.086) + \log_{10}(2)$,
where $\mathbf{x\,=\,\log_{10}}$(\avmuse).
The factor of 2 takes into account that \avmuse\ only probes the extinction along the 
line of sight, i.e., the front of the foreground screen (see above), and the other factor is
from Eq.\,\eqref{eqn:avscreen}. 
Indeed, for small \tauv, Eq. \eqref{eqn:avmixed} reduces to \av\,$\approx$\, 1.086\,(\tauv/2).
The difference of the two quantities \tauv\ and $y$ from \avmuse\ is negligible at low \tauv,
but becomes significant ($\gtrsim$25\%) for \avmuse$\gtrsim 0.3-0.5$\,mag.
At \av\ higher than this, \tauv\ becomes clearly the better tracer of dust opacity since the assumption
of a foreground screen progressively fails at high dust columns.
Our preference for \tauv\ is also supported by the comparison with Galactic GMCs as
discussed in the next Section (see Fig. \ref{fig:avco}).

\subsection{CO brightness temperature and visual extinction at 40\,pc resolution}
\label{sec:coav_tauv}

Figure \ref{fig:avco} shows the 12-m data plotted in the left panel against \avmuse, calculated
from \ebv\, 
and in the right against \tauv\ inferred from Eq.\,\eqref{eqn:lambert}. 
Both data sets have been rebinned to 2\arcsec\ (corresponding to $\approx$\,40\,pc,
see Table \ref{tab:sample}).
Like the trend of \ico\ with \mytau\ at 250\,pc resolution,
\ico\ at 40\,pc also varies systematically with \av\ and \tauv,
albeit with significant scatter.

Figure \ref{fig:avco} shows that
the 40\,pc CO-emitting regions have higher 
brightness temperatures 
than the $\sim$250\,pc regions shown in (the right panel of) Fig. \ref{fig:gastau}.
Indeed, the surface brightness of CO in our targets at the two resolutions sampled here is very different.
Comparison of Fig. \ref{fig:avco} and Fig. \ref{fig:gastau} illustrates 
that \ico\ of the brightest CO-emitting regions 
that are visible at high dust columns, traced by either \mytau\ or \tauv, 
are a factor of $\sim$10 higher at 40\,pc than 250\,pc.
This is consistent with beam dilution of a given molecular gas surface density 
because the ratio of the beam area is $(250/40)^2 \sim 40$.
Such behavior is not surprising because we expect CO
to be highly clumped on these scales, much more so than \hi\
\citep{leroy13}.
It also tells us that for probing CO emission in low-metallicity galaxies,
large beams, with single-dish observations at the
extreme end, are not as effective as beams on scales of (at least) a few tens of parcecs;
they are unable to trace the high-density CO columns. 
This means that single-dish efforts to detect CO at low metallicity may be doomed to fail
at the outset because of beam dilution, a point that was also made by \citet{rubio93}
in a pioneering paper using SEST observations of the SMC.
We will explore this further in Sect. \ref{sec:comparison}.

Also shown in Fig. \ref{fig:avco} are 
the CO velocity-integrated brightness temperatures of the GMCs in our Galaxy studied at $\sim$1\,pc resolution by \citet{lee18}.
They infer \av\ from \ebv\ (also using \av/\ebv\,=\,3.1) derived from the dust
optical depth at 850\,\micron\ $\tau_{850}$ as measured by \textit{Planck}.
(Sub)Millimeter/far-infrared optical depths such as $\tau_{850}$ (and even \mytau) 
make emission at these wavelengths 
optically-thin tracers of dust, better probing the entire line of sight through
dust and gas.
Comparison of the \citet{lee18} GMCs with our CO data is enlightening.
If we assume that \avmuse\ (left panel in Fig. \ref{fig:avco}) 
measured from the Balmer decrement is identical to \av\ inferred from $\tau_{850}$,
at a given \av, we would deduce that the GMC CO 1\,pc column densities are lower than those found
for our low-metallicity starbursts on $\sim$40\,pc scales.
However, in the right panel where \tauv\ is plotted rather than \avmuse,
for a given \tauv,
the CO integrated brightness temperatures of the GMCs (with their \av\ taken to be equivalent to our \tauv)
coincide very well with the CO emission observed at 40\,pc resolution.
This is an indication that the assumption that the obscuring
dust measured by \avmuse\ is mixed with the ionized gas, with an optical depth of \tauv,
is better than assuming that the dust is a foreground screen.
This is true for the low-metallicity starbursts studied here, but the differences
would even be more profound for galaxies with larger dust content.

However, that the 40\,pc integrated brightness temperatures of our targets are similar the GMCs in the 
MW observed at 1\,pc resolution by \citet{lee18} is surprising.
The GMCs would have been expected to show significantly
higher surface brightness within the much smaller beam.
Instead, our finding would imply that the CO clumps within the 40\,pc beam are of much higher surface brightness
than the GMCs measured by \citet{lee18}, consistent with the
idea that the only CO-emitting regions that remain at low metallicity are the dense cores
that are sufficiently shielded to avoid photodissociation. 


As discussed by \citet{lee18}, there are at least two features in the \ico\ versus \av\ (\tauv)
relation that would be expected from a theoretical point of view.
The first is a CO-formation visual extinction threshold, below which 
CO emission plummets because of photodissociation of the CO molecules
\citep[e.g.,][]{vandishoeck88,bell06,visser09,glover11,shetty11}.
This would be reflected in the presence of carbon 
emitting as C$^+$ or \ci\
in parts of the molecular clouds where CO is insufficiently shielded
\citep[e.g.,][]{hollenbach97,papa04,wolfire10}.

Another feature expected theoretically is a flattening at high \av\ (\tauv) where CO emission
saturates, becoming optically thick \citep[e.g.,][]{glover11,shetty11}.
Such a feature is observed in some GMC ensembles in the MW,
where there is clear plateau for \av$\gtrsim$1\,mag \citep[e.g.,][]{lombardi06,pineda08}.

\begin{figure*}[t!]
\hbox{
\includegraphics[angle=0,width=0.495\linewidth]{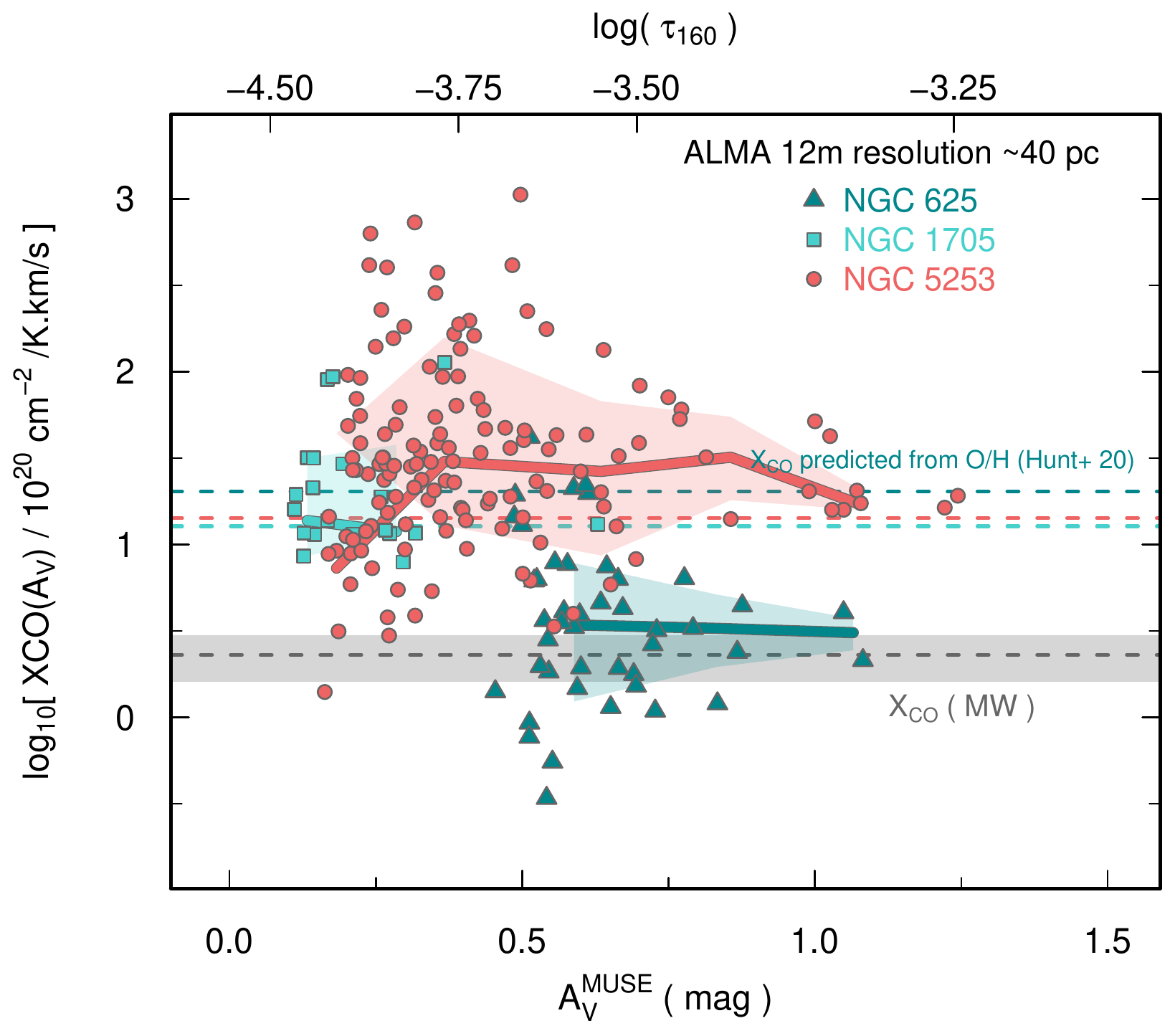}
\hspace{0.01\linewidth}
\includegraphics[angle=0,width=0.495\linewidth]{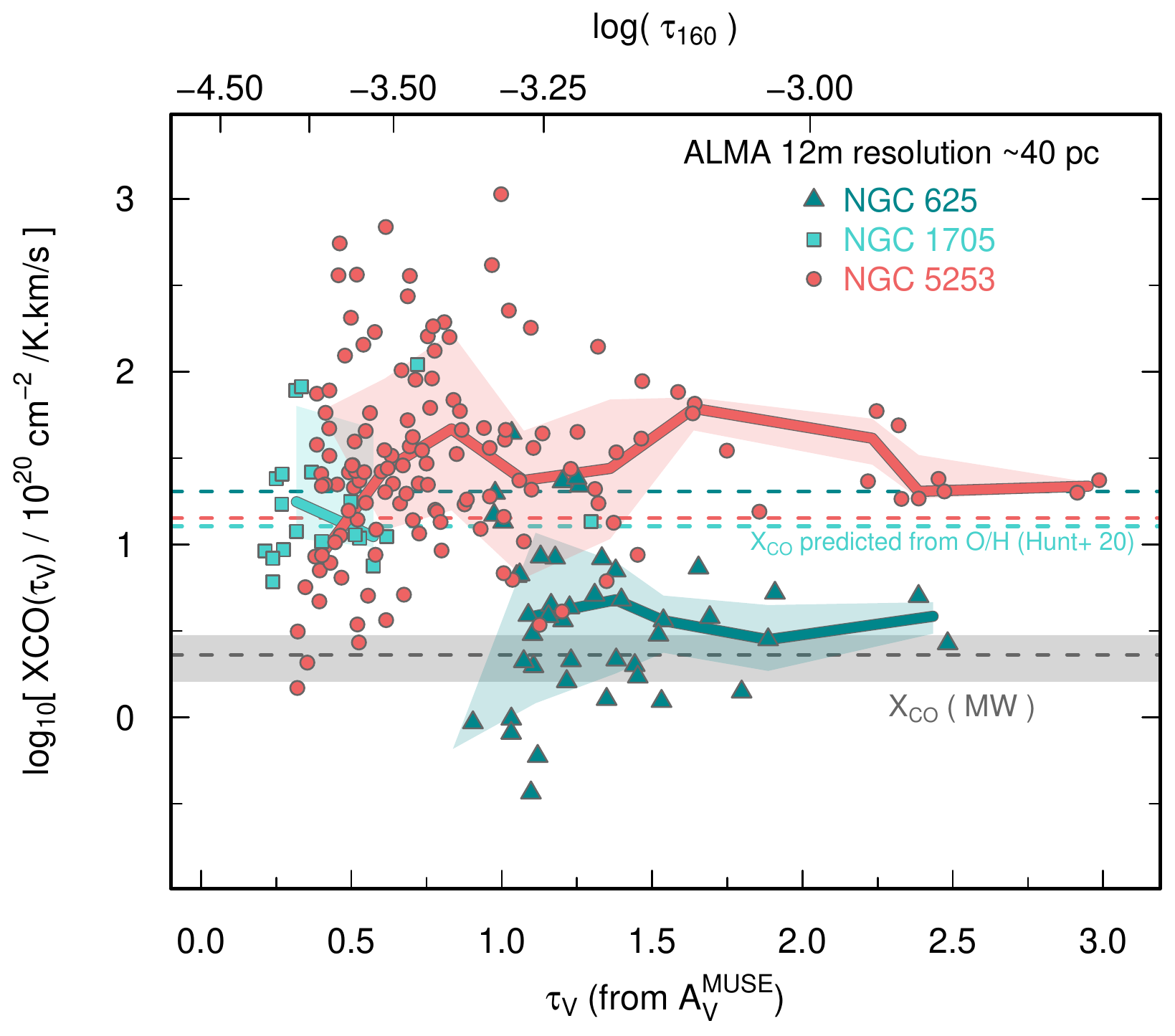}
}
\caption{\xco\ estimated from Eq.\,\eqref{eqn:h2av} is plotted as a function of \av\ in the left panel,
and of \tauv\ in the right.
As in Fig. \ref{fig:xcotau}, we include as horizontal dashed lines,
\xco\ that would be estimated from the metallicity dependence found by \citet{hunt20},
as \xco\,$\propto$\,$Z^{-1.55}$.
The MW \xco\,=\,$2.3\,\times\,10^{20}$\cmtwo/\kkms\ is also given together with the uncertainties. 
Heavy curves show the medians for each galaxy binned in \avmuse\ (left)
and \tauv\ (right) with shaded regions showing the 1$\sigma$ standard deviations.
\label{fig:xcoav}
}
\end{figure*}

To test for these features in the 40\,pc resolution \ico,\ \av\ (\tauv) relation for low-metallicity
starbursts, we have fit the points in Fig. \ref{fig:avco} with a function similar to that of
Eq.\,\eqref{eqn:radtranhi} of the form: 
\begin{equation}
I_\mathrm{CO}\,=\,\mathrm{W0}_\mathrm{CO}\ \left( 1 - e^{-k_\mathrm{CO}\,(A_\mathrm{V} - A_V^\mathrm{CO,thresh})} \right)\quad , 
\label{eqn:radtranco}
\end{equation}
where \woco\ is the 
velocity-integrated CO brightness temperature at saturation;
\kco\ defines the 
transition threshold for the dust optical depth \tauv\ where saturation sets on;
and \avcothresh\ gives the onset for CO emission.
As before, only points with signal-to-noise (S/N)\,$\geq$\,3 are included in the fits.
The best fits to this function are shown as curves in Fig. \ref{fig:avco};
in the right panel we have substituted \tauv\ for \av\ in Eq.\,\eqref{eqn:radtranco}.
The results of the fits and the visual inspection of the plots show that
the saturation feature at high \tauv\ is not visible in our data;
\woco\ and \kco\ are ill-determined both for the plots using \av\ and those with \tauv.
Instead, the \avcothresh\ parameter is somewhat better determined with:
\begin{equation}
A_V^\mathrm{CO,thresh}\,=\,\begin{cases}
0.09\,\pm\,0.06\,\mathrm{mag} & \text{NGC\,625} \\
0.10\,\pm\,0.02\,\mathrm{mag} & \text{NGC\,1705} \\
0.11\,\pm\,0.05\,\mathrm{mag} & \text{NGC\,5253} 
\end{cases}
\label{eqn:avcothresh}
\end{equation}
The results for \avcothresh\ using \tauv, rather than \av, have somewhat higher values, although with
slightly larger uncertainty.
In any case, our 40\,pc CO data show no evidence for a flattening at high \tauv,
but some tentative indication of an \av\ CO threshold of $\sim 0.1$\,mag.
This is a much lower value than given by the simulations of \citet[e.g.,][]{glover16} that
predict a steep reduction for \av$\la$\,1\,mag; however, their simulations
are at much finer (0.06\,pc) resolution than our observations, so may not be directly
comparable. 
Our value of \avcothresh\ $\approx$\,0.1\,mag is also roughly ten times lower than
found by \citet{pineda08} with \avcothresh\ generally $\gtrsim$\,1\,mag for their
sample of (solar metallicity) MW GMCs. 

\subsection{The \xco\ factor and visual optical depth}
\label{sec:xco_tauv}

We can now infer the \xco\ conversion factor at $\approx$\,40\,pc
resolution from \tauv\ (or \av), similar to what
we have done at lower resolution ($\approx$\,250\,pc) with \mytau\ in the previous sections.
We write:
\begin{equation}
N_\mathrm{H2}^\mathrm{A_V}\,=\,\frac{1}{2}\,\left( \frac{N_H}{A_V} \right) \left[2\,A_V^\mathrm{MUSE}  - A_V^\mathrm{HIcrit}\ \right],
\label{eqn:h2av}
\end{equation}
where $N_H$/\av\ is the total column density of hydrogen atoms in either atomic or molecular form
relative to visual extinction \av.
This approach follows \citet{lee18}, but they probe \av\ with dust emission, so we must
make different assumptions for the dependence of \Nhtwo\ on \avmuse.
The multiplicative factor of 2 for \avmuse\ in Eq.\,\eqref{eqn:h2av} is because our estimate  of
\avhicrit\ relies on \mytau\ which probes the entire
line of sight, thus sampling the shielding layer on both the front and back sides of the medium;
conversely \avmuse\ is calculated assuming a foreground screen, thus sampling only the front side
\citep[c.f.,][]{lee18}.
The \avhicrit\ term 
takes into account the transition from \hi\ to \htwo\ at high dust and gas columns
\citep[e.g.,][]{bigiel08,krumholz09,sternberg14};
it can be estimated from \khi\ in the \hi\ formulation of radiative transfer in Eq. \eqref{eqn:radtranhi},
after converting the units to \av\ mag$^{-1}$.
The analogous equation for \tauv\ would be:
\begin{equation}
N_\mathrm{H2}^\mathrm{\tau_V}\,=\,\frac{1}{2}\,\left( \frac{N_H}{\tau_V} \right) \left[\tau_V - \tau_V^\mathrm{HIcrit}\ \right]\ ,
\label{eqn:h2tauv}
\end{equation}
where $\tau_V^\mathrm{HIcrit}$ is defined as 1.086\,\avhicrit,
and $N_H$/\tauv\ is the equivalent of $N_H$/\av\ but in units of \tauv, under the assumption that the denominator measures
the extinction in the dust along the line of sight.
Equation \eqref{eqn:h2tauv} could be seen as inconsistent with our conclusion that the screen 
geometry for dust extinction is not an accurate description; 
however, 
$ \tau_V^\mathrm{HIcrit}$
describes the impact of the conversion of \htwo\ to \hi\ on the extinction dependence
of \Nhtwo, thus is not strictly related to the geometry of the dust relative to the ionized gas.

$N_H$/\av\ is a critical factor in our determination of \xco\ from \av\ on $\sim$40\,pc scales, and it relies on several assumptions.
Along diffuse lines of sight in the MW, \citet{bohlin78} find
$N_H$/\av\,=\,$1.87\,\times\,10^{21}$\,\cmtwo\,mag$^{-1}$. 
Assuming \av/\mytau\,=\,2180 
\citep{hensley21},
we can also estimate $N_H$/\av\ from \deltadgr\ at $\sim$250\,pc resolution 
[see Fig. \ref{fig:gastau}, Eq.\,\eqref{eqn:taudgr}] measured for our targets.
We find:
\begin{equation}
N_H/A_V\,=\,\begin{cases}
(6.9\,\pm\,0.2)\,\times\,10^{21}\,\mathrm{cm}^{-2}\,\mathrm{mag}^{-1} & \text{NGC\,625} \\
(1.6\,\pm\,0.5)\,\times\,10^{22}\,\mathrm{cm}^{-2}\,\mathrm{mag}^{-1} & \text{NGC\,1705} \\
(1.4\,\pm\,0.4)\,\times\,10^{22}\,\mathrm{cm}^{-2}\,\mathrm{mag}^{-1} & \text{NGC\,5253} 
\end{cases}
\label{eqn:gasav}
\end{equation}
As shown in the right panel of Fig. \ref{fig:xcotau}, there is a relatively large spread in these estimates,
ranging from $\sim$0.2\,dex to 0.5\,dex. 
In any case, at face value, the total gas-to-extinction ratios we measure for these low-metallicity
galaxies 
are $\approx\,4-8$ times higher than the MW value \citep[e.g.,][]{bohlin78}, 
qualitatively consistent with a linear metallicity scaling. 

Another critical factor in our 40\,pc formulation of \Nhtwo\ is \avhicrit, which 
is directly related to \khi\ in Eq.\,\eqref{eqn:radtranhi} as is the inflection where 
atomic gas transitions to \htwo.
Thus, 
with \av/\mytau, we
calculate values of \avhicrit\ for the low-metallicity starbursts:
\begin{equation}
A_V^\mathrm{HIcrit}\,=\,\begin{cases}
0.90\,\pm\,0.21\, \mathrm{mag} & \text{NGC\,625} \\
0.20\,\pm\,0.05\, \mathrm{mag} & \text{NGC\,1705} \\
0.32\,\pm\,0.04\,\mathrm{mag} & \text{NGC\,5253} 
\end{cases}
\label{eqn:avhicrit}
\end{equation}
For NGC\,1705 and NGC\,5253, the \avhicrit\ values are fairly low,
as can also be seen by the curves in Fig. \ref{fig:gastau}, while 
for NGC\,625, the value is roughly 3 times higher, although also 4-5 times more uncertain.
The values for NGC\,1705 and NGC\,5253 of 0.2-0.3\,mag are consistent with \avhicrit\,=\,0.2\,mag
for a fixed radiation field found by \citet{draine78}, 
and by \citet{krumholz09}, depending on dust properties.

Finally, both Eqs.\,\eqref{eqn:gasav} and \eqref{eqn:avhicrit} rely on the conversion of \av\ to \mytau.
This conversion is uncertain, and different groups adopt a range of values.
\citet{leroy09} use \av/mag\,=\,1910\,\mytau,
while \citet{lee15} adopt \av/mag\,=\,2200\,\mytau.
The higher dust temperatures and lower dust optical depth found by \citet{planck14}
lead to \av/mag\,=\,3246\,\mytau, assuming \ebv/\av\,=\,3.1 as done here.
Here,  we adopt \av/mag\,=\,2180\,\mytau\ \citep{hensley21}, 
close to the \citet{lee15} value, but still somewhat subject to uncertainty.

These uncertainties in the ingredients for the calculation of \Nhtwo\
make our inference of \xco\ subject to various caveats.
Nevertheless, results are encouraging. 
Figure \ref{fig:xcoav} shows \xco\ computed 
using \Nhtwo\ inferred from Eqs.\,\eqref{eqn:h2av} and \eqref{eqn:h2tauv};
the left panel shows \xco\ versus \av\ and the right versus \tauv.
Also shown in the figure are the expectations from the global metallicity dependence of \xco\ from
\citet[][see also \citealt{accurso17}]{hunt20}. 
There is possibly less scatter in the right panel (\tauv), at least for NGC\,5253,
but overall the two plots show mutually consistent values of \xco. 
While the median curve for NGC\,5253 falls roughly where expected for consistency with
a global metallicity dependence of \xco, NGC\,625 falls fairly close to the MW.
This is unexpected given that NGC\,625 has the lowest O/H of our sample, although the 
global O/H variations among the three targets are not large.
The reason why NGC\,625 shows \xco\ at 40\,pc resolution close to the Galactic value
is the high value of \avhicrit\ in Eq.\,\eqref{eqn:avhicrit}.
Given the relatively low signal-to-noise on the measurement of \avhicrit\ for NGC\,625,
we have experimented with lower values of \avhicrit$\,\approx 0.2-0.3$, and find that NGC\,625 would fall closer
to the global metallicity-dependent \xco\ expectations, consistent with the other galaxies.
Nevertheless, this would be a deviation of $\ga\,3\sigma$, so we continue to consider
the nominal \avhicrit\ value for NGC\,625.

\begin{figure}[t!]
\centerline{
\includegraphics[angle=0,width=0.99\linewidth]{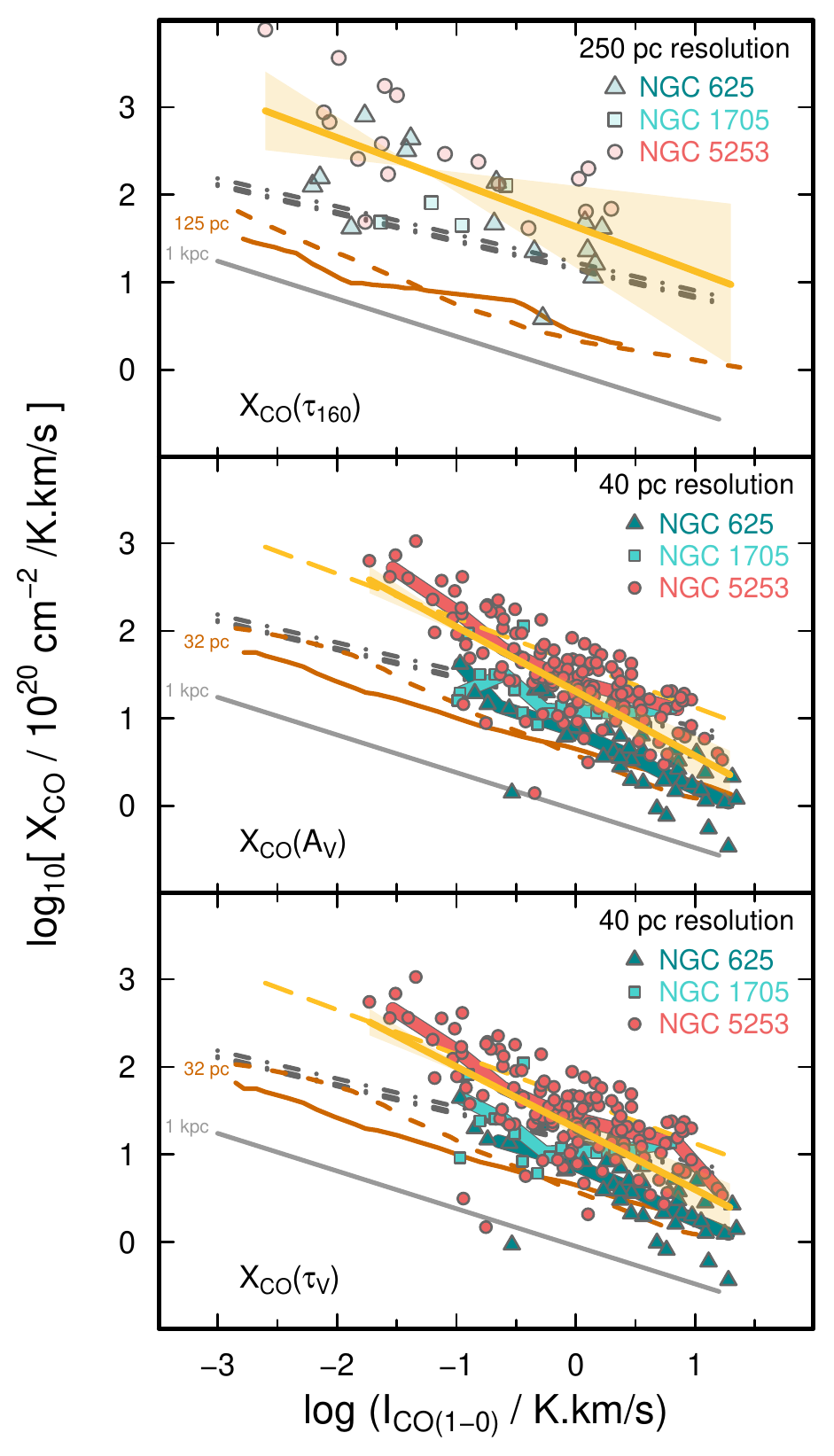}
\vspace{-\baselineskip}
}
\caption{\xco\ estimated from Eq.\,\eqref{eqn:h2dgr} versus \ico\ in the top panel
(250\,pc resolution);
\xco\ from Eq.\,\eqref{eqn:h2av} in the middle (40\,pc resolution);
and from Eq.\,\eqref{eqn:h2tauv} in the lower (40\,pc resolution).
Also shown as dark-grey dot-dashed lines are the predictions from \citet{nara12} for the different
metallicities of our targets. 
The 1\,kpc trend from \citet{hu22}, valid for all metallicities, is shown as a light-grey
solid line (labeled ``1\,kpc''). 
The brown solid (dashed) curves correspond to 0.3\,\zsun\ (\zsun) metallicities
at 125\,pc resolution (top panel) and at 32\,pc (middle and lower ones).
The 0.3\,\zsun\ curves are the best approximation to the metallicities of our target starbursts.
The robust fit in logarithmic space to our data is shown as a dark-yellow line
with 1$\sigma$ excursions by the yellow shaded area;
in the lower panels, the dotted dark-yellow line reproduces the 250\,pc best fit shown in the upper panel. 
The trends seen in our data follows the predictions of \citet{hu22}, in terms of excursion, but we
find a steeper slope and an overall higher normalization. 
We attribute this to the theoretical \huwco\ not being the exact same quantity as observed \ico,
so we approximate \huwco\ as \ico, which is a better approximation at high \ico, when \xco\
is low, but which is far from true at low \ico, when \xco\ is high.
See text for more details.
\label{fig:xcoico}
}
\end{figure}

\section{Comparison of large- and small-scale \xco}
\label{sec:comparison}

Figures \ref{fig:xcotau} and \ref{fig:xcoav} present two different pictures
of the behavior of \xco. 
At $\sim$250\,pc resolution, conversion factors \xco\ show large scatter,
and are generally higher than what would be predicted from a simple metallicity dependence. 
Instead, with 12-m data at $\sim$40\,pc resolution, \xco\ values are consistent with, or even lower than,
the predicted metallicity dependence, and, as at 250\,pc, there is large scatter.



The large scatter and inconsistencies between the two data sets at different resolutions
could also point to a dependence on additional parameters beyond metallicity.
In particular, the comparison between Figs. \ref{fig:gastau} and \ref{fig:avco} 
shows that the larger ACA beam, rebinned here to $\sim$250\,pc, does not sample
the brighter CO emission that emerges, instead, at $\sim$40\,pc with the 12-m array.
In the larger beam, there are more CO-faint regions than bright regions; 
consequently,  \xco\ is larger. 

Trends of the \htwo\ conversion factor \xco\ with the observed CO velocity-integrated brightness temperature \ico\ 
are also expected theoretically.
Measured \ico\ is mainly governed by mean density along the line-of-sight
\citep[e.g.,][]{gong18}, since CO forms at higher volume densities ($n\,\ga 100-200$\,\cmthree)
than \htwo\ ($n\,\sim 10-100$\,\cmthree).
This implies that, for environments with similar metallicities, \ico\ can be taken as a 
proxy of column or volume densities \citep[e.g.,][]{hu22}, and related to \xco, 
either inferred from simulations, or measured empirically, as we have done here. 
It is likely that the underlying reason why at 40\,pc the fainter regions have
larger \xco\ stems from the radiative transfer of CO; the faint CO emission regions are heavily
sub-thermally excited and may even become optically thin due to the low volume/column densities
they are associated with. 
Sub-thermal excitation would make
them less luminous in the CO line per hydrogen atom,
compared to the higher density/column density gas with 
excitation temperatures approaching kinetic temperatures 
\citep[e.g.,][]{hu22}.

We have tested this hypothesis by plotting in Fig. \ref{fig:xcoico} \xco\ 
inferred above at the two different resolutions
against \ico\ of the relevant data sets.
An anti-correlation of \xco\ with \ico\ emerges clearly from both sets of data, with 
Pearson correlation coefficients of $-0.65$ (250\,pc) and $-0.7$ (40\,pc).
This correlation is almost certainly causing at least part of the scatter in our \xco\ estimates
shown in Figs. \ref{fig:xcotau} and \ref{fig:xcoav}.
We have performed a robust fit in logarithmic space to \xco\ with \ico\ shown
as a yellow regression line in Fig. \ref{fig:xcoico}:

\medskip
\noindent
$\log_{10}\,X_\mathrm{CO}\,=\,$ 
\vspace{-0.5\baselineskip}
\begin{equation}
\quad \begin{cases}
(-0.51\,\pm\,0.12)\,\log_{10}\,I_{\mathrm{CO}} + (1.64\,\pm\,0.15) & \text{250\,pc, $\tau_{160}$} \\
(-0.74\,\pm\,0.05)\,\log_{10}\,I_{\mathrm{CO}} + (1.31\,\pm\,0.03) & \text{40\,pc, $A_V$}\\ 
(-0.70\,\pm\,0.05)\,\log_{10}\,I_{\mathrm{CO}} + (1.30\,\pm\,0.03) & \text{40\,pc, $\tau_V$} 
\end{cases}
\label{eqn:xcoicofit}
\end{equation}
The best-fit parameters for \xco\ estimated from the trends with \av\ and \tauv\ are similar, and 
also show a comparable mean dispersion.
The slope and intercept of the 250\,pc resolution data differ (at a $\sim 2\sigma$ level) 
from those of the 40\,pc fits, with the larger resolution data having flatter slope and a
larger intercept.
We expect these relations to be accurate for galaxies of around this metallicity, $\sim\,\frac{1}{3}$\zsun, 
but for higher and lower abundances, and different resolutions,
more data are needed to establish similar empirical relations.

\citet{nara12} examined the dependence of \xco\
on metallicities, gas temperatures, and velocity dispersion of the clouds.
Using post-processing of the simulated galaxies by \citet{nara11} to determine
the physical and chemical state of the molecular gas, 
\citet{nara12} also find that \xco\ is well correlated with \ico, and is a function of 
physical conditions within the clouds.
They find a simple power-law dependence of \xco\ on \ico\ and metallicity (shown 
as dot-dashed lines in Fig. \ref{fig:xcoico}), 
with an \ico\ power-law index (slope) of $-0.32$.
Our observations are inconsistent with this slope, as it is significantly lower than
the power-law index of $-0.7$ (40\,pc) found in our data [Eq.\,\eqref{eqn:xcoicofit}];
the $-0.5$ slope of the 250\,pc data is somewhat more consistent
with the Narayanan et al. result, although still $\ga\,2\sigma$ steeper.

To study CO emission at pc-scale resolutions, and measure \xco,
\citet{gong18,gong20} conducted solar-metallicity magneto-hydrodynamic galactic disk simulations 
with post-processing that includes a chemical network and radiative transfer.
Their analysis finds a slight anti-correlation between \xco\ and \ico, but with a very
shallow power-law slope of $-0.011\,\pm\,0.005$.
Such a shallow slope is comparable to that of $-0.05$ found empirically by \citet{remy17}
for clouds in the MW anti-center, but much flatter than the well-defined dependence
for our low-metallicity starbursts.

Recent work by \citet{hu22}, based on hydrodynamical simulations, chemistry, and radiative transfer,
predict that \xco\ depends not only on metallicity but also on the spatial resolution
of the maps and on the intensity of the emission line \ico. 
In  Fig. \ref{fig:xcoico},
we have reproduced their 32\,pc and 125\,pc curves for \zsun\ (dashed curves) and 0.3\,\zsun\ (solid ones)
and their kpc-resolution trend that is valid for all metallicities.
Our best-fit power-law slope of $\sim -0.5$ for 250\,pc resolution is not far from their value
of $-0.43$ (valid at 1\,kpc for roughly all metallicities). 
At higher resolution,
the individual trends found by \citet{hu22} for solar and 0.3\,\zsun\ metallicities with region sizes from 32\,pc
to 125\,pc (shown by brown curves in Fig. \ref{fig:xcoico}) show 
similar slopes, somewhat shallower than what we measure.
Nevertheless, the agreement between our observations and the predictions of \citet{hu22} is reasonably good
especially at high \ico.

Because of the expectation of more CO-dark gas in larger beams and at low metallicity,
it is somewhat surprising that the Hu et al. simulations find that \xco\ is smaller with increasing
beam size and decreasing metallicity.
This can be understood because of the distinction between the theoretical \huwco\ in \citet{hu22} and 
the observed quantity \ico.
The cosmic microwave background (CMB) temperature, \tbg, must be subtracted to calibrate
observed antenna temperature, \ico; thus \ico\ will be identically 0 for excitation temperature
\Tex\,=\,2.73\,K, 
while \huwco\ is only zero when \Nco\ is zero, whatever \Tex.
This implies that their CO-dark gas radiates even at low \htwo\ density, but in the
observed signal \ico, it does not. 
Their conclusion therefore is expected to underestimate the amount of CO-dark gas.

The results of \citet{hu22} suggest that observed velocity-integrated \tb\ below $\sim$10\,\kkms\ are optically
thin, and thus directly trace CO column density \Nco\ (for \Nco\,$\la\,10^{16}$\,\cmtwo).
This would imply that virtually all our observations are optically thin since there are very few
resolution elements brighter than this limit.
Most of the variation of \xco\ is expected to occur in the optically-thin regime, as it is regulated
by CO abundance and its formation by the transition from atomic carbon.
Our observations are apparently consistent with these expectations, 
since over a roughly 3 orders of magnitude change in \ico, \xco\ also varies by a similar factor.
Nevertheless, at the lowest \ico\ (lowest column densities),
at both the 250\,pc and 40\,pc resolutions \xco\ exceeds by a factor of 10 or so 
the predictions of the Hu et al. simulations.
Also, the difference we find between 250\,pc and 40\,pc resolution seems not to be a vertical shift,
but rather a continuation to lower surface brightnesses as shown in Fig. \ref{fig:xcoico}.
Since our targets are all at roughly the same 1/3\,\zsun\ abundance, we are unable to test
observational variation with metallicity.
In any case, overall, our observations agree with \citet{hu22} that 
\xco\ is a multi-variate function of three observables, \ico, metallicity, and beam size.

\section{Discussion and conclusions}
\label{sec:conclusions}

Perhaps the most significant caveat in our methodology is the assumption of a single
DGR across each of the galaxies, independently of spatial scale.
Rigorously speaking, \xco\ and the DGR should be measured independently within each
spatial element, as done by \citet{sandstrom13} on $\sim$kpc scales.
In the study of the SMC by \citet{bolatto11}, on $\sim$200\,pc scales,
the DGR varies by $\sim$0.4\,dex along the 
SMC Bar, the Wing, and within the star-forming regions N83/N84.
The assumption of a constant DGR is intimately related to our estimate of \xco,
at least where CO emission dominates over \hi.
If we assume a constant metallicity-dependent value for \xco, 
as shown in the right panel of Fig. \ref{fig:xcotau},
we infer decreasing DGRs toward higher dust column densities.
This effect may be real, but for our targets, with relatively few 250\,pc resolution elements, 
it is impossible to assess spatial variations in the DGR,
and disentangle them from variations in \xco.

\subsection{Additional considerations}
\label{sec:considerations}

Putting this caveat aside, our analysis (Figs. \ref{fig:xcotau}, \ref{fig:xcoav})
suggests that \xco\ is not correlated with either
dust optical depth (\mytau) or visual extinction/optical depth (\av, \tauv).
This would be in disagreement with the individual cloud models by \citet{glover11} and 
\citet{shetty11} who find a threshold effect for \av$\la$1\,mag.
However, our observations do not resolve individual clouds, so that the discrepancy
could be simply a question of spatial resolution.
As pointed out by \citet{gong18}, a resolution of $\la$2\,pc is 
necessary to resolve the average \xco\ within molecular clouds for conditions similar
to the solar neighborhood; at lower metallicity, given the small $\sim$\,pc scales,
possibly even better resolution would be required.

Various recent simulations find that \xco\ is related to \ico\ and also to spatial
scale, that is to say the region over which the parameters are averaged \citep[e.g.,][]{gong18,gong20,hu22};
observationally, spatial scale corresponds to beam size, or in our case the scale
imposed by rebinning.
We would argue that these parameters, \ico\ and spatial scale, are partially dependent, 
as shown in Fig. \ref{fig:xcoico};
relative to the 40\,pc data, the data at 
250\,pc extend to fainter CO emission, but do not reach the brighter CO emission. 
It is very difficult to achieve high values of \ico\ with large beams as can also be
appreciated by the comparison of Figs. \ref{fig:gastau} and \ref{fig:avco}.

Moreover, as might be intuitively expected,
the fraction of CO-faint or dark gas is expected to increase with increasing beam size \citep{gong18,hu22}.
Part of this depends on the \ico\ 
lower limit (imposed on the simulations, or
observationally by signal-to-noise), and part is due to the erosion of the CO-emitting
volumes by photodissociation.
Photodissociation results in regions that are dominated by carbon in atomic or ionized
phases, \ci\ and \cii, rather than CO, and these regions can be much more extended
than the CO emission.

\subsection{Summary}
\label{sec:summary}

We have derived \xco\ in three low-metallicity starbursts, NGC\,625, NGC\,1705, and NGC\,5253, 
using ALMA 12-m data at $\sim$40\,pc resolution and ACA observations at $\sim$250\,pc resolution, together
with a dust-based method to infer \Nhtwo. 
Our work can be summarized as follows:
\begin{itemize}
\item
We fit the flux in \hers\ PACS maps to two-temperature MBBs in order to quantify
dust optical depth \mytau, and compare it with \hi\ maps and CO maps measured with ACA, 
all rebinned to $\sim$250\,pc resolution.
\item
By approximating the trends of \hi\ and CO versus \mytau\ with linear functions 
and ones that reflect radiative transfer, we have quantified the DGR for each galaxy and the
level at which \hi\ is converted to \htwo.
This makes it possible to estimate \xco\ at 250\,pc resolution and compare it
to what would be expected from global metallicity estimates for \xco.
We find that the \xco\ values at 250\,pc resolution can be as much as an order
of magnitude larger than what would be expected from metallicity trends.
There is also large scatter implying, possibly, spatial variation of \xco\ at
this resolution.
\item
At $\sim$40\,pc resolution, roughly the native 12-m ALMA beam,
we compare CO measured with the ALMA 12-m array and \av, estimated from 
the VLT/MUSE maps of the Balmer decrement.
We eschew the assumption of \av\ as a foreground screen, and instead infer \tauv\
from \av\ by assuming that the 
extinguishing dust is well mixed with the ionized gas.
Overall, this turns out to be the better assumption,
as supported by the closer agreement of our data with the Galactic GMCs (see Fig. \ref{fig:avco}).
By fitting the CO versus \av\ or \tauv\ trends, we have been able to identify in our
data an \av\ threshold for CO emission of $\sim$0.1\,mag for all three galaxies.
This is significantly smaller than the $\sim$1\,mag \av\ threshold found in
simulations \citep[e.g.,][]{glover16} or observations of MW molecular clouds
\citep[e.g.,][]{pineda08}.
\item
Incorporating the DGR measured at 250\,pc resolution, and the \av\ transition from \hi\
to \htwo\ at high dust optical depths \avhicrit, we are able to infer \xco\ on 40\,pc
scales.
Results are generally consistent with the expectations from a global metallicity-dependence
of \xco, except for NGC\,625 which shows values 
typically encountered in environments closer to solar metallicity.
This almost certainly stems from the higher \avhicrit\ found for this galaxy ($\sim$\,0.9\,mag
relative to $\sim\,$0.2--0.3\,mag), and when we estimate \xco\ using these lower values
also for NGC\,625, \xco\ becomes consistent with metallicity expectations.
As for the 250\,pc \xco\ analysis, there is large scatter.
\item
Finally, we compare \xco\ to CO brightness temperature \ico, predicted by simulations to be closely related.
Figure \ref{fig:xcoico} shows that the two quantities are highly anti-correlated, 
with power-law slopes steeper than those found by the simulations 
of \citet{nara12} and \citet{gong18}, 
although possibly consistent
with the slope of $\sim -0.4$ found by \citet{hu22} across a range of resolutions.
Relative to \citet{nara12} and \citet{hu22}, because of the difference in
power-slopes between the data and the simulations, the
data show a higher normalization of the relation between \xco\ and \ico\
for faint CO emission levels.
Despite these differences in details, our observations confirm that \xco\
depends not only on metallicity, but, 
also on beam size, and CO surface brightness \ico.
Such behavior can be attributed to the increasing fraction of CO-dark \htwo\ gas with 
lower spatial resolution (larger beams) and lower surface-brightness emission.
\end{itemize}

Our analysis is limited to three low-metallicity starbursts all of the same metallicity,
$\sim 0.3$\,\zsun.
Thus, we are sampling only a narrow range of sub-solar metal abundance.
Future work would profit from a broader spread in metallicities and spatial resolution, and inclusion of metal-poor
dwarf galaxies that are forming stars more quiescently than the starbursts studied here.

\begin{acknowledgements}
We would like to thank the referee for thoughtful, constructive comments
that greatly improved the paper.
This paper makes use of the following ALMA data: ADS/JAO.ALMA\#2018.1.00219.S. 
ALMA is a partnership of ESO (representing its member states),
NSF (USA) and NINS (Japan), together with NRC (Canada), MOST and ASIAA (Taiwan), 
and KASI (Republic of Korea), in cooperation with the Republic of Chile. 
The Joint ALMA Observatory is operated by ESO, AUI/NRAO and NAOJ.
This work was partly done using GNU Astronomy Utilities (Gnuastro,
ascl.net/1801.009) version 0.15. Work on Gnuastro has been funded by the
Japanese Ministry of Education, Culture, Sports, Science, and Technology (MEXT)
scholarship and its Grant-in-Aid for Scientific Research (21244012, 24253003),
the European Research Council (ERC) advanced grant 339659-MUSICOS, European
Union’s Horizon 2020 research and innovation programme under Marie
Sklodowska-Curie grant agreement No 721463 to the SUNDIAL ITN, and from the
Spanish Ministry of Economy and Competitiveness (MINECO) under grant number
AYA2016-76219-P. 
SGB acknowledges support from the research project PID2019-106027GA-C44 of the Spanish Ministerio de Ciencia e 
Innovaci{\'o}n, and
GV acknowledges support from ANID program FONDECYT Postdoctorado 3200802.
\end{acknowledgements}

\bibliographystyle{aa}
\bibliography{alma_dwarfs}

\appendix
\section{Details of the fits to the PACS data to derive \mytau}
\label{sec:taudetails}

\begin{figure}[h!]
\includegraphics[width=0.9\linewidth]{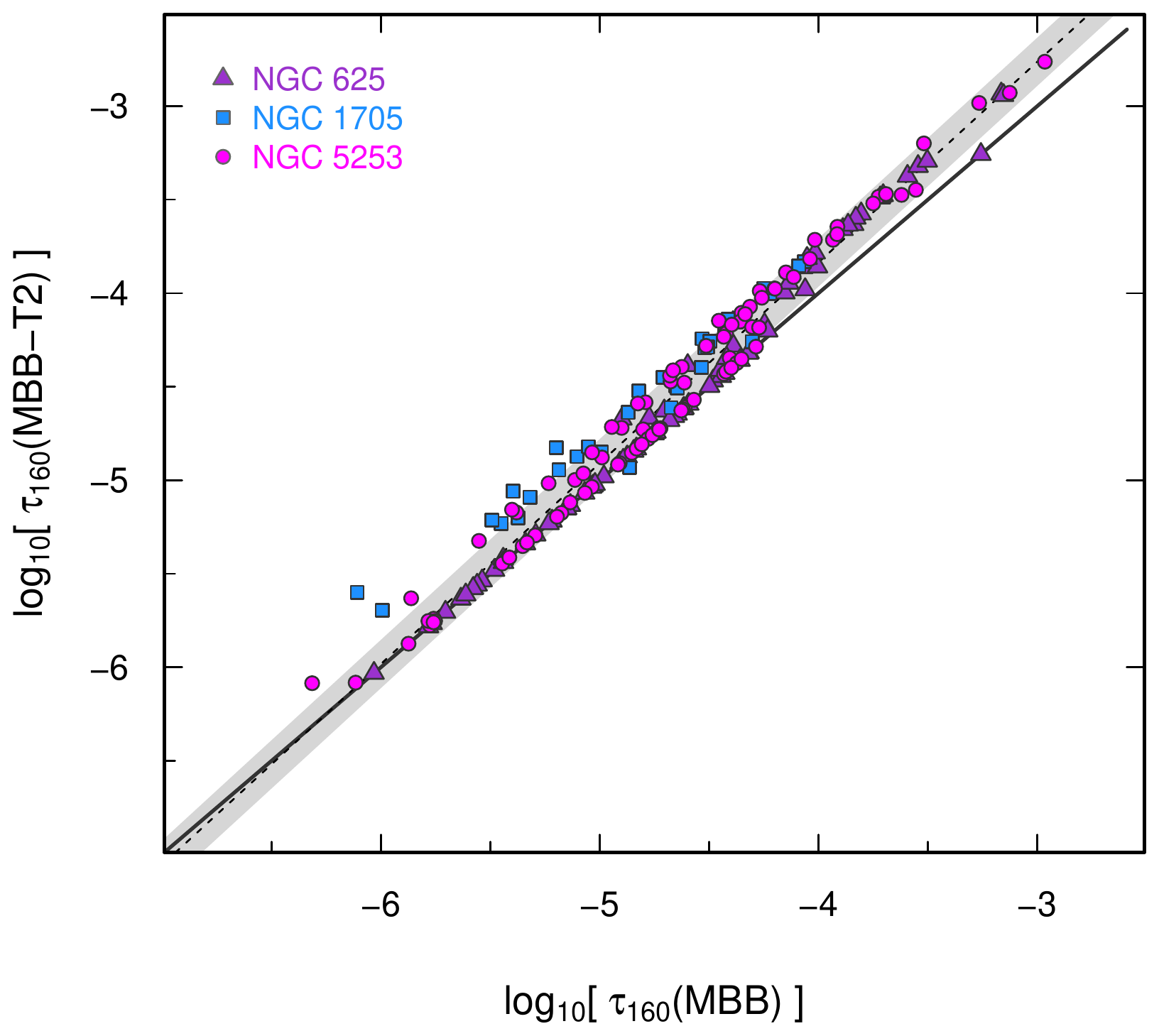}
\caption{\mytau\ from MBB-T2 fits plotted against MBB \mytau. 
$\log\,\tau_{160}(\mathrm{MBB-T2})\,=\,(1.073\,\pm\,0.010)\ \log\,\tau_{160}(\mathrm{MBB}) + (0.46\,\pm\,0.05)$
The scatter of the comparison is $\sim$0.13\,dex; 
the shaded region denotes $\pm\,1\sigma$ relative to the best fit.
In general, \mytau(MBB-T2) is larger than \mytau(MBB), as shown by the non-unit power-law slope and positive
intercept.
\label{fig:comptau} 
}
\end{figure}

The 160\,\micron\ optical depths \mytau\ of the two approaches (namely modified single-temperature
blackbodies, MBB, and modified two-temperature blackbodies, MBB-2T) are compared in Fig. \ref{fig:comptau}.
The scatter of the comparison (with imposed $\gamma\,\geq\,0$) is $\sim$0.13\,dex, 
with \mytau(MBB-T2) larger than \mytau(MBB), as shown by the non-unit power-law slope and positive
intercept of the best (robust) fit.

\begin{figure*}[h!]
\includegraphics[width=0.9\linewidth]{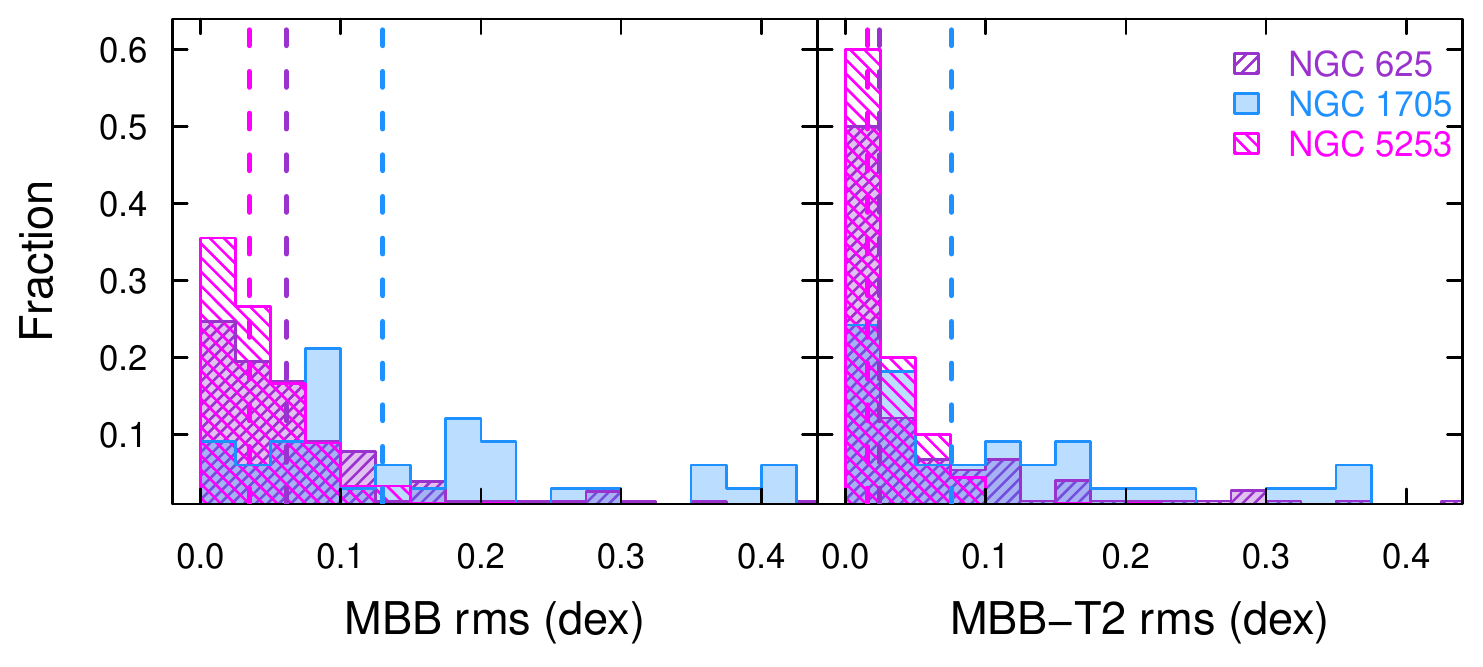}
\caption{Distributions of rms values from MBB (left) and MBB-T2 (right) fits.
Median values for each galaxy are shown by vertical dashed lines.
The rms values of the MBB-T2 fits is generally excellent, possibly due to the extra fitted parameter
and the consequent zero degrees of freedom. 
There are 3 regions each in NGC\,625 and NGC\,5253 not included here
because their rms values exceed 0.5\,dex.
\label{fig:histrms} 
}
\end{figure*}

The distributions of the rms (calculated as the root-mean-square difference in $\log_{10}$ space between the model
and the data) of the two fits are reported in Fig. \ref{fig:histrms}.
Given that the expected uncertainty in the individual PACS photometry is $\sim$10\%,
the rms values of both MBB and MBB-T2 fits are quite good, and, except for NGC\,1705, well within the expected
error.
However, the MBB-T2 fits are overall superior to MBB, possibly due to the 
extra fitted parameter and the consequent zero degrees of freedom. 
Three regions in both NGC\,625 and NGC\,5253 are not included in Fig. \ref{fig:histrms}, because of 
rms $>\,0.5$ due to the very low 70\,\micron\ fluxes and the consequent poor fits.

\begin{figure*}[h!]
\includegraphics[width=0.9\linewidth]{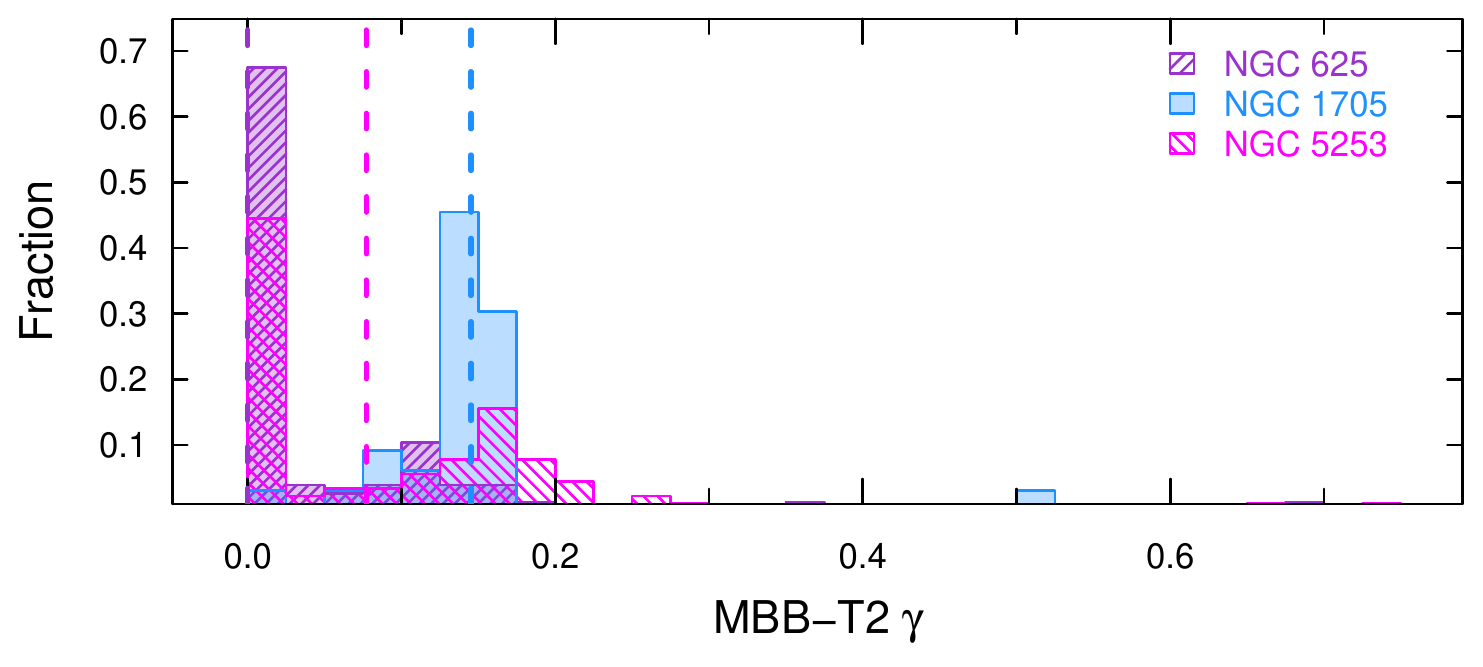}
\caption{Distribution of fitted $\gamma$ from MBB-T2 fits. 
The vertical dashed lines show the median fraction of IR emission due to the warmer component:
(0.00, 0.14, 0.08) for NGC\,(625, 1705, 5253).
\label{fig:histgamma} 
}
\end{figure*}

The overall $\gamma$ distribution (restricted to $0\,\leq\,\gamma\,\leq\,1$) is shown in Fig. \ref{fig:histgamma}.
There are (45, 0, 33) fits for NGC\,(625, 1705, 5253), respectively,
that would have preferred either $\gamma\,<\,0$ (58\%, 0, 37\%).
The median $\gamma\,=\,0.0$ for NGC\,625 is associated with the relatively high
fraction (45/77\,=\,58\%) of regions that would have have been best fit with these anomalous $\gamma$ values.

\end{document}